\documentclass[%
  reprint,
  superscriptaddress,
nofootinbib,
 amsmath,amssymb,
 aps,
 prc
floatfix
]{revtex4-2}

\usepackage{graphicx}
\usepackage{dcolumn}
\usepackage{bm}
\usepackage[version=4]{mhchem}
\usepackage{siunitx}
\usepackage{wasysym}
\usepackage{subfig}
\usepackage{multirow}
\usepackage{siunitx}
\DeclareSIUnit\speedoflight{c}
\usepackage{caption}
    \makeatletter
\def\@fnsymbol#1{\ensuremath{\ifcase#1\or \dagger\or \ddagger\or
   \mathsection\or \mathparagraph\or \|\or **\or \dagger\dagger
   \or \ddagger\ddagger \else\@ctrerr\fi}}
    \makeatother

\begin{document}

\title{A Measurement of Proton, Deuteron, Triton and $\alpha$-Particle Emission after Nuclear Muon Capture on Al, Si and Ti with the AlCap Experiment}

\author{Andrew Edmonds}
\affiliation{Boston University, Boston, Massachusetts, USA}
\affiliation{Lawrence Berkeley National Laboratory, Berkeley, California, USA}
\affiliation{University College London, London, UK}

\author{John Quirk}
\affiliation{Boston University, Boston, Massachusetts, USA}

\author{Ming-Liang Wong}
\altaffiliation[Present address: ]{M. Smoluchowski Institute of Physics, Jagiellonian University, Krak\'{o}w, Poland}
\affiliation{Osaka University, Osaka, Japan}

\author{Damien Alexander}
\affiliation{University of Houston, Houston, Texas, USA}

\author{Robert H. Bernstein}
\affiliation{Fermi National Accelerator Laboratory, Batavia, USA}

\author{Aji Daniel}
\affiliation{University of Houston, Houston, Texas, USA}

\author{Eleonora Diociaiuti}
\affiliation{INFN, Laboratori Nazionali di Frascati, Frascati, Italy}

\author{Raffaella Donghia}
\affiliation{INFN, Laboratori Nazionali di Frascati, Frascati, Italy}

\author{Ewen L. Gillies}
\affiliation{Imperial College London, London, UK}

\author{Ed V. Hungerford}
\affiliation{University of Houston, Houston, Texas, USA}

\author{Peter Kammel}
\affiliation{University of Washington, Seattle, Washington, USA}

\author{Benjamin E. Krikler}
\altaffiliation[Present address: ]{Bristol University, Bristol, UK}
\affiliation{Imperial College London, London, UK}

\author{Yoshitaka Kuno}
\affiliation{Osaka University, Osaka, Japan}

\author{Mark Lancaster}
\altaffiliation[Present address: ]{University of Manchester, Manchester, UK}
\affiliation{University College London, London, UK}

\author{R. Phillip Litchfield}
\altaffiliation[Present address: ]{University of Glasgow, Glasgow, UK}
\affiliation{University College London, London, UK}

\author{James P. Miller}
\affiliation{Boston University, Boston, Massachusetts, USA}

\author{Anthony Palladino}
\affiliation{Boston University, Boston, Massachusetts, USA}

\author{Jose Repond}
\affiliation{Argonne National Laboratory, Lemont, Illinois, USA}

\author{Akira Sato}
\affiliation{Osaka University, Osaka, Japan}

\author{Ivano Sarra}
\affiliation{INFN, Laboratori Nazionali di Frascati, Frascati, Italy}

\author{Stefano Roberto Soleti}
\altaffiliation[Present address: ]{Lawrence Berkeley National Laboratory, Berkeley, California, USA}
\affiliation{INFN, Laboratori Nazionali di Frascati, Frascati, Italy}

\author{Vladimir Tishchenko}
\affiliation{Brookhaven National Laboratory, Upton, New York, USA}

\author{Nam H. Tran}
\affiliation{Boston University, Boston, Massachusetts, USA}
\affiliation{Osaka University, Osaka, Japan}

\author{Yoshi Uchida}
\affiliation{Imperial College London, London, UK}

\author{Peter Winter}
\affiliation{Argonne National Laboratory, Lemont, Illinois, USA}

\author{Chen Wu}
\affiliation{Osaka University, Osaka, Japan}

\collaboration{AlCap Collaboration}

\date{\today}

\newcommand{\Xray}{$X$-ray}
\newcommand{\Xrays}{$X$-rays}
\newcommand\EvdE{$E_{1}$ vs. $E_{1} + E_{2}$}
\newcommand\EvdEThree{$E_{1} + E_{2}$ vs. $E_{1} + E_{2} + E_{3}$}
\newcommand\dEdx{$dE/dx$}

\newcommand\AlXRayE{\SI{346.828(2)}}
\newcommand\AlXRayI{\SI{79.8(8)}}
\newcommand\AlLifetime{\SI{864(2)}}
\newcommand\AlLifetimeNoErr{\SI{864}}
\newcommand\AlCaptureFraction{\SI{60.9(14)}}

\newcommand\AlHundredNTotalTMEs{\num{254e6}}
\newcommand\AlHundredNAnalysedTMEs{\num{239e6}}
\newcommand\AlHundredGeHiGainNXRays{\num{70.6(5)e3}}
\newcommand\AlHundredGeHiGainNStoppedMuons{\num{134(1)e6}}
\newcommand\AlHundredGeHiGainNCapturedMuons{\num{81.7(6)e6}}
\newcommand\AlHundredNTotalTMEsTab{\num{254}} 
\newcommand\AlHundredNAnalysedTMEsTab{\num{239}} 
\newcommand\AlHundredGeHiGainNXRaysTab{\num{70.6(5)}} 
\newcommand\AlHundredGeHiGainNStoppedMuonsTab{\num{134(1)}} 
\newcommand\AlHundredGeHiGainNCapturedMuonsTab{\num{81.7(6)}} 

\newcommand\AlNTotalTMEs{\num{466e6}}
\newcommand\AlNAnalysedTMEs{\num{421e6}}
\newcommand\AlFiftyGeHiGainNXRays{\num{86.1(5)e3}}
\newcommand\AlFiftyGeHiGainNStoppedMuons{\num{163(1)e6}}
\newcommand\AlFiftyGeHiGainNCapturedMuons{\num{99.4(6)e6}}
\newcommand\AlNTotalTMEsTab{\num{466}} 
\newcommand\AlNAnalysedTMEsTab{\num{421}} 
\newcommand\AlFiftyGeHiGainNXRaysTab{\num{86.1(5)}} 
\newcommand\AlFiftyGeHiGainNStoppedMuonsTab{\num{163(1)}} 
\newcommand\AlFiftyGeHiGainNCapturedMuonsTab{\num{99.4(6)}} 

\newcommand\AlFiftyNRawProton{\num{72.2(3)e3}}
\newcommand\AlFiftyNRawDeuteron{\num{24.9(2)e3}}
\newcommand\AlFiftyNRawTriton{\num{7.49(9)e3}}
\newcommand\AlFiftyNRawAlpha{\num{3.13(6)e3}}
\newcommand\AlFiftyNRawProtonTab{\num{72.2(3)}}
\newcommand\AlFiftyNRawDeuteronTab{\num{24.9(2)}}
\newcommand\AlFiftyNRawTritonTab{\num{7.49(9)}}
\newcommand\AlFiftyNRawAlphaTab{\num{3.13(6)}}

\newcommand\AlHundredNRawProton{\num{47.2(2)e3}}
\newcommand\AlHundredNRawDeuteron{\num{15.9(1)e3}}
\newcommand\AlHundredNRawTriton{\num{4.27(7)e3}}
\newcommand\AlHundredNRawAlpha{\num{1.32(4)e3}}
\newcommand\AlHundredNRawProtonTab{\num{47.2(2)}}
\newcommand\AlHundredNRawDeuteronTab{\num{15.9(1)}}
\newcommand\AlHundredNRawTritonTab{\num{4.27(7)}}
\newcommand\AlHundredNRawAlphaTab{\num{1.32(4)}}

\newcommand\AlProtonExpoFitT{$6.5 \pm 0.1$}
\newcommand\AlDeuteronExpoFitT{$8.5 \pm 0.3$}
\newcommand\AlTritonExpoFitT{$6.9 \pm 0.3$}
\newcommand\AlAlphaExpoFitT{$4.0 \pm 0.5$}

\newcommand\SiXRayE{400.177(5)}
\newcommand\SiXRayI{$80.3(8)$}
\newcommand\SiLifetime{$756(1)$}
\newcommand\SiLifetimeNoErr{756 ns}
\newcommand\SiCaptureFraction{$65.8(13)$}

\newcommand\SibNTotalTMEs{\num[round-precision=3, round-mode=figures, scientific-notation=engineering]{153993939}}
\newcommand\SibNAnalysedTMEs{\num[round-precision=3, round-mode=figures, scientific-notation=engineering]{39442965}}
\newcommand\SibNTotalTMEsTab{\num[round-precision=3, round-mode=figures, scientific-notation=engineering]{153.994}}
\newcommand\SibNAnalysedTMEsTab{\num[round-precision=3, round-mode=figures, scientific-notation=engineering]{39.443}}

\newcommand\SibGeLoGainNXRays{\num[round-precision=3, round-mode=figures, scientific-notation=engineering]{14769.847653\pm136.368431}}
\newcommand\SibGeLoGainNStoppedMuons{\num[round-precision=3, round-mode=figures, scientific-notation=engineering]{31281181.226495 \pm 640140.086763}}
\newcommand\SibGeLoGainNCapturedMuons{\num[round-precision=3, round-mode=figures, scientific-notation=engineering]{20583017.247033 \pm 421212.177090}}
\newcommand\SibGeLoGainNXRaysTab{\num[round-precision=3, round-mode=figures]{14.769848}(\num[round-precision=1, round-mode=figures]{1.363684})}
\newcommand\SibGeLoGainNStoppedMuonsTab{\num[round-precision=3, round-mode=figures]{31.281181}(\num[round-precision=1,round-mode=figures]{6.401401})}
\newcommand\SibGeLoGainNCapturedMuonsTab{\num[round-precision=3, round-mode=figures, scientific-notation=engineering]{20583017.247033 \pm 421212.177090}}
\newcommand\SibGeHiGainNXRays{\num[round-precision=3, round-mode=figures, scientific-notation=engineering]{15524.430110\pm123.894887}}
\newcommand\SibGeHiGainNStoppedMuons{\num[round-precision=3, round-mode=figures, scientific-notation=engineering]{31435835.353477 \pm 610456.930518}}
\newcommand\SibGeHiGainNCapturedMuons{\num[round-precision=3, round-mode=figures, scientific-notation=engineering]{20684779.662588 \pm 401680.660281}}
\newcommand\SibGeHiGainNXRaysTab{\num[round-precision=3, round-mode=figures]{15.524430}(\num[round-precision=1, round-mode=figures]{1.238949})}
\newcommand\SibGeHiGainNStoppedMuonsTab{\num[round-precision=3, round-mode=figures]{31.435835}(\num[round-precision=1,round-mode=figures]{6.104569})}
\newcommand\SibGeHiGainNCapturedMuonsTab{\num[round-precision=3, round-mode=figures, scientific-notation=engineering]{20684779.662588 \pm 401680.660281}}
\newcommand\SibGeLoGainNXRaysAlternate{\num[round-precision=3, round-mode=figures, scientific-notation=engineering]{1173.429689\pm58.286930}}
\newcommand\SibGeLoGainNStoppedMuonsAlternate{\num[round-precision=3, round-mode=figures, scientific-notation=engineering]{32762723.061770 \pm 2437617.593194}}
\newcommand\SibGeLoGainNCapturedMuonsAlternate{\num[round-precision=3, round-mode=figures, scientific-notation=engineering]{21557871.774645 \pm 1603952.376322}}
\newcommand\SibGeLoGainNXRaysTabAlternate{\num[round-precision=3, round-mode=figures]{1.173430}(\num[round-precision=1, round-mode=figures]{0.582869})}
\newcommand\SibGeLoGainNStoppedMuonsTabAlternate{\num[round-precision=3, round-mode=figures]{32.762723}(\num[round-precision=1,round-mode=figures]{24.376176})}
\newcommand\SibGeLoGainNCapturedMuonsTabAlternate{\num[round-precision=3, round-mode=figures, scientific-notation=engineering]{21557871.774645 \pm 1603952.376322}}
\newcommand\SibGeHiGainNXRaysAlternate{\num[round-precision=3, round-mode=figures, scientific-notation=engineering]{1260.988727\pm53.875068}}
\newcommand\SibGeHiGainNStoppedMuonsAlternate{\num[round-precision=3, round-mode=figures, scientific-notation=engineering]{35207413.635062 \pm 2462940.848151}}
\newcommand\SibGeHiGainNCapturedMuonsAlternate{\num[round-precision=3, round-mode=figures, scientific-notation=engineering]{23166478.171871 \pm 1620615.078083}}
\newcommand\SibGeHiGainNXRaysTabAlternate{\num[round-precision=3, round-mode=figures]{1.260989}(\num[round-precision=1, round-mode=figures]{0.538751})}
\newcommand\SibGeHiGainNStoppedMuonsTabAlternate{\num[round-precision=3, round-mode=figures]{35.207414}(\num[round-precision=1,round-mode=figures]{24.629408})}
\newcommand\SibGeHiGainNCapturedMuonsTabAlternate{\num[round-precision=3, round-mode=figures, scientific-notation=engineering]{23166478.171871 \pm 1620615.078083}}

\newcommand\SibVetoEff{\num[round-precision=2, round-mode=figures]{0.74 \pm 0.07}}
\newcommand\SibTimeCutEff{\num[round-precision=3, round-mode=figures]{0.941 \pm 0.002}}

\newcommand\SibNRawProton{\num[round-precision=3, round-mode=figures, scientific-notation=engineering]{15651.758\pm125.107}}
\newcommand\SibNRawProtonTab{\num[round-precision=3, round-mode=figures]{15.652}(\num[round-precision=1, round-mode=figures]{1.251})}
\newcommand\SibNRawDeuteron{\num[round-precision=3, round-mode=figures, scientific-notation=engineering]{2749.000\pm52.431}}
\newcommand\SibNRawDeuteronTab{\num[round-precision=3, round-mode=figures]{2.749}(\num[round-precision=1, round-mode=figures]{5.243})}
\newcommand\SibNRawTriton{\num[round-precision=3, round-mode=figures, scientific-notation=engineering]{574.000\pm23.958}}
\newcommand\SibNRawTritonTab{\num[round-precision=3, round-mode=figures]{0.574}(\num[round-precision=2, round-mode=figures]{23.958})}
\newcommand\SibNRawAlpha{\num[round-precision=3, round-mode=figures, scientific-notation=engineering]{265.000\pm16.279}}
\newcommand\SibNRawAlphaTab{\num[round-precision=3, round-mode=figures]{0.265}(\num[round-precision=2, round-mode=figures]{16.279})}

\newcommand\SibProtonBestRange{3 -- 17}
\newcommand\SibProtonBestRateBare{69.36}
\newcommand\SibProtonBestStatErrBare{0.74}
\newcommand\SibProtonBestHighSystErrBare{2.03}
\newcommand\SibProtonBestLowSystErrBare{1.86}
\newcommand\SibProtonBestRateFull{$\SibProtonBestRateBare \pm \SibProtonBestStatErrBare ~\text{(stat.)} \pm \SibProtonBestHighSystErrBare~\text{(syst.)}$}
\newcommand\SibProtonBestStatFracUncertainty{$\pm1.06\%$}
\newcommand\SibProtonBestSystFracUncertainty{$^{+2.92\%}_{-2.69\%}$}
\newcommand\SibProtonBestFracUncertainty{\SibProtonBestStatFracUncertainty ~(stat.) \SibProtonBestSystFracUncertainty ~(syst.)}
\newcommand\SibDeuteronBestRange{5 -- 17}
\newcommand\SibDeuteronBestRateBare{9.80}
\newcommand\SibDeuteronBestStatErrBare{0.22}
\newcommand\SibDeuteronBestHighSystErrBare{0.24}
\newcommand\SibDeuteronBestLowSystErrBare{0.28}
\newcommand\SibDeuteronBestRateFull{$\SibDeuteronBestRateBare \pm \SibDeuteronBestStatErrBare ~\text{(stat.)} \pm \SibDeuteronBestLowSystErrBare~\text{(syst.)}$}
\newcommand\SibDeuteronBestStatFracUncertainty{$\pm2.29\%$}
\newcommand\SibDeuteronBestSystFracUncertainty{$^{+2.47\%}_{-2.89\%}$}
\newcommand\SibDeuteronBestFracUncertainty{\SibDeuteronBestStatFracUncertainty ~(stat.) \SibDeuteronBestSystFracUncertainty ~(syst.)}
\newcommand\SibTritonBestRange{5 -- 17}
\newcommand\SibTritonBestRateBare{2.07}
\newcommand\SibTritonBestStatErrBare{0.09}
\newcommand\SibTritonBestHighSystErrBare{0.10}
\newcommand\SibTritonBestLowSystErrBare{0.09}
\newcommand\SibTritonBestRateFull{$\SibTritonBestRateBare \pm \SibTritonBestStatErrBare ~\text{(stat.)} \pm \SibTritonBestHighSystErrBare~\text{(syst.)}$}
\newcommand\SibTritonBestStatFracUncertainty{$\pm4.34\%$}
\newcommand\SibTritonBestSystFracUncertainty{$^{+4.96\%}_{-4.44\%}$}
\newcommand\SibTritonBestFracUncertainty{\SibTritonBestStatFracUncertainty ~(stat.) \SibTritonBestSystFracUncertainty ~(syst.)}
\newcommand\SibAlphaBestRange{15 -- 20}
\newcommand\SibAlphaBestRateBare{0.59}
\newcommand\SibAlphaBestStatErrBare{0.03}
\newcommand\SibAlphaBestHighSystErrBare{0.10}
\newcommand\SibAlphaBestLowSystErrBare{0.07}
\newcommand\SibAlphaBestRateFull{$\SibAlphaBestRateBare \pm \SibAlphaBestStatErrBare ~\text{(stat.)} \pm \SibAlphaBestHighSystErrBare~\text{(syst.)}$}
\newcommand\SibAlphaBestStatFracUncertainty{$\pm4.71\%$}
\newcommand\SibAlphaBestSystFracUncertainty{$^{+16.57\%}_{-11.00\%}$}
\newcommand\SibAlphaBestFracUncertainty{\SibAlphaBestStatFracUncertainty ~(stat.) \SibAlphaBestSystFracUncertainty ~(syst.)}
\newcommand\protonRatioBare{50.2}
\newcommand\protonRatioStatErrBare{4.9}
\newcommand\protonRatioHighSystErrBare{9.9}
\newcommand\protonRatioLowSystErrBare{11.0}
\newcommand\protonRatioHighTotalErrBare{11.0}
\newcommand\protonRatioLowTotalErrBare{12.1}
\newcommand\protonRatioNoErr{\protonRatioBare\%}
\newcommand\protonRatioFull{$(\protonRatioBare \pm \protonRatioStatErrBare~\text{(stat.)} ^{+\protonRatioHighSystErrBare}_{-\protonRatioLowSystErrBare}~\text{(syst.)})\%$}
\newcommand\protonRatioCondensed{$\protonRatioBare ^{+\protonRatioHighTotalErrBare}_{-\protonRatioLowTotalErrBare}\%$}
\newcommand\deuteronRatioBare{30.0}
\newcommand\deuteronRatioStatErrBare{3.1}
\newcommand\deuteronRatioHighSystErrBare{3.6}
\newcommand\deuteronRatioLowSystErrBare{4.0}
\newcommand\deuteronRatioHighTotalErrBare{4.7}
\newcommand\deuteronRatioLowTotalErrBare{5.0}
\newcommand\deuteronRatioNoErr{\deuteronRatioBare\%}
\newcommand\deuteronRatioFull{$(\deuteronRatioBare \pm \deuteronRatioStatErrBare~\text{(stat.)} ^{+\deuteronRatioHighSystErrBare}_{-\deuteronRatioLowSystErrBare}~\text{(syst.)})\%$}
\newcommand\deuteronRatioCondensed{$\deuteronRatioBare ^{+\deuteronRatioHighTotalErrBare}_{-\deuteronRatioLowTotalErrBare}\%$}
\newcommand\tritonRatioBare{4.3}
\newcommand\tritonRatioStatErrBare{0.9}
\newcommand\tritonRatioHighSystErrBare{1.4}
\newcommand\tritonRatioLowSystErrBare{1.3}
\newcommand\tritonRatioHighTotalErrBare{1.7}
\newcommand\tritonRatioLowTotalErrBare{1.6}
\newcommand\tritonRatioNoErr{\tritonRatioBare\%}
\newcommand\tritonRatioFull{$(\tritonRatioBare \pm \tritonRatioStatErrBare~\text{(stat.)} ^{+\tritonRatioHighSystErrBare}_{-\tritonRatioLowSystErrBare}~\text{(syst.)})\%$}
\newcommand\tritonRatioCondensed{$\tritonRatioBare ^{+\tritonRatioHighTotalErrBare}_{-\tritonRatioLowTotalErrBare}\%$}
\newcommand\alphaRatioBare{15.5}
\newcommand\alphaRatioStatErrBare{1.2}
\newcommand\alphaRatioHighSystErrBare{3.8}
\newcommand\alphaRatioLowSystErrBare{2.7}
\newcommand\alphaRatioHighTotalErrBare{4.0}
\newcommand\alphaRatioLowTotalErrBare{2.9}
\newcommand\alphaRatioNoErr{\alphaRatioBare\%}
\newcommand\alphaRatioFull{$(\alphaRatioBare \pm \alphaRatioStatErrBare~\text{(stat.)} ^{+\alphaRatioHighSystErrBare}_{-\alphaRatioLowSystErrBare}~\text{(syst.)})\%$}
\newcommand\alphaRatioCondensed{$\alphaRatioBare ^{+\alphaRatioHighTotalErrBare}_{-\alphaRatioLowTotalErrBare}\%$}
\newcommand\SibProtonExpoFitT{$3.7 \pm 0.1$}
\newcommand\SibProtonExpoFitRate{$0.0926 \pm 0.0053$}
\newcommand\SibDeuteronExpoFitT{$8.4 \pm 1.0$}
\newcommand\SibDeuteronExpoFitRate{$0.0174 \pm 0.0008$}
\newcommand\SibTritonExpoFitT{$5.3 \pm 0.7$}
\newcommand\SibTritonExpoFitRate{$0.0041 \pm 0.0018$}
\newcommand\SibAlphaExpoFitT{$2.2 \pm 0.5$}
\newcommand\SibAlphaExpoFitRate{$0.2343 \pm 0.2877$}
\newcommand\SibSumExpoFitT{$4.5 \pm 0.2$}
\newcommand\SibSumExpoFitRate{$0.1117 \pm 0.0050$}

\newcommand\SiLNTotalTMEs{\num[round-precision=3, round-mode=figures, scientific-notation=engineering]{140195769}}
\newcommand\SiLNAnalysedTMEs{\num[round-precision=3, round-mode=figures, scientific-notation=engineering]{4912947}}
\newcommand\SiLNTotalTMEsTab{\num[round-precision=3, round-mode=figures, scientific-notation=engineering]{140.1958}}
\newcommand\SiLNAnalysedTMEsTab{\num[round-precision=3, round-mode=figures, scientific-notation=engineering]{4.9129}}

\newcommand\SiLGeLoGainNXRays{\num[round-precision=3, round-mode=figures, scientific-notation=engineering]{11311.1342\pm113.2591}}
\newcommand\SiLGeLoGainNStoppedMuons{\num[round-precision=3, round-mode=figures, scientific-notation=engineering]{23955943.7939 \pm 498947.2131}}
\newcommand\SiLGeLoGainNCapturedMuons{\num[round-precision=3, round-mode=figures, scientific-notation=engineering]{15763011.0164 \pm 328307.2662}}
\newcommand\SiLGeLoGainNXRaysTab{\num[round-precision=3, round-mode=figures]{11.3111}(\num[round-precision=1, round-mode=figures]{1.1326})}
\newcommand\SiLGeLoGainNStoppedMuonsTab{\num[round-precision=3, round-mode=figures]{23.9559}(\num[round-precision=1, round-mode=figures]{4.9895})}
\newcommand\SiLGeLoGainNCapturedMuonsTab{\num[round-precision=3, round-mode=figures]{15.7630}(\num[round-precision=1, round-mode=figures]{3.2831})}
\newcommand\SiLGeHiGainNXRays{\num[round-precision=3, round-mode=figures, scientific-notation=engineering]{11781.0097\pm113.0294}}
\newcommand\SiLGeHiGainNStoppedMuons{\num[round-precision=3, round-mode=figures, scientific-notation=engineering]{23855682.8875 \pm 480359.6368}}
\newcommand\SiLGeHiGainNCapturedMuons{\num[round-precision=3, round-mode=figures, scientific-notation=engineering]{15697039.3400 \pm 316076.6410}}
\newcommand\SiLGeHiGainNXRaysTab{\num[round-precision=3, round-mode=figures]{11.7810}(\num[round-precision=1, round-mode=figures]{1.1303})}
\newcommand\SiLGeHiGainNStoppedMuonsTab{\num[round-precision=3, round-mode=figures]{23.8557}(\num[round-precision=1, round-mode=figures]{4.8036})}
\newcommand\SiLGeHiGainNCapturedMuonsTab{\num[round-precision=3, round-mode=figures]{15.6970}(\num[round-precision=1, round-mode=figures]{3.1608})}
\newcommand\SiLGeLoGainNXRaysAlternate{\num[round-precision=3, round-mode=figures, scientific-notation=engineering]{818.8405\pm41.7344}}
\newcommand\SiLGeLoGainNStoppedMuonsAlternate{\num[round-precision=3, round-mode=figures, scientific-notation=engineering]{22862422.4361 \pm 1720926.4129}}
\newcommand\SiLGeLoGainNCapturedMuonsAlternate{\num[round-precision=3, round-mode=figures, scientific-notation=engineering]{15043473.9629 \pm 1132369.5797}}
\newcommand\SiLGeLoGainNXRaysTabAlternate{\num[round-precision=3, round-mode=figures]{0.8188}(\num[round-precision=1, round-mode=figures]{0.4173})}
\newcommand\SiLGeLoGainNStoppedMuonsTabAlternate{\num[round-precision=3, round-mode=figures]{22.8624}(\num[round-precision=1, round-mode=figures]{17.2093})}
\newcommand\SiLGeLoGainNCapturedMuonsTabAlternate{\num[round-precision=3, round-mode=figures]{15.0435}(\num[round-precision=1, round-mode=figures]{11.3237})}
\newcommand\SiLGeHiGainNXRaysAlternate{\num[round-precision=3, round-mode=figures, scientific-notation=engineering]{878.0550\pm36.3595}}
\newcommand\SiLGeHiGainNStoppedMuonsAlternate{\num[round-precision=3, round-mode=figures, scientific-notation=engineering]{24515720.1029 \pm 1695499.6679}}
\newcommand\SiLGeHiGainNCapturedMuonsAlternate{\num[round-precision=3, round-mode=figures, scientific-notation=engineering]{16131343.8277 \pm 1115638.7815}}
\newcommand\SiLGeHiGainNXRaysTabAlternate{\num[round-precision=3, round-mode=figures]{0.8781}(\num[round-precision=1, round-mode=figures]{0.3636})}
\newcommand\SiLGeHiGainNStoppedMuonsTabAlternate{\num[round-precision=3, round-mode=figures]{24.5157}(\num[round-precision=1, round-mode=figures]{16.9550})}
\newcommand\SiLGeHiGainNCapturedMuonsTabAlternate{\num[round-precision=3, round-mode=figures]{16.1313}(\num[round-precision=1, round-mode=figures]{11.1564})}

\newcommand\SiLNTotalCount{\num{460627.0000\pm678.6951}}
\newcommand\SiLNTotalCountTab{\num[round-precision=4,round-mode=figures]{460.6270}(\num[round-precision=1,round-mode=figures]{6.7870})}

\newcommand\allRecoil{$7.3\%$}

\newcommand\SiLActiveRange{1.4 -- 26}
\newcommand\SiLActiveRate{\num[round-precision=3, round-mode=figures]{170.975312876116561}(\num[round-precision=1, round-mode=figures]{33.236013936027383})}

\newcommand\TiProtonMinE{3.5}
\newcommand\TiProtonMaxE{20.0}
\newcommand\TiProtonBestRange{3.5 -- 20.0}
\newcommand\TiProtonBestRateBare{26.48}
\newcommand\TiProtonBestStatErrBare{0.35}
\newcommand\TiProtonBestHighSystErrBare{0.80}
\newcommand\TiProtonBestLowSystErrBare{0.80}
\newcommand\TiProtonBestRateFull{$\TiProtonBestRateBare \pm \TiProtonBestStatErrBare ~\text{(stat.)} \pm \TiProtonBestLowSystErrBare~\text{(syst.)}$}
\newcommand\TiProtonBestStatFracUncertainty{$\pm1.31\%$}
\newcommand\TiProtonBestSystFracUncertainty{$^{+3.01\%}_{-3.01\%}$}
\newcommand\TiProtonBestFracUncertainty{\TiProtonBestStatFracUncertainty ~(stat.) \TiProtonBestSystFracUncertainty ~(syst.)}
\newcommand\TiProtonTable{26.48(35)(80)}
\newcommand\TiProtonTableFirst{26.48(35~stat.)(80~syst.)}
\newcommand\TiDeuteronMinE{4.5}
\newcommand\TiDeuteronMaxE{17.0}
\newcommand\TiDeuteronBestRange{4.5 -- 17.0}
\newcommand\TiDeuteronBestRateBare{5.02}
\newcommand\TiDeuteronBestStatErrBare{0.10}
\newcommand\TiDeuteronBestHighSystErrBare{0.20}
\newcommand\TiDeuteronBestLowSystErrBare{0.20}
\newcommand\TiDeuteronBestRateFull{$\TiDeuteronBestRateBare \pm \TiDeuteronBestStatErrBare ~\text{(stat.)} \pm \TiDeuteronBestLowSystErrBare~\text{(syst.)}$}
\newcommand\TiDeuteronBestStatFracUncertainty{$\pm1.96\%$}
\newcommand\TiDeuteronBestSystFracUncertainty{$^{+3.96\%}_{-3.96\%}$}
\newcommand\TiDeuteronBestFracUncertainty{\TiDeuteronBestStatFracUncertainty ~(stat.) \TiDeuteronBestSystFracUncertainty ~(syst.)}
\newcommand\TiDeuteronTable{5.02(10)(20)}
\newcommand\TiDeuteronTableFirst{5.02(10~stat.)(20~syst.)}
\newcommand\TiTritonMinE{5.0}
\newcommand\TiTritonMaxE{17.0}
\newcommand\TiTritonBestRange{5.0 -- 17.0}
\newcommand\TiTritonBestRateBare{1.36}
\newcommand\TiTritonBestStatErrBare{0.05}
\newcommand\TiTritonBestHighSystErrBare{0.07}
\newcommand\TiTritonBestLowSystErrBare{0.07}
\newcommand\TiTritonBestRateFull{$\TiTritonBestRateBare \pm \TiTritonBestStatErrBare ~\text{(stat.)} \pm \TiTritonBestLowSystErrBare~\text{(syst.)}$}
\newcommand\TiTritonBestStatFracUncertainty{$\pm3.43\%$}
\newcommand\TiTritonBestSystFracUncertainty{$^{+4.86\%}_{-4.86\%}$}
\newcommand\TiTritonBestFracUncertainty{\TiTritonBestStatFracUncertainty ~(stat.) \TiTritonBestSystFracUncertainty ~(syst.)}
\newcommand\TiTritonTable{1.36(5)(7)}
\newcommand\TiTritonTableFirst{1.36(5~stat.)(7~syst.)}
\newcommand\TiAlphaMinE{15.0}
\newcommand\TiAlphaMaxE{20.0}
\newcommand\TiAlphaBestRange{15.0 -- 20.0}
\newcommand\TiAlphaBestRateBare{0.45}
\newcommand\TiAlphaBestStatErrBare{0.02}
\newcommand\TiAlphaBestHighSystErrBare{0.05}
\newcommand\TiAlphaBestLowSystErrBare{0.05}
\newcommand\TiAlphaBestRateFull{$\TiAlphaBestRateBare \pm \TiAlphaBestStatErrBare ~\text{(stat.)} \pm \TiAlphaBestLowSystErrBare~\text{(syst.)}$}
\newcommand\TiAlphaBestStatFracUncertainty{$\pm4.38\%$}
\newcommand\TiAlphaBestSystFracUncertainty{$^{+11.12\%}_{-11.12\%}$}
\newcommand\TiAlphaBestFracUncertainty{\TiAlphaBestStatFracUncertainty ~(stat.) \TiAlphaBestSystFracUncertainty ~(syst.)}
\newcommand\TiAlphaTable{0.45(2)(5)}
\newcommand\TiAlphaTableFirst{0.45(2~stat.)(5~syst.)}
\newcommand\SiProtonMinE{4.0}
\newcommand\SiProtonMaxE{20.0}
\newcommand\SiProtonBestRange{4.0 -- 20.0}
\newcommand\SiProtonBestRateBare{52.46}
\newcommand\SiProtonBestStatErrBare{0.62}
\newcommand\SiProtonBestHighSystErrBare{1.73}
\newcommand\SiProtonBestLowSystErrBare{1.82}
\newcommand\SiProtonBestRateFull{$\SiProtonBestRateBare \pm \SiProtonBestStatErrBare ~\text{(stat.)} \pm \SiProtonBestLowSystErrBare~\text{(syst.)}$}
\newcommand\SiProtonBestStatFracUncertainty{$\pm1.19\%$}
\newcommand\SiProtonBestSystFracUncertainty{$^{+3.30\%}_{-3.47\%}$}
\newcommand\SiProtonBestFracUncertainty{\SiProtonBestStatFracUncertainty ~(stat.) \SiProtonBestSystFracUncertainty ~(syst.)}
\newcommand\SiProtonTable{52.5(6)(18)}
\newcommand\SiProtonTableFirst{52.5(6~stat.)(18~syst.)}
\newcommand\SiDeuteronMinE{5.0}
\newcommand\SiDeuteronMaxE{17.0}
\newcommand\SiDeuteronBestRange{5.0 -- 17.0}
\newcommand\SiDeuteronBestRateBare{9.80}
\newcommand\SiDeuteronBestStatErrBare{0.22}
\newcommand\SiDeuteronBestHighSystErrBare{0.38}
\newcommand\SiDeuteronBestLowSystErrBare{0.41}
\newcommand\SiDeuteronBestRateFull{$\SiDeuteronBestRateBare \pm \SiDeuteronBestStatErrBare ~\text{(stat.)} \pm \SiDeuteronBestLowSystErrBare~\text{(syst.)}$}
\newcommand\SiDeuteronBestStatFracUncertainty{$\pm2.29\%$}
\newcommand\SiDeuteronBestSystFracUncertainty{$^{+3.87\%}_{-4.15\%}$}
\newcommand\SiDeuteronBestFracUncertainty{\SiDeuteronBestStatFracUncertainty ~(stat.) \SiDeuteronBestSystFracUncertainty ~(syst.)}
\newcommand\SiDeuteronTable{9.80(22)(41)}
\newcommand\SiDeuteronTableFirst{9.80(22~stat.)(41~syst.)}
\newcommand\SiTritonMinE{6.0}
\newcommand\SiTritonMaxE{17.0}
\newcommand\SiTritonBestRange{6.0 -- 17.0}
\newcommand\SiTritonBestRateBare{1.70}
\newcommand\SiTritonBestStatErrBare{0.08}
\newcommand\SiTritonBestHighSystErrBare{0.10}
\newcommand\SiTritonBestLowSystErrBare{0.10}
\newcommand\SiTritonBestRateFull{$\SiTritonBestRateBare \pm \SiTritonBestStatErrBare ~\text{(stat.)} \pm \SiTritonBestHighSystErrBare~\text{(syst.)}$}
\newcommand\SiTritonBestStatFracUncertainty{$\pm4.91\%$}
\newcommand\SiTritonBestSystFracUncertainty{$^{+6.06\%}_{-5.99\%}$}
\newcommand\SiTritonBestFracUncertainty{\SiTritonBestStatFracUncertainty ~(stat.) \SiTritonBestSystFracUncertainty ~(syst.)}
\newcommand\SiTritonTable{1.70(8)(10)}
\newcommand\SiTritonTableFirst{1.70(8~stat.)(10~syst.)}
\newcommand\SiAlphaMinE{15.0}
\newcommand\SiAlphaMaxE{20.0}
\newcommand\SiAlphaBestRange{15.0 -- 20.0}
\newcommand\SiAlphaBestRateBare{0.57}
\newcommand\SiAlphaBestStatErrBare{0.03}
\newcommand\SiAlphaBestHighSystErrBare{0.10}
\newcommand\SiAlphaBestLowSystErrBare{0.07}
\newcommand\SiAlphaBestRateFull{$\SiAlphaBestRateBare \pm \SiAlphaBestStatErrBare ~\text{(stat.)} \pm \SiAlphaBestHighSystErrBare~\text{(syst.)}$}
\newcommand\SiAlphaBestStatFracUncertainty{$\pm4.81\%$}
\newcommand\SiAlphaBestSystFracUncertainty{$^{+17.52\%}_{-11.85\%}$}
\newcommand\SiAlphaBestFracUncertainty{\SiAlphaBestStatFracUncertainty ~(stat.) \SiAlphaBestSystFracUncertainty ~(syst.)}
\newcommand\SiAlphaTable{0.57(3)(10)}
\newcommand\SiAlphaTableFirst{0.57(3~stat.)(10~syst.)}
\newcommand\AlProtonMinE{3.5}
\newcommand\AlProtonMaxE{20.0}
\newcommand\AlProtonBestRange{3.5 -- 20.0}
\newcommand\AlProtonBestRateBare{26.64}
\newcommand\AlProtonBestStatErrBare{0.28}
\newcommand\AlProtonBestHighSystErrBare{0.77}
\newcommand\AlProtonBestLowSystErrBare{0.77}
\newcommand\AlProtonBestRateFull{$\AlProtonBestRateBare \pm \AlProtonBestStatErrBare ~\text{(stat.)} \pm \AlProtonBestLowSystErrBare~\text{(syst.)}$}
\newcommand\AlProtonBestStatFracUncertainty{$\pm1.06\%$}
\newcommand\AlProtonBestSystFracUncertainty{$^{+2.88\%}_{-2.88\%}$}
\newcommand\AlProtonBestFracUncertainty{\AlProtonBestStatFracUncertainty ~(stat.) \AlProtonBestSystFracUncertainty ~(syst.)}
\newcommand\AlProtonTable{26.64(28)(77)}
\newcommand\AlProtonTableFirst{26.64(28~stat.)(77~syst.)}
\newcommand\AlDeuteronMinE{4.5}
\newcommand\AlDeuteronMaxE{17.0}
\newcommand\AlDeuteronBestRange{4.5 -- 17.0}
\newcommand\AlDeuteronBestRateBare{8.46}
\newcommand\AlDeuteronBestStatErrBare{0.09}
\newcommand\AlDeuteronBestHighSystErrBare{0.24}
\newcommand\AlDeuteronBestLowSystErrBare{0.24}
\newcommand\AlDeuteronBestRateFull{$\AlDeuteronBestRateBare \pm \AlDeuteronBestStatErrBare ~\text{(stat.)} \pm \AlDeuteronBestLowSystErrBare~\text{(syst.)}$}
\newcommand\AlDeuteronBestStatFracUncertainty{$\pm1.06\%$}
\newcommand\AlDeuteronBestSystFracUncertainty{$^{+2.88\%}_{-2.88\%}$}
\newcommand\AlDeuteronBestFracUncertainty{\AlDeuteronBestStatFracUncertainty ~(stat.) \AlDeuteronBestSystFracUncertainty ~(syst.)}
\newcommand\AlDeuteronTable{8.46(9)(24)}
\newcommand\AlDeuteronTableFirst{8.46(9~stat.)(24~syst.)}
\newcommand\AlTritonMinE{5.0}
\newcommand\AlTritonMaxE{17.0}
\newcommand\AlTritonBestRange{5.0 -- 17.0}
\newcommand\AlTritonBestRateBare{2.58}
\newcommand\AlTritonBestStatErrBare{0.04}
\newcommand\AlTritonBestHighSystErrBare{0.09}
\newcommand\AlTritonBestLowSystErrBare{0.09}
\newcommand\AlTritonBestRateFull{$\AlTritonBestRateBare \pm \AlTritonBestStatErrBare ~\text{(stat.)} \pm \AlTritonBestLowSystErrBare~\text{(syst.)}$}
\newcommand\AlTritonBestStatFracUncertainty{$\pm1.74\%$}
\newcommand\AlTritonBestSystFracUncertainty{$^{+3.53\%}_{-3.53\%}$}
\newcommand\AlTritonBestFracUncertainty{\AlTritonBestStatFracUncertainty ~(stat.) \AlTritonBestSystFracUncertainty ~(syst.)}
\newcommand\AlTritonTable{2.58(4)(9)}
\newcommand\AlTritonTableFirst{2.58(4~stat.)(9~syst.)}
\newcommand\AlAlphaMinE{15.0}
\newcommand\AlAlphaMaxE{20.0}
\newcommand\AlAlphaBestRange{15.0 -- 20.0}
\newcommand\AlAlphaBestRateBare{0.44}
\newcommand\AlAlphaBestStatErrBare{0.01}
\newcommand\AlAlphaBestHighSystErrBare{0.09}
\newcommand\AlAlphaBestLowSystErrBare{0.09}
\newcommand\AlAlphaBestRateFull{$\AlAlphaBestRateBare \pm \AlAlphaBestStatErrBare ~\text{(stat.)} \pm \AlAlphaBestLowSystErrBare~\text{(syst.)}$}
\newcommand\AlAlphaBestStatFracUncertainty{$\pm2.97\%$}
\newcommand\AlAlphaBestSystFracUncertainty{$^{+20.68\%}_{-20.68\%}$}
\newcommand\AlAlphaBestFracUncertainty{\AlAlphaBestStatFracUncertainty ~(stat.) \AlAlphaBestSystFracUncertainty ~(syst.)}
\newcommand\AlAlphaTable{0.44(1)(9)}
\newcommand\AlAlphaTableFirst{0.44(1~stat.)(9~syst.)}
\newcommand\TiProtonExpo{$4.1 \pm~0.1$}
\newcommand\TiDeuteronExpo{$9.8 \pm~1.0$}
\newcommand\TiTritonExpo{$7.3 \pm~0.8$}
\newcommand\TiAlphaExpo{$4.4 \pm~1.7$}
\newcommand\SiProtonExpo{$3.8 \pm~0.1$}
\newcommand\SiDeuteronExpo{$7.8 \pm~0.7$}
\newcommand\SiTritonExpo{$5.1 \pm~0.7$}
\newcommand\SiAlphaExpo{$2.0 \pm~0.3$}
\newcommand\AlProtonExpo{$4.8 \pm~0.2$}
\newcommand\AlDeuteronExpo{$8.1 \pm~0.3$}
\newcommand\AlTritonExpo{$6.6 \pm~0.4$}
\newcommand\AlAlphaExpo{$3.7 \pm~2.0$}

\newcommand\TiXRayE{\SI{931.57(40)}}
\newcommand\TiXRayI{\SI{75.1(14)}}
\newcommand\TiLifetime{\SI{329.3(13)}}
\newcommand\TiLifetimeNoErr{}
\newcommand\TiCaptureFraction{85.29(6)}

\newcommand\TiNTotalTMEs{\num{209e6}}
\newcommand\TiNAnalysedTMEs{\num{188e6}}
\newcommand\TiGeHiGainNXRays{\num{18.8(3)e3}}
\newcommand\TiGeHiGainNStoppedMuons{\num{80(7)e6}}
\newcommand\TiGeHiGainNCapturedMuons{\num{68(6)e6}}
\newcommand\TiNTotalTMEsTab{\num{209}} 
\newcommand\TiNAnalysedTMEsTab{\num{188}} 
\newcommand\TiGeHiGainNXRaysTab{\num{18.8(3)}} 
\newcommand\TiGeHiGainNStoppedMuonsTab{\num{96(3)}} 
\newcommand\TiGeHiGainNCapturedMuonsTab{\num{68(6)}} 

\newcommand\TiNRawProton{\num{53.4(2)e3}}
\newcommand\TiNRawDeuteron{\num{11.2(1)e3}}
\newcommand\TiNRawTriton{\num{2.84(5)e3}}
\newcommand\TiNRawAlpha{\num{1.76(4)e3}}
\newcommand\TiNRawProtonTab{\num{53.4(2)}}
\newcommand\TiNRawDeuteronTab{\num{11.2(1)}}
\newcommand\TiNRawTritonTab{\num{2.84(5)}}
\newcommand\TiNRawAlphaTab{\num{1.76(4)}}

\newcommand\TiProtonExpoFitT{$4.6 \pm 0.1$}
\newcommand\TiDeuteronExpoFitT{$8.6 \pm 0.8$}
\newcommand\TiTritonExpoFitT{$7.0 \pm 0.8$}
\newcommand\TiAlphaExpoFitT{$5.1 \pm 1.0$}

\newcommand\PbLifetime{\SI{75.4(10)}}
\newcommand\PbLifetimeNoErr{75.4 ns}

\begin{abstract}
\begin{description}
\item[Background] Heavy charged particles after nuclear muon capture are an important nuclear physics background to the muon-to-electron conversion experiments Mu2e and COMET, which will search for charged lepton flavor violation at an unprecedented level of sensitivity.
\item[Purpose] The AlCap experiment aimed to measure the yield and energy spectra of protons, deuterons, tritons, and $\alpha$-particles emitted after the nuclear capture of muons stopped in Al, Si, and Ti in the low energy range relevant for the muon-to-electron conversion experiments.
\item[Methods] Individual charged particle types were identified in layered silicon detector packages and their initial energy distributions were unfolded from the observed energy spectra.
\item[Results] The proton yields per muon capture were determined as $Y_{p}(\text{Al})$~=~\AlProtonTableFirst $\times 10^{-3}$ and $Y_{p}(\text{Ti})$~=~\TiProtonTable $\times 10^{-3}$ in the energy range \AlProtonMinE\ -- \AlProtonMaxE\ MeV, and as $Y_{p}(\text{Si})$~=~\SiProtonTable $\times 10^{-3}$ in the energy range \SiProtonMinE\ -- \SiProtonMaxE\ MeV. Detailed information on yields and energy spectra for all observed nuclei are presented in the paper.
\item[Conclusion] The yields in the candidate muon stopping targets, Al and Ti, are approximately half of that in Si, which was used in the past to estimate this background. The reduced background allows for less shielding and a better energy resolution in these experiments. It is anticipated that the comprehensive information presented in this paper will stimulate modern theoretical calculations of the rare process of muon capture with charged particle emission, and inform the design of future muon-to-electron conversion experiments.
\end{description}
\end{abstract}

\maketitle

\section{Introduction}
\label{sec:introduction}
In the coming decade, next-generation muon-beam experiments will probe charged lepton 
flavor violation with unprecedented sensitivity. The COMET (J-PARC)~\cite{COMETTDR} and 
Mu2e (Fermilab)~\cite{Mu2eTDR} experiments will search for the charged lepton flavor violating process of neutrinoless muon-to-electron conversion in aluminum 
($\mu^{-} + \ce{Al}\to e^{-} + \ce{Al}$) with a sensitivity of $10^{-16}$ -- a four order of magnitude improvement over the current experimental limit~\cite{Bertl:2006up}. In these experiments, aluminum targets will stop low-energy muons and high-resolution detectors will measure the momentum of particles emitted from the stopping target. The signal for muon-to-electron conversion is a monoenergetic electron with a momentum of \SI{105}{\mega\electronvolt/\speedoflight}. Incorporating neutrino oscillations into the standard model of particle physics leads to an undetectably small branching ratio ($O(10^{-52})$) and so an observation of such a signal would be an unambiguous sign of new physics.

Many standard model extensions predict enhanced rates of muon-to-electron conversion. When a negative muon is bound in the ground state of a muonic atom there is a significant overlap between the bound muonic $1s$ wavefunction and the nucleus. Depending on the model Lagrangian, this allows the conversion process to occur via the exchange of a photon between the nucleus and the muon (via higher order loops), or through short-range interactions~\cite{adegouvea2009, Kitano:2002mt}.

COMET and Mu2e must understand all of the processes that occur when a negative muon is captured in a stopping target.
First, within a few picoseconds after stopping, the muon will cascade down through excited atomic states to the $1s$ ground state. The transitions between energy levels produce characteristic X-rays that can be used to count the number of stopped muons. Once the muon has reached the ground state, the large overlap between the muon wavefunction and the nucleus modifies the muon decay process ($\mu^- \to e + \bar{\nu}_{e} + \nu_{\mu}$) and allows for nuclear capture ($\mu^-+\ce{Al}\to\nu_{\mu}+\ce{Mg}$) via the short-range, charged-current weak interaction. 
For the muon decay, the Coulomb interaction between the muon and the nucleus modifies the energy distribution of the emitted decay electrons. Because the nucleus can
absorb significant recoil momentum, the high-energy endpoint of the decay electron spectrum extends from \SI{52.8}{\MeV} up to the conversion signal energy, which is \SI{105}{\MeV} for muon-to-electron conversion in aluminum~\cite{czarnecki_michel_2014,Szafron:2016cbv}. 
During muon capture, energy corresponding to the muon mass (minus the atomic binding energy of \SI{450}{\kilo\electronvolt} for aluminum) is imparted to the capture products. The neutrino typically absorbs most of the 
energy, leaving \SI{20}{\MeV}, on average, to be distributed in the nucleus~\cite{Measday2001}. This excess energy results in the emission of protons, neutrons, deuterons, tritons, 
and $\alpha$-particles from the nucleus. 

To reach the required sensitivity, COMET Phase-I~\cite{COMETTDR} and Mu2e will stop a very large number of muons ($O(10^{18})$) in aluminum and so must design their experiments to reduce the impact of the large flux of background particles emitted from the target on their detectors. 
Both experiments will place their stopping targets on the axis of a solenoidal field and employ cylindrical detectors to measure the momentum of emitted electrons.  
The detectors will be kept at a large radius from the solenoid axis to avoid the intense flux of low momentum particles (\SI{<60}{MeV/c}), but maintain good acceptance for the higher-energy conversion electrons.

For heavy charged particles emitted after the nuclear muon capture process, only a small amount of kinetic energy is required for a momentum of \SI{60}{MeV/\speedoflight}. For example, a \SI{2}{MeV} proton has enough transverse momentum to hit the active detector elements. To reduce the hit rate of these highly-ionizing particles, a thin cylindrical absorber will be placed between the aluminum stopping target and the detector to absorb low-energy protons.
However, energy straggling in the absorber will degrade the energy resolution of the conversion electron signal and so the thickness of the absorber needs to be minimized. Therefore, in order to optimize the design of the absorber, it is essential to know the yield and energy spectrum of emitted 
heavy charged particles produced after nuclear muon capture.

There are several summaries of past experimental results~\cite{Measday2001,uberall:1974,singer:1974,mukhop:1977} which discuss the observation and
theoretical interpretation of the emission spectrum of protons, deuterons, and $\alpha$-particles after muon capture. The most advanced theoretical model of this process was developed by Lifshitz and Singer~\cite{Lifshitz:1980gi, Lifshitz:1978qj} applying nuclear structure methods available in the late 1970s. They used a simplified weak Hamiltonian, and a nuclear excitation function of quasifree particles with an empirical nucleon-momentum distribution that allowed for both pre-equilibrium and compound-nucleus emission. Their main goal was to describe the process over a wide range of nuclei. They found that the calculated number of emitted particles per muon capture was consistent with the data available at the time.

Unfortunately, previous experimental results are not directly relevant to COMET and Mu2e. The results are either from materials other than aluminum~\cite{Sobottka1968, Budyashov1971,vil:1971}, had no particle identification~\cite{heusser1972radioisotope, wyttenbach1978probabilities}, or the spectrum was only measured at high energies~\cite{Krane1979}. A recent measurement of the proton and deuteron momentum spectra from muon capture on aluminum from TWIST~\cite{Gaponenko:2019efc} has improved the picture. 

\begin{figure}[ht]
  \centering
  \includegraphics[width=1.0\columnwidth]{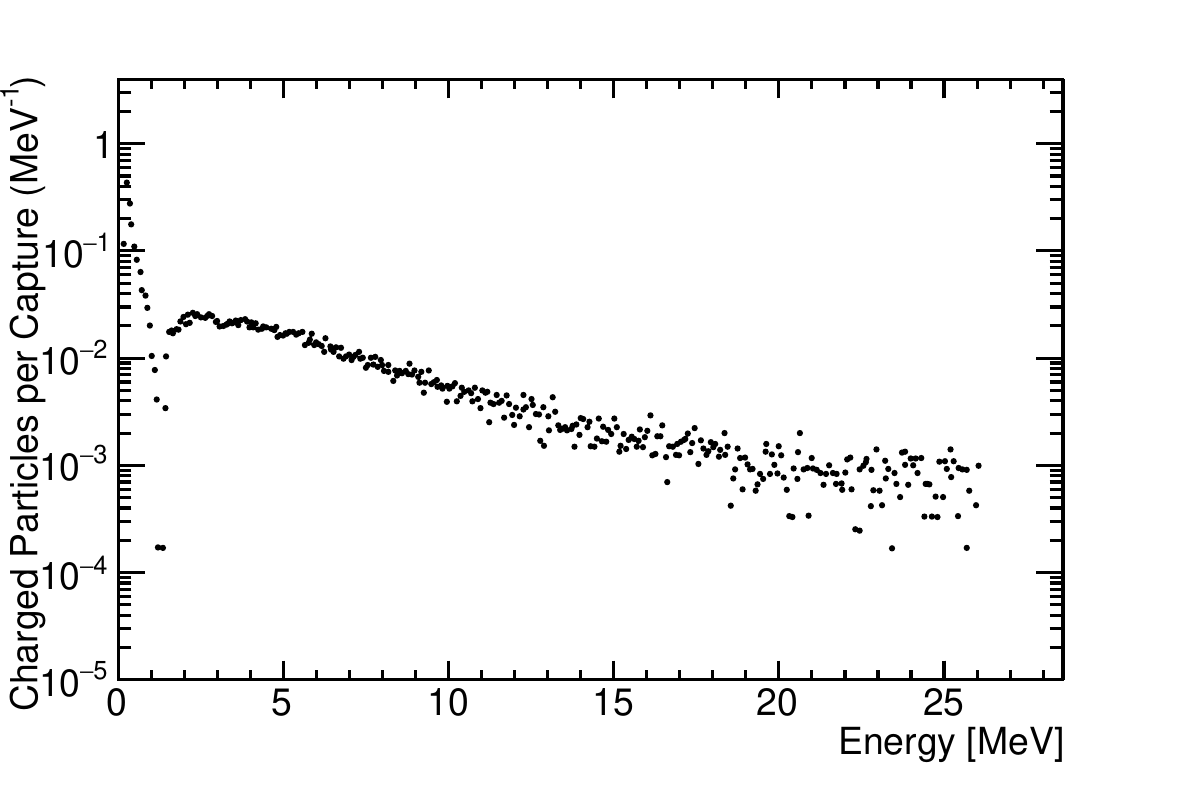}
  \caption{Sobottka-Wills data of charged particle emission after nuclear muon capture on silicon (digitized from Ref.~\cite{Sobottka1968}). The data below the valley at \SI{1.4}{MeV} is due to nuclear recoil, mainly from neutron emission, and the data above is due to charged particle emission. The total number of charged particles emitted per muon capture between \SI{1.4}{MeV} and \SI{26}{MeV} is \num{0.15(2)}. Sobottka and Wills note that the valley has a large statistical uncertainty and that the exponential decay parameter is 4.6~MeV.}
  \label{fig:intro:sispec}
\end{figure}

In the past, muon-to-electron conversion experiments used a fit to the silicon spectrum from Sobottka and Wills~\cite{Sobottka1968} (reproduced in Fig.~\ref{fig:intro:sispec}) scaled to aluminum based on estimates from radiochemical experiments. This fit was first developed for the MECO experiment~\cite{meco2001} and consists of a rapid rise above a threshold energy, and an exponentially-decaying tail to higher energies. The values of these parameters are known for silicon but charged particle emission for other materials could be significantly different. 

To measure the energy spectra of heavy charged particles emitted after nuclear muon capture, members of COMET and 
Mu2e collaborated on the AlCap experiment. AlCap's overall program consisted of three work packages: i) measurements of the emission of heavy charged particles following nuclear muon capture; ii) measurements of neutron yields after muon capture; and, iii) measurements of muonic X-rays and gamma rays 
required for the normalization of the stopping rate in COMET and Mu2e.

This paper describes the AlCap experiment and focuses on the results from data collected in 2015 for work package i). Specifically, it presents the yields and energy 
spectra of protons, deuterons, tritons, and $\alpha$-particles following nuclear muon capture on aluminum, silicon, and titanium. 


\section{The AlCap Experiment}
\label{sec:alcap}
\subsection{Experimental Strategy}
\label{sec:expt-strategy}
One of the goals of the AlCap experiment is to measure the energy spectra of heavy charged particles emitted after nuclear muon capture. Once the muon has reached the atomic ground state, 
nuclear muon capture and muon decay in orbit determine the total muon disappearance rate $\lambda_{t}=\lambda_{c} + \lambda_{d}$. Accordingly, the number of captured muons, $N_{c}$, is given by:
\begin{equation}
  N_{c} = N_{\mu}\; \dfrac{\lambda_{c}}{\lambda_{c} + \lambda_{d}},
  \label{eqn:n-cap}
\end{equation}
where $N_{\mu}$ is the number of stopped muons, $\lambda_{c}$ is the nuclear capture rate and $\lambda_{d}$ is the muon decay rate. These rates are well-known from measurements of the total disappearance rates $\lambda_{t}$~\cite{Suzuki1987}. The muon decay rate, $\lambda_{d}$, is close to the free muon decay rate \cite{Tishchenko:2012ie}, but is reduced by the Huff factor \cite{ABSE_Huff, Watanabe1993165}, a small nuclear dependent correction factor that accounts for bound state corrections arising from Coulomb corrections and relativistic effects.

AlCap determines the yield of particles of type $\eta$ emitted per unit energy and per captured muon 
\begin{equation}
    Y_{\eta}^{i}(E) = \dfrac{d N_{\eta}^{i}(E)}{d E\; N_c}\;.
  \label{eqn:rate}
\end{equation}
Experimentally, the yield 
\begin{equation}
    Y_{\eta}^{m}(E) = Y_{\eta}^{i}(E)\; \varepsilon_\eta(E)
  \label{eqn:rate1}
\end{equation}
is observed, 
where the indices $i$ and $m$ distinguish between the yield of 
\textit{initially} emitted particles after muon capture and those \textit{measured} with an energy dependent detector efficiency $\varepsilon_\eta(E)$, respectively. 

Using a germanium detector, AlCap measured the number of captured muons by counting the number of characteristic $2p$--$1s$ X-rays, $N_{\text{$2p$--$1s$}}$. The number of stopped muons, $N_{\mu}$, can be determined by using the equation
\begin{equation}
  N_{\text{$2p$--$1s$}} = N_{\mu}\; I_{\text{$2p$--$1s$}}\; \varepsilon_{\text{$2p$--$1s$}},
  \label{eqn:norm}
\end{equation}
where $I_{\text{$2p$--$1s$}}$ is the relative intensity of the $2p$--$1s$ X-ray (per stopped muon), and $\varepsilon_{\text{$2p$--$1s$}}$ is the efficiency of detecting the X-ray. Both the X-ray intensity and capture fractions are known from the literature (see Table~\ref{tab:muonic-atoms}) and so $N_{c}$ is determined by combining equations (\ref{eqn:n-cap}) and (\ref{eqn:norm}).

\begin{table}[htbp]
  \caption{Table of muonic atom information gathered from external sources for the AlCap stopping target materials.}
  \label{tab:muonic-atoms}
  \begin{center}
    \resizebox{1.0\columnwidth}{!}{
      \begin{tabular}{cccccc}
        \hline
        \hline
        \multirow{2}{*}{Material} & \multirow{2}{*}{Lifetime [ns]} & Capture  & \multicolumn{2}{c}{$2p-1s$ \Xray} & \multirow{2}{*}{Refs.} \\
        \cline{4-5}
                                  &                                & Fraction [\%] &  Energy [keV] & Intensity [\%]       & \\
        \hline
        Al & \AlLifetime & \AlCaptureFraction & \AlXRayE & \AlXRayI & \cite{Suzuki1987, Measday2007}\\
        Si & \SiLifetime & \SiCaptureFraction & \SiXRayE & \SiXRayI & \cite{Suzuki1987, Measday2007} \\
        Ti & \TiLifetime & \TiCaptureFraction & \TiXRayE & \TiXRayI & \cite{Suzuki1987, ENGFER1974509, PhysRevLett.18.1179}\\
        \hline
        \hline
      \end{tabular}
    }
  \end{center}
\end{table}

\begin{figure}[!htbp]
  \centering
  \includegraphics[width=\columnwidth]{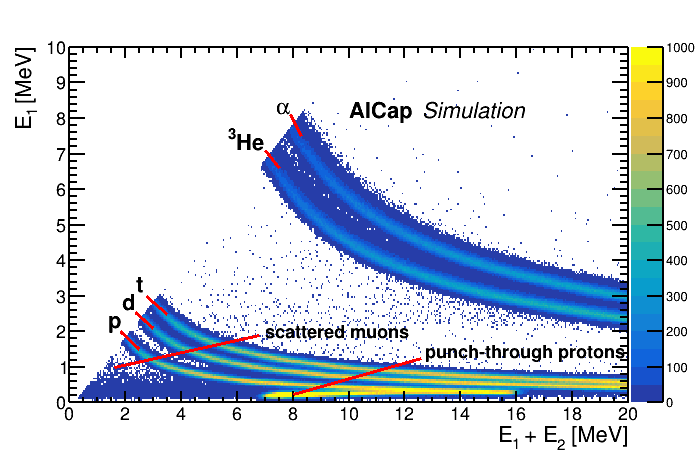}
  \caption{Demonstration of the particle identification method using the AlCap simulation. By plotting the energy deposited in the thin detector, $E_{1}$, against the total energy deposited in both layers, $E_{1}+E_{2}$, different bands corresponding to different particle types appear. The band of punch-through protons is vetoed by a third layer situated behind the second layer.}
  \label{fig:pid-demo}
\end{figure}

Silicon detector packages measured the energy of the emitted charged particles. These detector packages consisted of one thin and two thick layers of silicon diode detectors working as a \dEdx\ -- $E$ telescope. When the energy deposited in the first (thin) layer, $E_{1}$ (corresponding to \dEdx), is plotted against the total energy deposited in both layers, $E_{1}+E_{2}$, a band structure appears with each band characterizing a distinct particle type (Fig.~\ref{fig:pid-demo}). However, protons with sufficiently high energy will penetrate (punch through) both detectors (Fig.~\ref{fig:pid-demo}, ``punch-through protons''). A third layer of silicon was installed which could optionally be used to veto punch-through protons or measure their energy ($E_{1}+E_{2}+E_{3}$) if stopped by the third layer.

Measuring these spectra to much lower energies than available in the literature produces additional challenges. Low energy particles have a short range (less than \SI{100}{\micro\meter} in aluminum for a \SI{3}{\mega\electronvolt} proton) and an energy loss which changes rapidly as a function of energy. The energy lost in the target and other material between the target and detectors must be minimized. This requires very thin stopping targets and silicon detectors, all enclosed in a vacuum chamber. An intense low-energy muon beam with a narrow momentum width is needed to obtain a sufficient number of stopped muons in these thin targets. 

Even after minimizing energy loss, the measured spectrum has to be corrected for energy that is lost in the target and any passive detector layers. The measured yield, $Y^{m}(E)$, is given as a generalization of Eq.\eqref{eqn:rate1} is
\begin{align}
  \label{eqn:response-matrix}
	Y^{m}(E_{m}) &= \int Y^{i}(E_i) R(E_{m}, E_{i})\; dE_{i}, \\
    \bf{y^{m}} &= \bf{R} \,\bf{y^i}
\end{align}
where $R(E_{m} , E_i )$ is a response matrix that
 maps the initial energy $E_i$ to the measured energy $E_{m}$.
 The second equation expresses the same relation in matrix form
 more suitable to the experimental situation where both  $\bf{y^i}$ and $\bf{y^{m}}$ are vectors of energy bins.
 To construct the response matrix, the experiment was simulated to calculate the expected energy loss. The simulation incorporated the geometric acceptance, $\varepsilon$, of the detector packages into the response matrix.

\subsection{Experimental Setup}
\subsubsection{Muon Beam}
\begin{figure*}[!htbp]
  \centering
  \includegraphics[width=\textwidth]{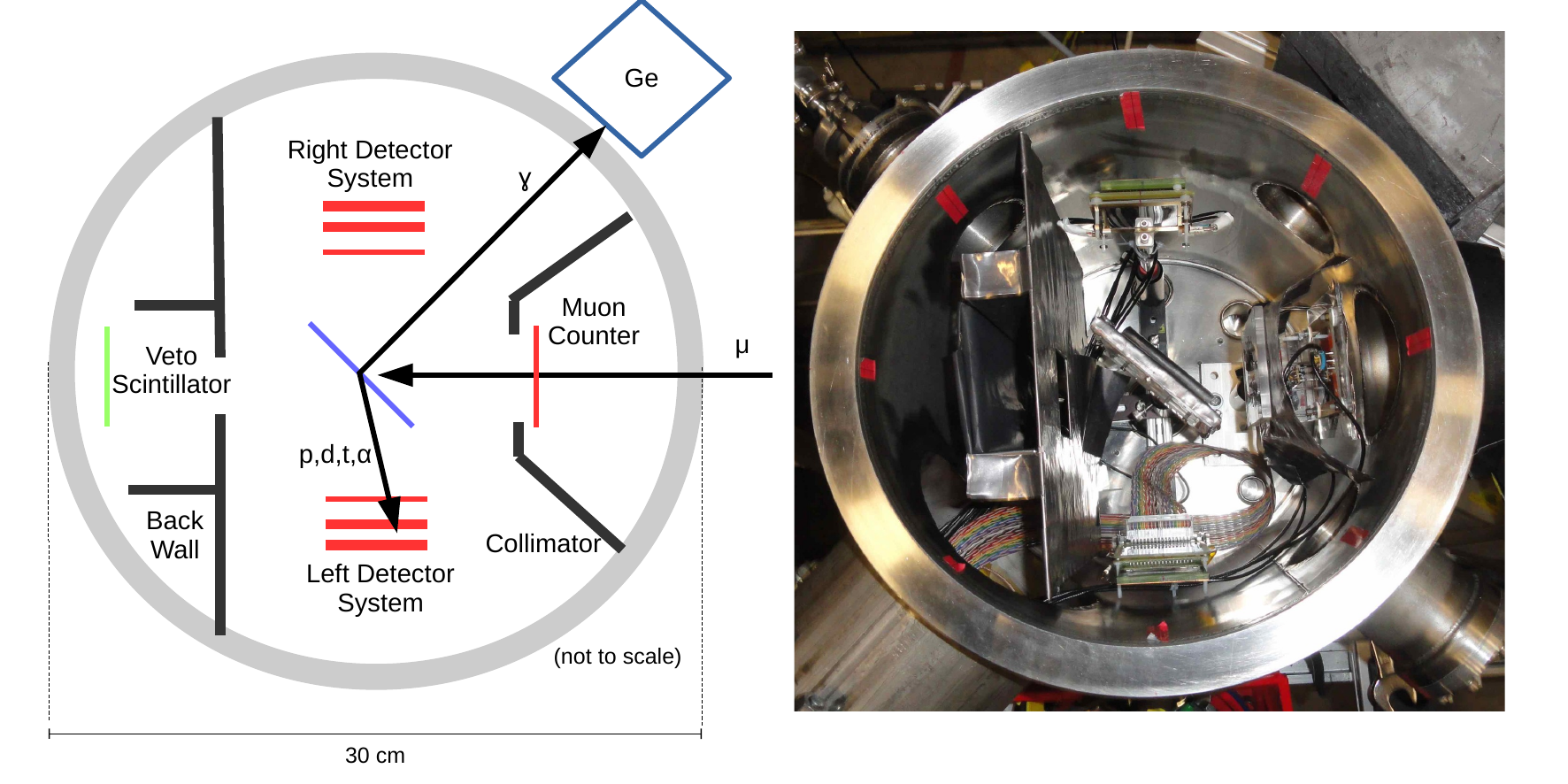}
  \caption{
  Diagram (left) and photo (right) of the AlCap apparatus. In the diagram, the gray circle shows the outer wall of the vacuum chamber (\SI{30}{cm} in diameter and \SI{60}{cm} in height), the black lines indicate lead shielding, the red lines indicate silicon detectors, and the blue line is the muon stopping target. The veto scintillator was not used in the analysis.}
  \label{fig:experimental-setup:alcap}
\end{figure*}

The experiment was conducted with the negative muon beam provided by the $\pi$E1 beamline at the High Intensity Proton Accelerator at the Paul Scherrer Institut (PSI)~\cite{piE1}. This beam met the energy, intensity, and momentum width requirements outlined in the previous section. The beam was tuned to obtain muon momenta between \SI{25.9}{\mega\electronvolt/\speedoflight} and \SI{28.4}{\mega\electronvolt/\speedoflight} with a momentum width of 3\% FWHM. Muon stopping rates were typically between 5 and 6 kHz. The pion contamination was negligible since a  \SI{26}{MeV/c} pion has a probability $O(10^{-5})$ to survive the 16~m long beamline. A measurement of the beam profile at the stopping target is given in Section~\ref{sec:simulation}.

\subsubsection{Vacuum Chamber and Muon Counter}
\renewcommand{\arraystretch}{1.4}
\begin{table*}[htbp]
  \begin{center}
  \captionsetup{justification=centering}
    \caption{Overview of the AlCap datasets.}
\label{tab:datasets}
\begin{tabular}{ccccccccccccc}
  \hline
  \hline
  & & \multirow{2}{*}{Material} & Thickness & Orientation & \multicolumn{2}{c}{Number of Muon} & Number of              & Number of   & \multicolumn{4}{c}{Number of Raw Detected Particles} \\
  \cline{10-13}
  & &                           & [$\mu$m] & [deg] & \multicolumn{2}{c}{Events [$\times10^6$]}  & X-Rays  & Muon Stops & Protons & Deuterons & Tritons & Alphas            \\
  \cline{6-7}
  & &          &                 &           & Total & Selected       & [$\times10^3$]          & [$\times10^6$]      &    [$\times10^3$]     &   [$\times10^3$]         &     [$\times10^3$]     &     [$\times10^3$]               \\
  \hline
  \multirow{3}{*}{\rotatebox[origin=c]{90}{Main}} & \multirow{3}{*}{\rotatebox[origin=c]{90}{Analyses}}
    & Al & 50 & 45 & \AlNTotalTMEsTab & \AlNAnalysedTMEsTab & \AlFiftyGeHiGainNXRaysTab & \AlFiftyGeHiGainNStoppedMuonsTab& \AlFiftyNRawProtonTab & \AlFiftyNRawDeuteronTab & \AlFiftyNRawTritonTab & \AlFiftyNRawAlphaTab \\
  & & Ti & 50 & 45 & \TiNTotalTMEsTab & \TiNAnalysedTMEsTab & \TiGeHiGainNXRaysTab & \TiGeHiGainNStoppedMuonsTab & \TiNRawProtonTab & \TiNRawDeuteronTab & \TiNRawTritonTab & \TiNRawAlphaTab \\
  & & Si & 52 & 45 & \SibNTotalTMEsTab &  \SibNAnalysedTMEsTab & \SibGeHiGainNXRaysTab & \SibGeHiGainNStoppedMuonsTab & \SibNRawProtonTab & \SibNRawDeuteronTab & \SibNRawTritonTab & \SibNRawAlphaTab \\

  \hline
  \multirow{2}{*}{\rotatebox[origin=c]{90}{Cross-}} & \multirow{2}{*}{\rotatebox[origin=c]{90}{checks}}
    & Al & 100 & 45 & \AlHundredNTotalTMEsTab & \AlHundredNAnalysedTMEsTab & \AlHundredGeHiGainNXRaysTab & \AlHundredGeHiGainNStoppedMuonsTab & \AlHundredNRawProtonTab & \AlHundredNRawDeuteronTab & \AlHundredNRawTritonTab & \AlHundredNRawAlphaTab \\
  & & Si & 1500 & 90 & \SiLNTotalTMEsTab & \SiLNAnalysedTMEsTab & \SiLGeHiGainNXRaysTab & \SiLGeHiGainNStoppedMuonsTab & \multicolumn{4}{c}{\SiLNTotalCountTab} \\

  \hline
  \hline
  \end{tabular}
\end{center}
\end{table*}

The AlCap apparatus, pictured in Fig.~\ref{fig:experimental-setup:alcap}, consisted of a stainless steel vacuum chamber  containing ion-implanted silicon detectors. The vacuum was less than \SI{3e-4}{mbar}. A thin (\SI{52}{\micro\meter}), \SI{5x5}{cm}, quadrant-segmented silicon detector (SiT) was positioned inside the vacuum chamber at the beam entrance of the chamber to tag incoming muons and provide a measure of the beam energy via $dE/dx$. Locating the entrance detector inside the vacuum reduced the distance between it and the stopping target, greatly lessening the effects of muon beam scattering and achieved stopping fractions (with uncertainties) of \SI{35(2)}{\percent}, \SI{38(4)}{\percent}, and \SI{20(2)}{\percent}, for the aluminum, titanium, and silicon targets, respectively.

\subsubsection{Targets}
The muon stopping target was centered in the chamber and oriented at \SI{45}{\degree} with respect to the incoming muon beam. Three target materials were used: aluminum, silicon, and titanium.

Aluminum is the primary choice of stopping target material for both COMET and Mu2e. The AlCap experiment used a \SI{50}{\micro\meter}-thick, \SI{>99}{\percent} pure aluminum foil target with natural isotopic abundance (100\% $^{27}$Al) for the main analysis. As a cross-check, data was taken with a thicker \SI{100}{\micro\meter}-thick aluminum target. This increased the muon stopping rate at the cost of larger corrections for energy loss. 

Titanium is an alternative stopping target material for COMET and Mu2e because the rate for muon-to-electron conversion is expected to be higher in titanium than in aluminum. 
However, this advantage is largely canceled by the shorter lifetime of muonic titanium, thereby reducing the number of muonic atoms available 
during the measurement period, which is delayed compared to the incident pulsed beam in order to avoid large backgrounds that arrive promptly at the stopping target.
AlCap made measurements on a \SI{>99}{\percent} pure, \SI{50}{\micro\meter}-thick titanium target of natural isotopic abundance (74\% $^{48}$Ti).

Silicon is the only material where charged particle spectra at the low energies relevant for COMET and Mu2e background estimates
have previously been measured~\cite{Sobottka1968}. AlCap used a \SI{52}{\micro\meter}-thick silicon detector of natural isotopic abundance (92\% $^{28}$Si) as a target for our main analysis. 
As a cross-check between the new experimental method and the previous measurement, AlCap used a \SI{1.5}{mm}-thick active silicon detector
oriented at \SI{90}{\degree} to the muon beam and measured capture products in the same detector. 
Finally, special runs with active, thin segmented silicon detectors as the target measured the energy and profile of the muon beam for different beam momenta. These data were used to tune the muon beam simulation (see Section~\ref{sec:simulation}).

The datasets are summarized in Table~\ref{tab:datasets}.

\subsubsection{Detectors}

Silicon wafer detectors measured the energy of the emitted charged particles. They were organized into two three-layer detector packages and were placed symmetrically beam-left and beam-right of the target. This symmetric geometry made the sum of measured events insensitive to the stopping depth of muons since a shift in the stopping depth in either direction reduces the path length for particles striking one arm, while increasing it for the opposite arm. Thus the energy deposition in the target averages in the sum of both arms. The layered setup for the detectors allowed for particle identification as described earlier.

Each detector package was composed of a thin (\SI{52}{\micro\meter}) detector (SiL1 and SiR1 for beam left and right, respectively), and two thick (\SI{1500}{\micro\meter}) detectors (SiL2, SiR2, SiL3, and SiR3). All had \SI{5x5}{cm} cross-sections, a \SI{0.5}{\micro\meter} dead layer, and a \SI{0.5}{\micro\meter} aluminum window~\cite{micron}. The thin silicon detectors were segmented to reduce noise due to detector capacitance: SiR1 was divided into four quadrants, and SiL1 into 16 vertical strips. Unfortunately, high-voltage problems and high noise rates in SiL3 and the first and last of the 16 strips in SiL1 resulted in these channels being excluded from the analysis, reducing the left arm's acceptance

The targets used for the silicon target datasets were used as detectors in the other datasets. SiL1 was the target for the main analysis and beam profile measurements, and SiL2 was the target for the cross-check analysis. For the main analysis, this meant that there was only one detector arm and so the advantages of the symmetric setup were lost, resulting in a larger systematic uncertainty.

A germanium detector measured the emitted $2p$--$1s$ X-rays for counting the number of stopped muons. A Canberra GC2018 coaxial high-purity germanium detector (HPGe) was placed outside of the chamber and upstream of the target. It was pointed towards the target through a port in the vacuum chamber with a \SI{2}{\milli\metre}-thick aluminum window.

\subsubsection{Shielding}

Muons that do not stop in the target could stop in the chamber wall or other parts of the infrastructure. 
Since the muonic atom lifetime of these materials is similar to that of the target materials, heavy charged particles emitted from these could be mistaken as coming from the target. To mitigate this, lead shielding was placed at all potential scattering locations. Muonic lead has a much shorter lifetime (\PbLifetime{ns}~\cite{Suzuki1987}) than the target materials (Table~\ref{tab:muonic-atoms}). 
Therefore, acceptance of charged particles was delayed until after the arrival of an incoming muon (as defined by the entrance counter) such that the vast majority of muonic lead atoms had decayed, while most muonic target atoms had not.
The lead shielding included a lead backstop, lead collimator, and lead casing on the target mount. 

\subsubsection{Readout and Data Acquisition}
\label{sec:daq}
A trigger-less, MIDAS-based~\cite{MIDAS, midas2}, data acquisition system collected data in 10-minute runs, composed of \SI{100}{ms} blocks. Amplified signals from every detector that went above a specified threshold were read out and analyzed by waveform digitizers (WFDs). The deadtime between the blocks was negligible. The sample rates varied depending on the digitizer employed, and were operated at stepped-down multiples of a \SI{500}{MHz} global clock.

The silicon detectors were connected to either an MSI-8 or an MPRS-16 Mesytec amplifier module~\cite{mesytec-msi8, mesytec-mpr16}. Each included a built-in charge-sensitive preamplifier, shaping amplifier, and timing filter amplifier. The thick silicon detectors (SiR2, SiR3, and SiL2) and channels from the quadrant detectors (SiR1 and SiT) were connected to MSI-8 modules. The 16 channels from SiL1 were connected to a similar MPRS-16 module. The germanium detector came with built-in preamplification, whose output was directed into an Ortec 673 shaping amplifier. Because the timing signals from the fast timing filter amplifiers of the silicon detectors had unacceptably high noise levels, the main analysis was based on the shaped amplifier output signals, which had excellent energy resolution at the expense of poorer time resolution (FWHM $O$(100 ns)).

The silicon shaping amplifier outputs were all connected to 8 channel, 12-bit Struck SIS 3300 VME-based WFD boards~\cite{struck}. The germanium shaping amplifier output was connected to an 8-channel, 14-bit CAEN V1724~\cite{caen}. All boards could sample up to \SI{100}{MHz}. To completely capture the pulses, the WFDs sampled the silicon detectors for \SI{5}{\micro\second} and the germanium detector for \SI{19}{\micro\second} after the pulse had gone above threshold. The WFDs used for the silicon detectors had a \SI{1}{\micro\second} deadtime after sampling a pulse, and so any particle arriving at a detector within \SI{1}{\micro\second} of another particle would be lost. The hit rates in all the detector packages were sufficiently low (\SI{\approx100}{\Hz}) that this deadtime was not an issue.

\subsection{Detector Calibration and Performance}
\label{sec:calibration}
To calibrate the energy response of the silicon detectors, a \ce{^241Am} source with a known particle emission spectrum (\SI{5.5}{MeV} $\alpha$-particles), and a pulser that injected a known charge directly into the detector preamplifiers were used. The
latter provided equivalent signals of \SI{0.75}{MeV}, \SI{1.50}{MeV}, \SI{2.25}{MeV}, and \SI{3.00}{MeV} for calibration. Minimum ionization energies were determined using beam electrons scattered by the target or emitted by decaying stopped muons.

The energy scale and efficiency of the germanium detector were calibrated with a \ce{^152Eu} source placed at the target position, thus the solid angle acceptance was included in the measurement. 
The source provided gamma rays in the range  \SIrange[range-phrase={--}]{244}{1410}{keV}, with well-known energies and intensities for each gamma-line. The absolute activity of the source was known to 3\%. 
The efficiency was cross-checked with the thick silicon target dataset, where $2p$--$1s$ X-rays coincident with a stopped muon provided a direct measure of the efficiency at \SI{400}{keV}. The 
efficiencies (and energies) for the aluminum, silicon, and titanium $2p$--$1s$ X-rays were \SI{6.6\pm0.2}{\times 10^{-4}} (\AlXRayE\ keV), \SI{6.14\pm0.09}{\times 10^{-4}} (\SiXRayE\ keV), and \SI{2.6\pm0.1}{\times 10^{-4}} (\TiXRayE\ keV), respectively.

The measured energy resolutions of all detectors are tabulated in Table~\ref{tab:experimental-setup:resolutions}. The energy resolutions for the silicon detectors are taken from the  \ce{^241Am} calibration, and the energy resolution for the germanium detector is taken from the aluminum $2p$--$1s$ X-ray peak.

\begin{table}[htb]
  \centering
  \caption{Energy resolutions (FWHM) and detector thicknesses of the AlCap detectors. For segmented detectors, the worst resolution is listed.}
   \label{tab:experimental-setup:resolutions}
   \begin{tabular}{ccc}
     \hline
     \hline
    \multirow{2}{*}{Detector} & Thickness          & Energy  \\
                              &[\si{\micro\meter}] & Resolution [\si{keV}] \\
    \hline
    SiT  & \num{58}   & \num{115(8)}         \\
    SiR1 & \num{58}   & \num{99(3)}   \\
    SiR2 & \num{1545} & \num{75(1)}    \\
    SiR3 & \num{1500} & \num{146(3)}   \\
    SiL1 & \num{52}   & \num{75(3)}      \\
    SiL2 & \num{1500} & \num{73(1)}    \\
    HPGe   & \num{44000} & \num{2.63(1)}  \\
    \hline
    \hline
\end{tabular}
\end{table}


\section{Monte Carlo Simulation}
\label{sec:simulation}
\begin{figure*}[!]
  \centering
  \captionsetup[subfigure]{justification=centering}
  \subfloat[][$p_{\text{beam}} = 25.9$ MeV/c]{\includegraphics[width=0.33\textwidth]{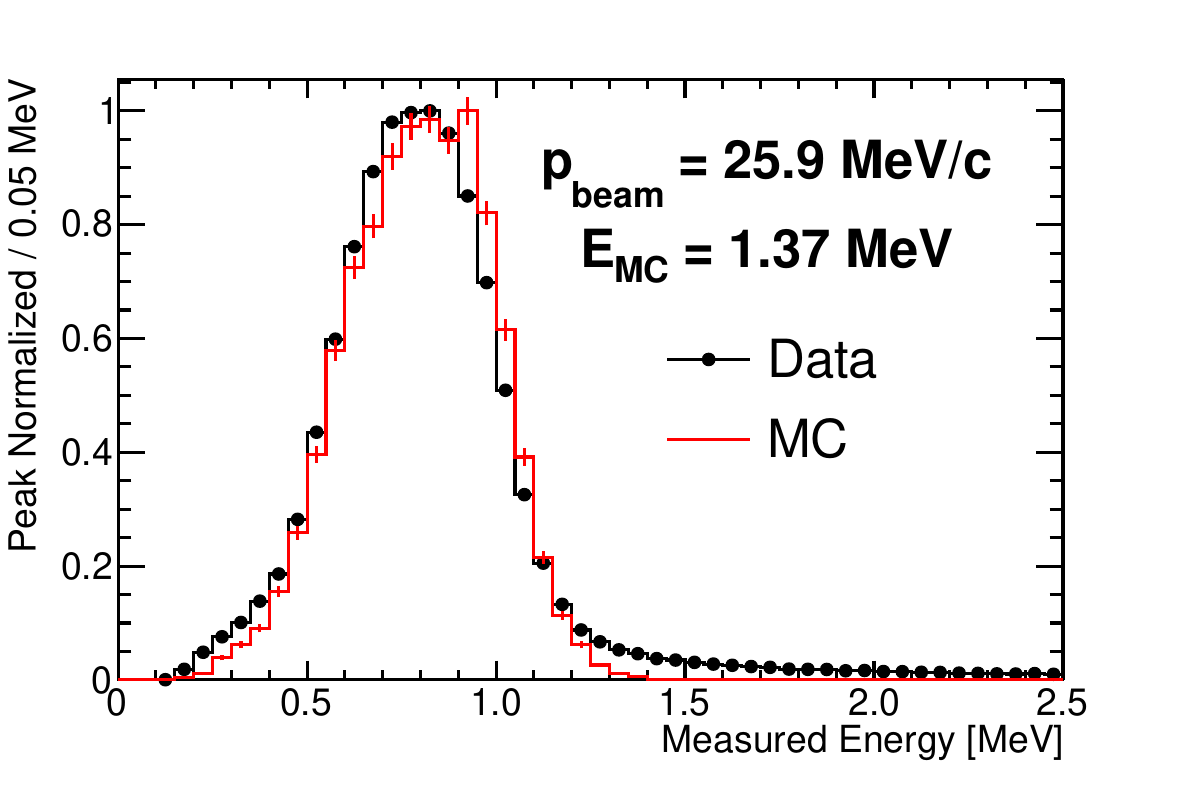}}
  \subfloat[][$p_{\text{beam}} = 26.2$ MeV/c]{\includegraphics[width=0.33\textwidth]{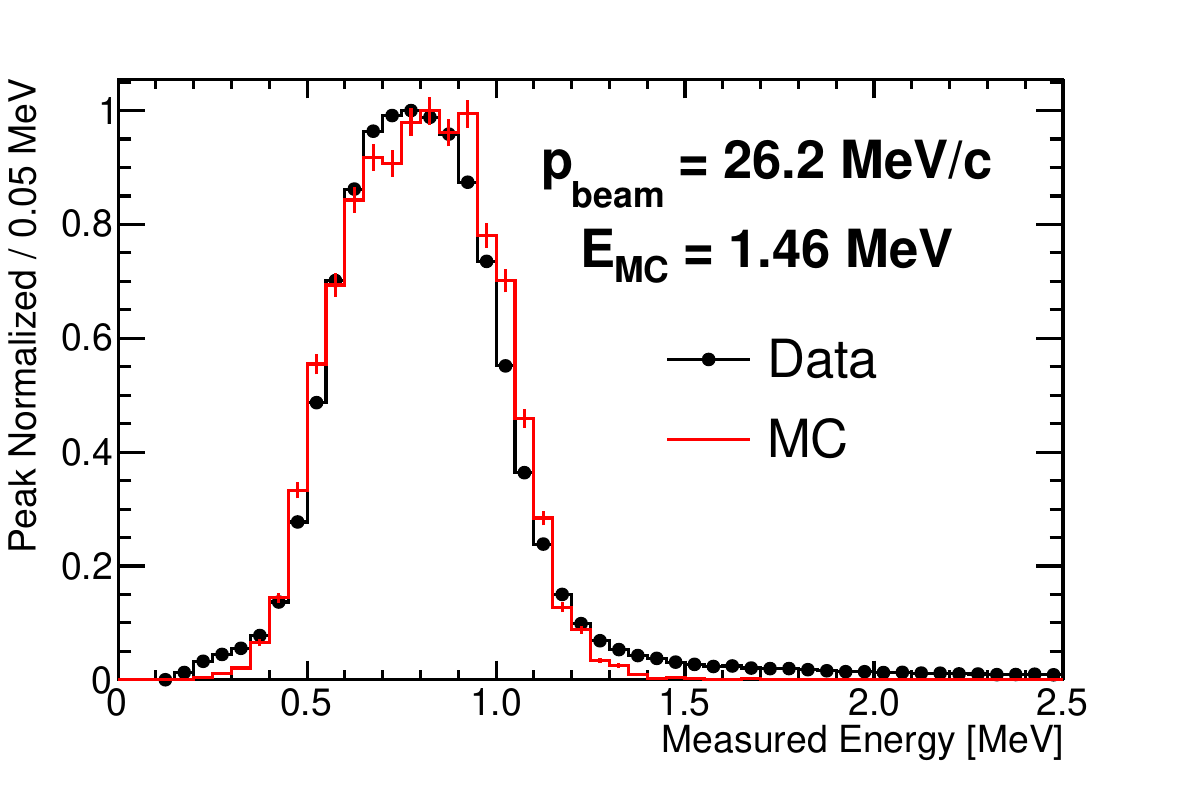}}
  \subfloat[][$p_{\text{beam}} = 26.7$ MeV/c]{\includegraphics[width=0.33\textwidth]{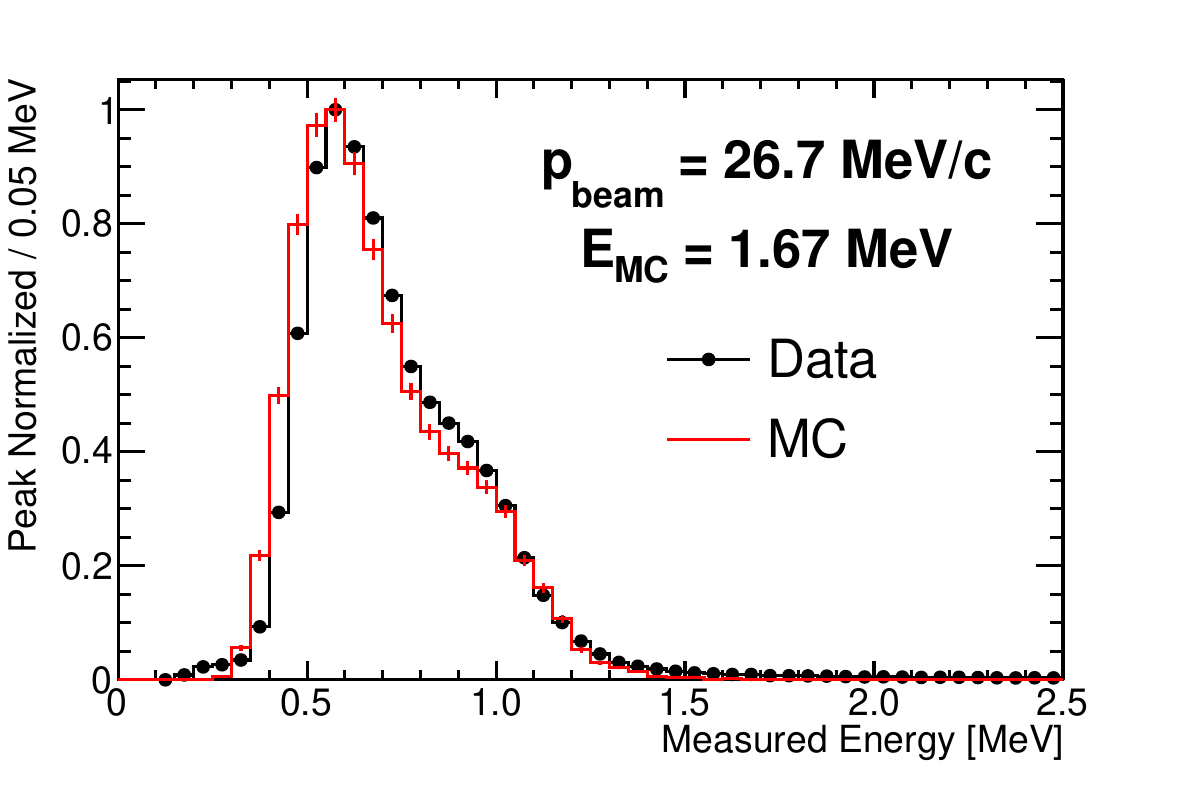}}
  \caption{Measured energy in a thin silicon target for different beam momenta from data (black) and simulation (red). The data points are from beam tuning scans with the muon momentum set to $p_{\text{beam}}$, and the simulation points are from muons generated with kinetic energy $E_{\text{MC}}$ upstream of the entrance window. As the beam momentum increases, more muons punch through the target. Since stopped muons deposit more energy than punch-through muons, the stopping peak is at higher energies than the punch-through peak and appears as a bulge at the highest energy shown here. Energies above \SI{1200}{\kilo\electronvolt} are from capture products not included in the first stage simulation.}
  \label{fig:beam-tune}
\end{figure*}

A Geant4-based (v10.5.p01)~\cite{Allison:2016lfl, Allison:2006ve, Agostinelli:2002hh} simulation was used to calculate both the energy lost in the target by heavy charged particles and the geometric acceptance of the silicon detector packages. The simulation was run in two stages. The first stage simulated the muon beam entering the chamber and stopping in the target, and the second stage simulated heavy charged particles traveling from the target to the silicon detectors. The input parameters of the muon beam simulation were tuned using data collected with the active 16-strip silicon detector as a stopping target using two orientations. This data provided the vertical and horizontal beam profiles at the stopping target after dispersion and scattering effects ($\sigma_{\text{horiz.}} \approx$\SI{15}{mm}, $\sigma_{\text{vert.}}\approx$\SI{20}{mm}). Beam profile data were also collected for different beam momenta. Figure~\ref{fig:beam-tune} shows the validation of the beam energy tuning.

The second stage of the simulation generated the heavy charged particles in the stopping target at the muon stopping positions. These particles were tracked as they lost energy leaving the target and deposited their remaining energy in the silicon detectors. From the simulation results, a response matrix is constructed by mapping the observed energy in the silicon detectors to the generated energy. This information, combined with measured detector resolutions and thresholds, was used to unfold the energy lost as charged particles left the target. 
Since the simulated heavy charged particles were emitted in the full \num{4\pi} solid angle, the response matrix also encoded the geometric acceptance of the detectors. 


\section{Data Preparation}
\label{sec:data-preparation}
\subsection{Digitized Pulse Analysis}
\label{sec:pulse-analysis}

\begin{figure*}[!htbp]
  \centering
  \captionsetup[subfigure]{justification=centering}
  \subfloat[][Standard Pulse Analysis\label{fig:pulse-analysis:standard}]{\includegraphics[width=1.0\columnwidth]{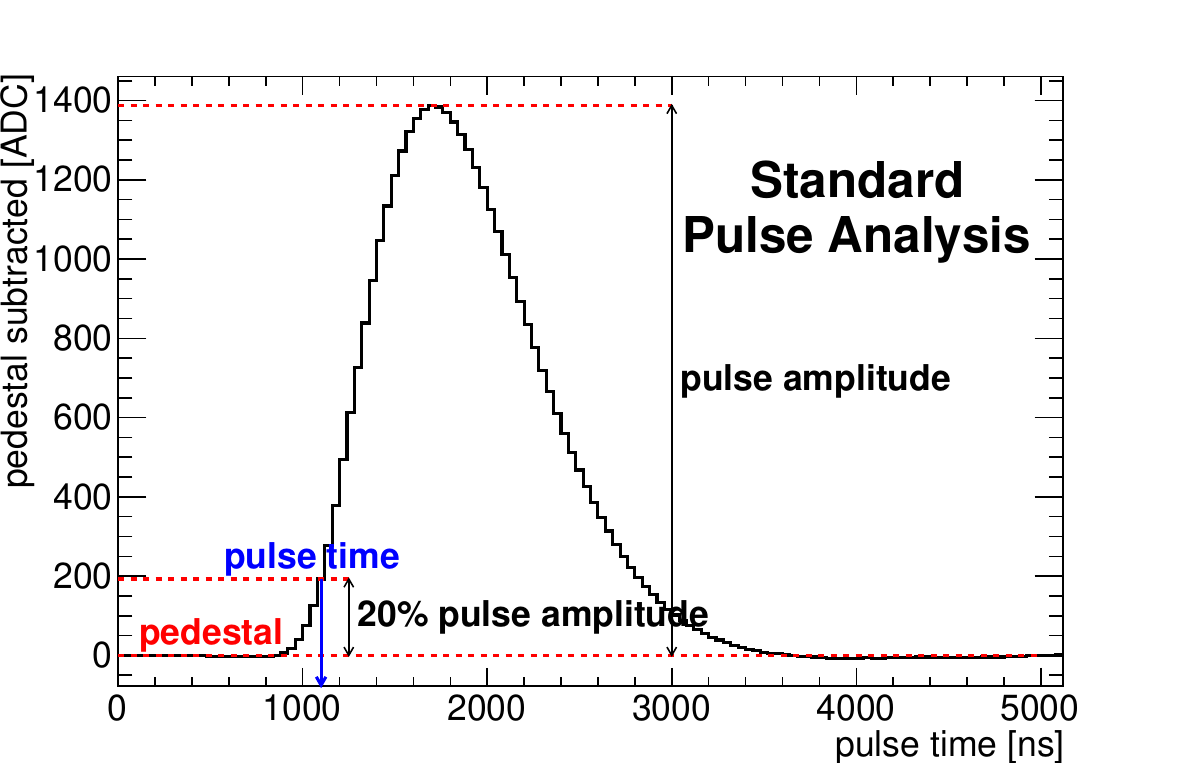}}
  \subfloat[][Template Fitter Pulse Analysis 
  \label{fig:pulse-analysis:template}]{\includegraphics[width=1.0\columnwidth]{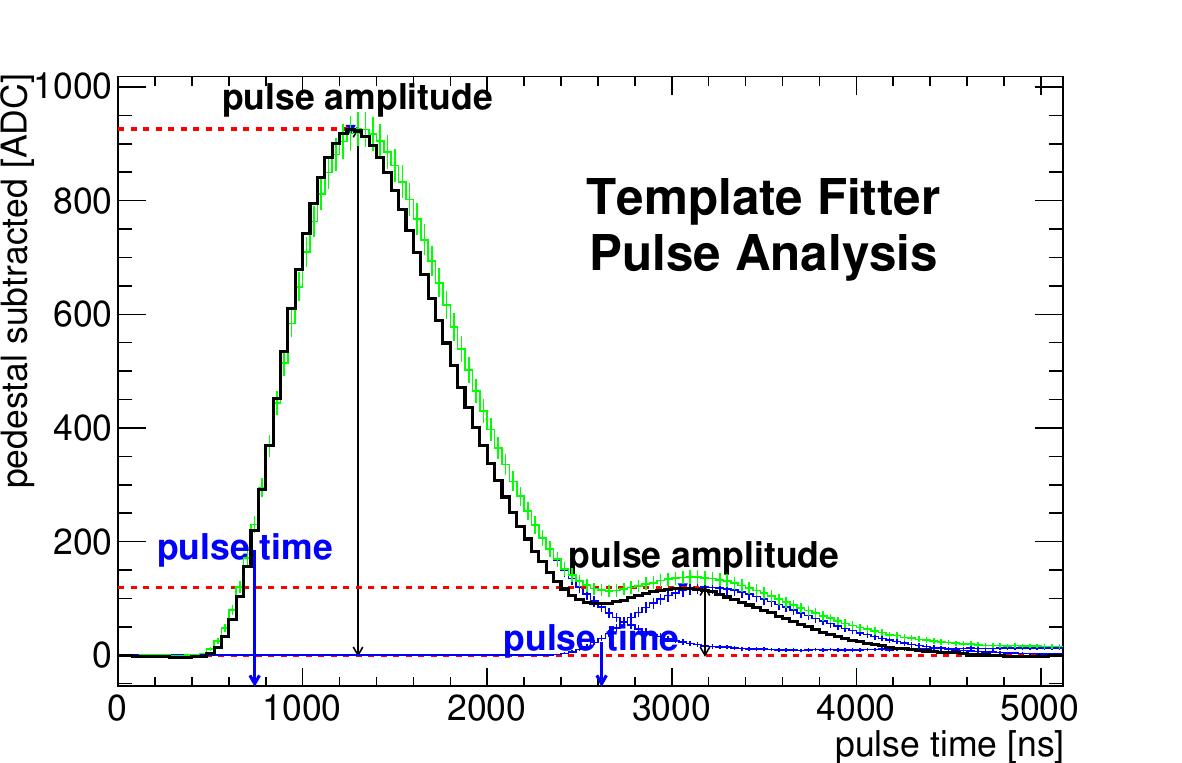}}
  \caption{Diagrams of the two pulse analysis techniques used in this paper. The standard pulse analysis (left) was used in the main analyses, and the template fitter pulse analysis (right) was used in the thick active-target silicon dataset in Section~\ref{sec:cross-checks:silicon}. In the standard pulse analysis, the amplitude and time of a digitized pulse (black) are defined as the peak minus pedestal, and the time the pulse crosses 20\% of its amplitude, respectively. In the template fitter pulse analysis, individual templates (blue) are combined and fitted (green) to the digitized pulse (black). The energy and time of the subpulses are defined in the same way as the standard pulse analysis.}
  \label{fig:pulse-analyses}
\end{figure*}

The energies and times of the detector hits were extracted by analyzing the digitized waveforms. 
The amplitude of the pulse (defined as the maximum WFD sample amplitude after pedestal subtraction) was proportional to the deposited energy. 
The time of the pulse was defined as the time when the pulse reached \SI{20}{\percent} of its maximum amplitude. 
Figure~\ref{fig:pulse-analysis:standard} shows a typical pulse from one of the silicon detectors with the amplitude and time definitions described.

\section{Data Analysis}
\label{sec:analysis}

\subsection{Muon Stops and X-ray Analysis}
\label{sec:normalization}
All analyses were based on \textit{muon events}. A muon event consisted of a hit in the entrance counter (the \textit{central muon}) as well as any hits in any of the other detectors which occurred within \SI{\pm20}{\micro\second} of the entering muon. A \textit{pile-up protection} cut of \SI{\pm10}{\micro\second} was applied so that accepted muon events had no other entrance hits within \SI{10}{\micro\second}. These times are much longer than the muonic atom lifetimes of the target materials ($O(100~\text{ns})$, see Table~\ref{tab:muonic-atoms}), but shorter than the time between arriving muons (typically $\approx$\SI{150}{\micro\second}). For the silicon target analysis, the muon event was also required to have at least one hit in the target.  

\begin{figure}[htbp]
  \centering
  \captionsetup[subfigure]{justification=centering}
  \subfloat[][Aluminum\label{fig:analysis:normalization-al50}]{\includegraphics[width=1.0\columnwidth]{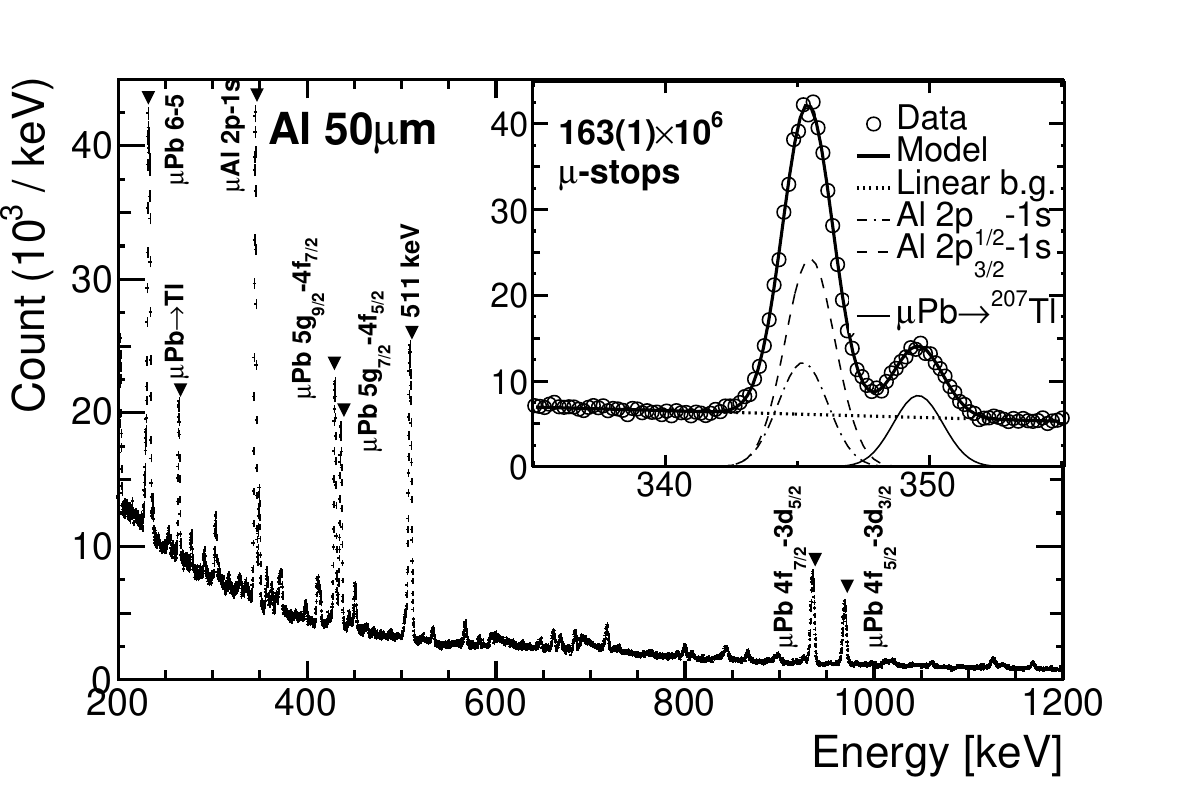}} \\
  \subfloat[][Silicon\label{fig:analysis:normalization-si16b}]{\includegraphics[width=1.0\columnwidth]{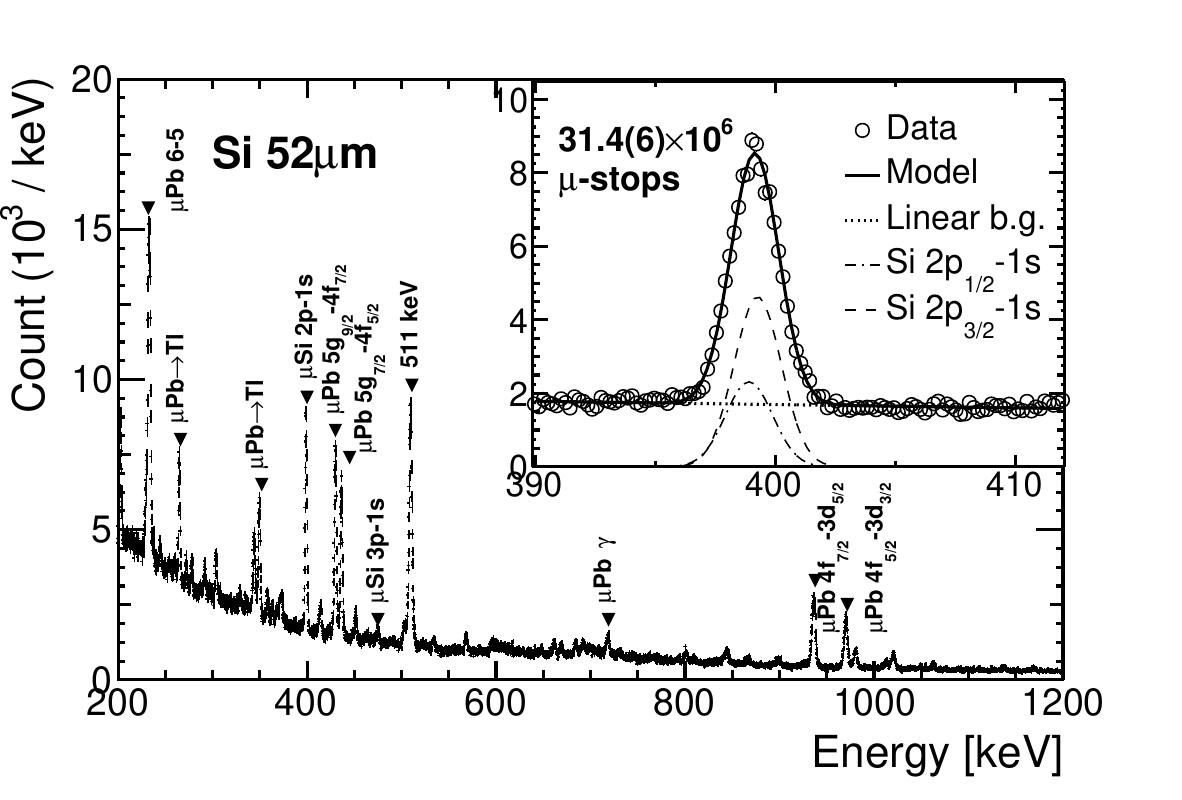}}\\
  \subfloat[][Titanium\label{fig:analysis:normalization-ti}]{\includegraphics[width=1.0\columnwidth]{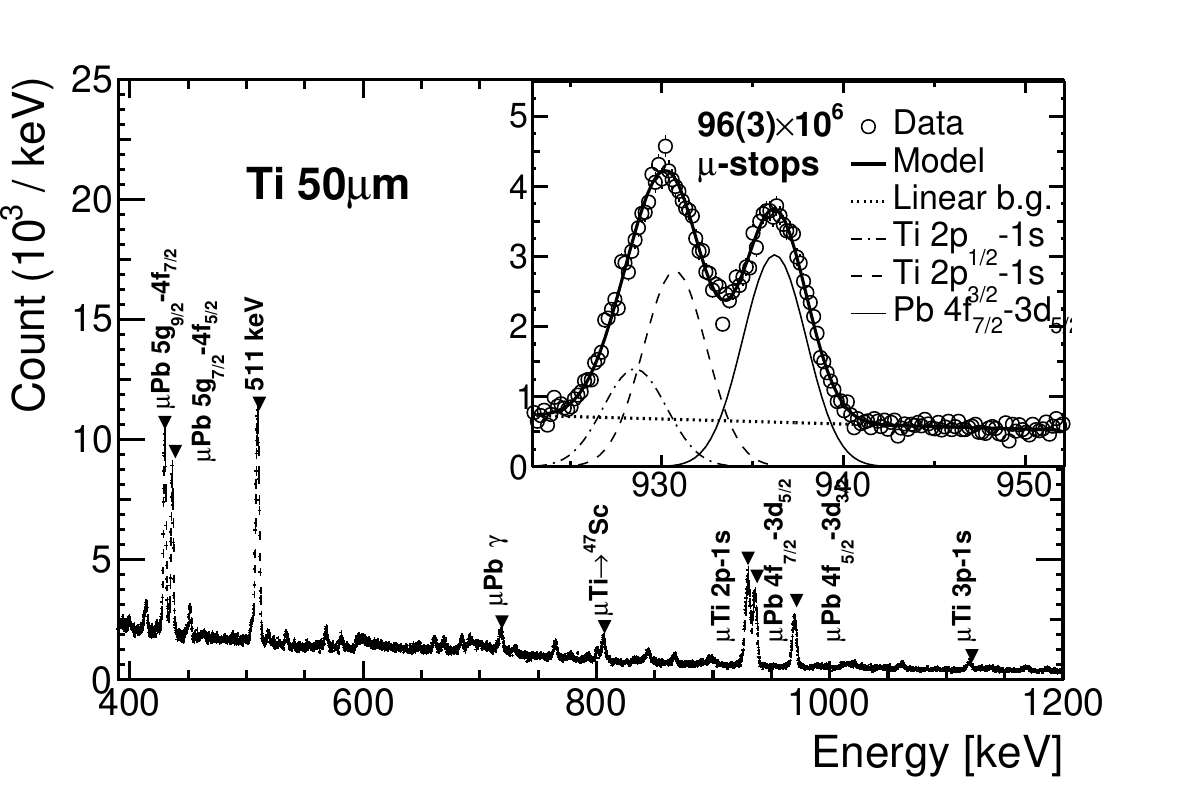}}
  \caption{X- and $\gamma$-ray spectra, used to determine the number of muon stops in the targets. The fine splitting of the $2p$ level was included. Nearby peaks in the Al and Ti datasets were due to muon stops in lead.}
  \label{fig:analysis:normalization}
\end{figure}

The number of muons stopping in the target was evaluated according to Eq.\eqref{eqn:norm} based on the characteristic X-rays observed with the germanium detector.
The $2p$--$1s$ peak was fitted to a double Gaussian shape (corresponding to the fine-structure splitting of the $2p$ level) with a linear background. For the aluminum and titanium analyses, there were additional nearby background lines that required a triple Gaussian fit. The X-ray spectra are shown in Fig.~\ref{fig:analysis:normalization}.

The number of captured muons is determined using Eq.\eqref{eqn:n-cap}. Since the capture rate is well-known, the uncertainty in the number of captured muons is dominated by the uncertainty in the number of stopped muons.

The total number of muon events, the number of muon events that passed the selection criteria and the number of stopped muons are compiled in Table~\ref{tab:datasets}.

\subsection{Charged Particle Analysis}
\subsubsection{Hit Selection in the Detector Arms}
\label{sec:hit-selection}
\begin{figure}[!htbp]
  \centering
  \includegraphics[width=1.0\columnwidth]{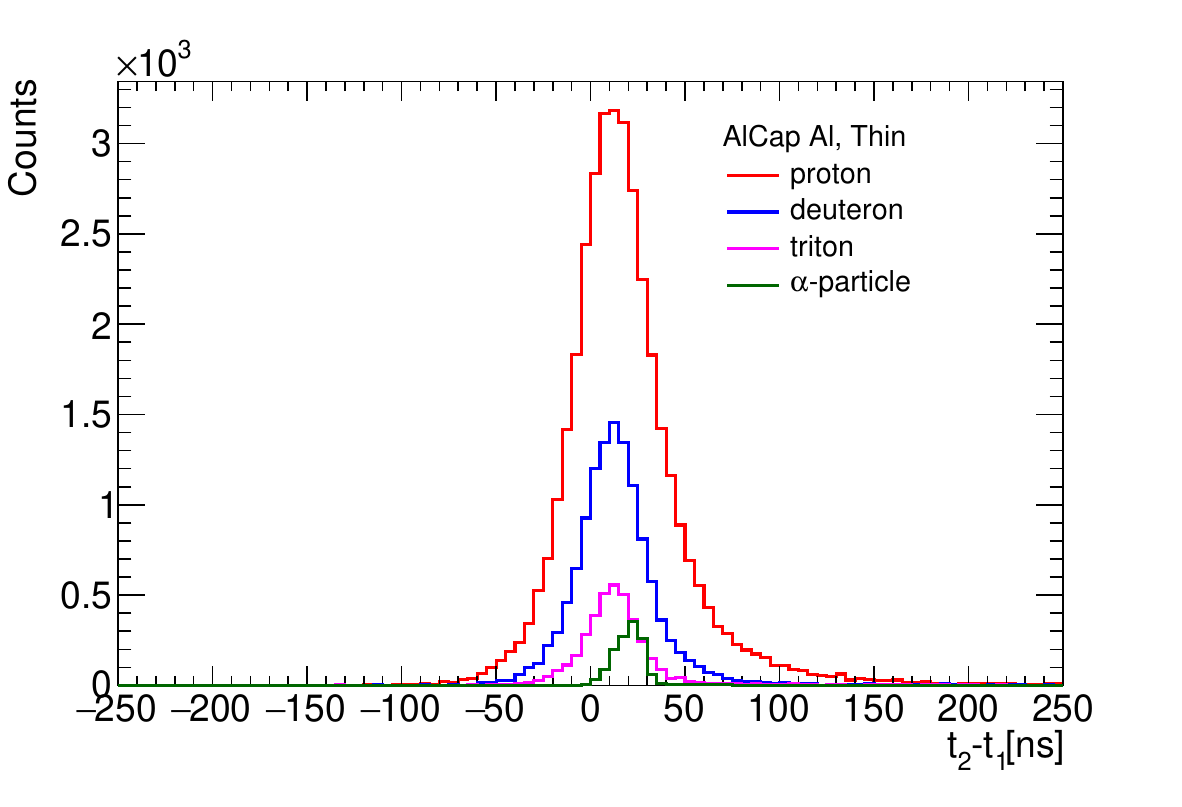}
  \caption{Time between hits in the first two layers for different charged particles as selected by the data-driven particle identification method.}
  \label{fig:analysis:coincidence-time}
\end{figure}

\begin{figure}[!htbp]
  \centering
    \subfloat[][\EvdE\ for selected hits (black) and those that are vetoed by the third layer (red). Other regions of interest 
    are (i) scattered muons, (ii) electrons and noise, (iii) punch-through protons and (iv) punch-through deuterons. 
    The $\alpha$-band is at higher $E_1$ energies outside the plot range.
    \label{fig:analysis:evde-plots-two-layer}]
    {\includegraphics[width=\columnwidth]{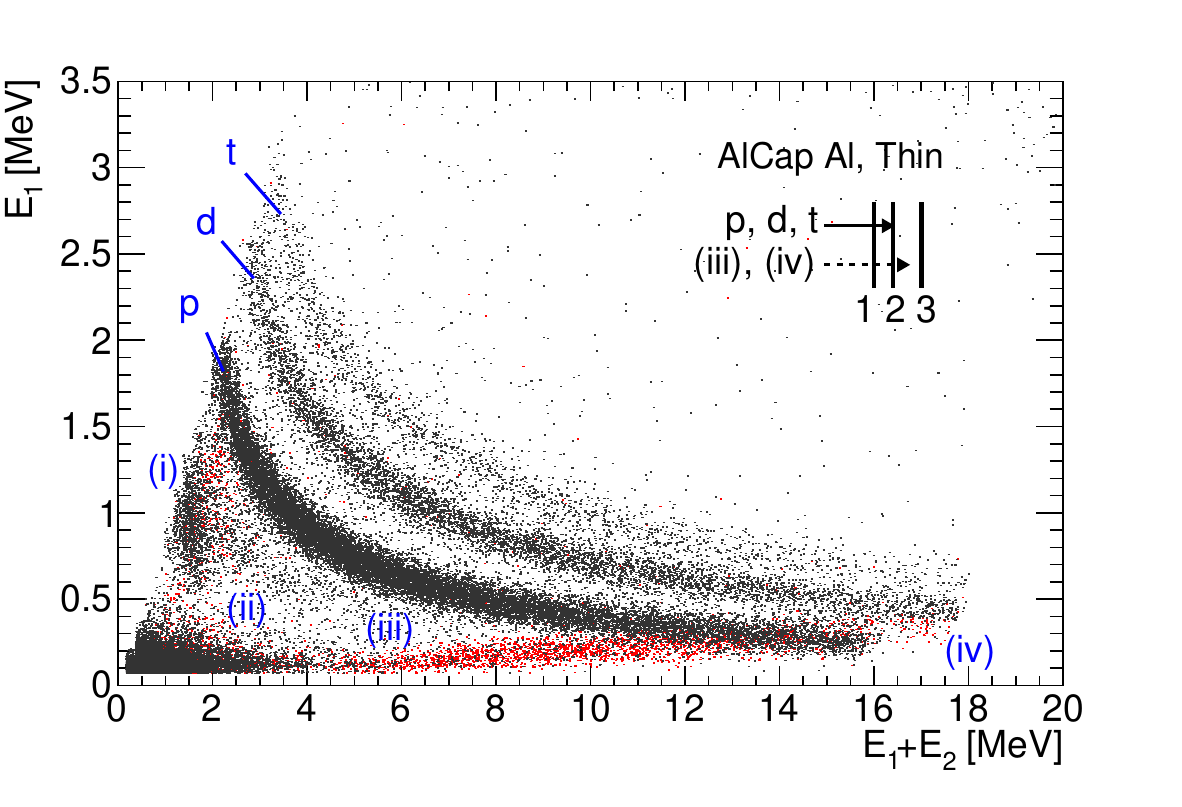} }\\
    \subfloat[][\EvdEThree\ for protons and deuterons ($p$ and $d$) punching though the second silicon layer (red points in top figure).  Other regions of interest are (v) noise, electrons and muons, (vi) punch-through protons and (vii) punch-through deuterons. 
    Only punch-through protons up to \SI{20}{MeV}, which fully stop in the third layer, are recovered in the analyses.\label{fig:analysis:evde-plots-three-layer}]{\includegraphics[width=\columnwidth]{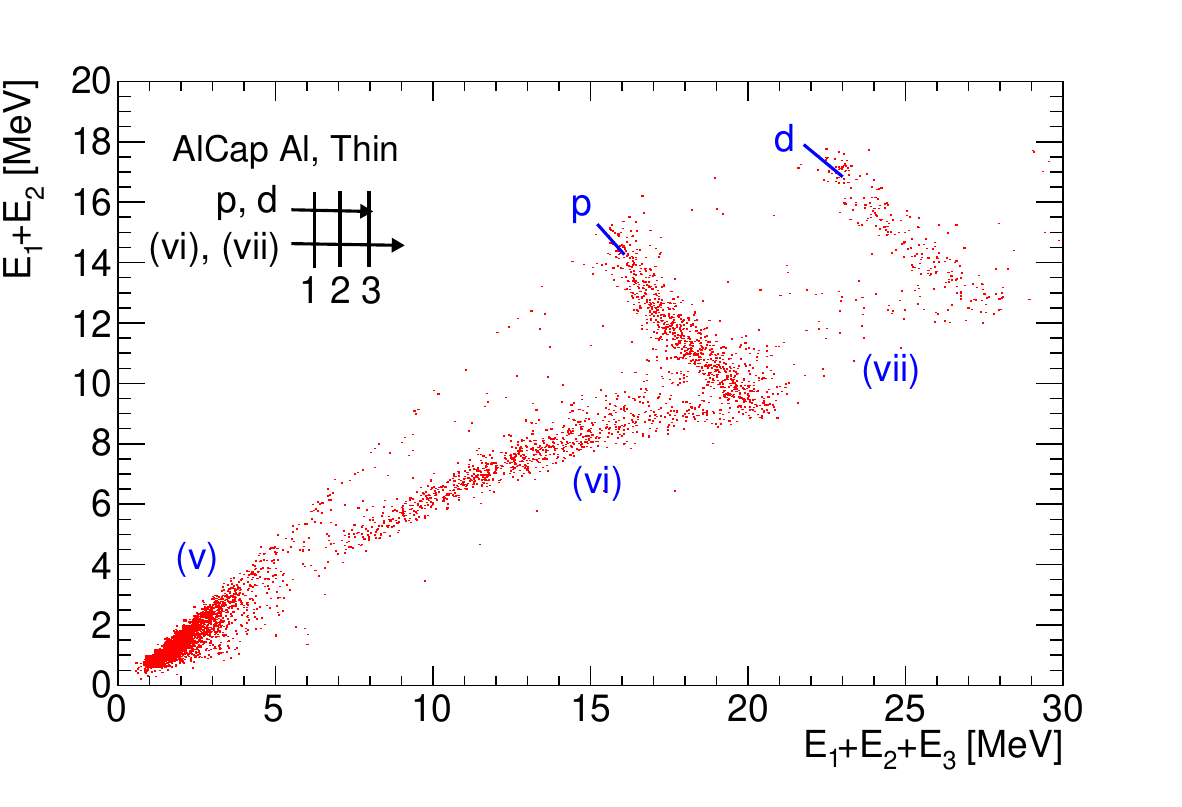} }
  \caption{\EvdE{} and \EvdEThree{} plots for the Al dataset demonstrating the third-layer veto.}
  \label{fig:analysis:evde-plots}
\end{figure}

Hits in the detector arms were selected if the time difference between the first and second layers was less than \SI{500}{ns} (Fig.~\ref{fig:analysis:coincidence-time}). Plotting \EvdE{} for the selected hits gives rise to bands corresponding to different particle types (Fig.~\ref{fig:analysis:evde-plots-two-layer}). Particle bands corresponding to protons, deuterons, tritons, and $\alpha$-particles are seen. There is no evidence for the emission of $^{3}$He nuclei. The lowest detectable energy is defined by the energy required to pass through the first detector layer and reach the second detector layer.
 
High-energy protons can punch through the second layer. These appear in the \EvdE{} plot at $E_1\approx\SI{0.2}{MeV}$ and $E_1+E_2>\SI{0.6}{MeV}$. These events were suppressed by a veto from the third layer which was \SibVetoEff{} efficient. 
The third-layer hit was used to extend the energy range of the proton analyses by creating an \EvdEThree{} plot (Fig.~\ref{fig:analysis:evde-plots-three-layer}) with coincident hits in all three layers.

For data where there was no third layer (i.e. in the left detector package and the thick aluminum cross-check dataset), there exists a noticeable bump between \SI{11}{MeV} and \SI{15}{MeV} due to the overlap between stopped and punch-through protons. A simulation-derived punch-through correction removed this background. In this correction, the simulation provides the fraction of protons that would punch through the second layer assuming a proton energy spectrum as given by the right-detector package. This method is verified by comparing the simulation-derived punch-through correction with the veto-corrected data from the right detector package. The simulation-derived punch-through correction was also used for the proton spectra between 17 MeV and 20 MeV to account for protons that punch through the third layer.

To suppress backgrounds due to muons stopping in the lead shielding, an arrival time cut is applied on the hits. Only detector hits that arrived between \SI{0.4}{\micro\second} and \SI{10}{\micro\second} after the muon passed through the entrance counter were accepted. Within this time window, \SI{62.9}{\percent} of aluminum muonic atoms and \SI{29.7}{\percent} of titanium muonic atoms either decay or capture. For the silicon analysis, no time cut was used since the muon event selection required a coincident signal in the active target, which removed any lead background.

\subsubsection{Particle Identification}
\label{sec:band-extraction}
Two different techniques were used to extract the particle bands from the \EvdE{} plots. The first method used a data-driven process to define the cuts, and the second used power-law functions to delineate the bands. Both methods gave consistent results.

\begin{figure}[!htbp]
  \centering
  \subfloat[][
  Geometric cuts obtained by the data-driven method for $2\sigma$ width. On the right inset, an example of a projection from the region between the dotted lines is shown along with its fit.\label{fig:analysis:rotated-pid-cut}]{
\hspace*{-.47in}
\includegraphics[width=\columnwidth]{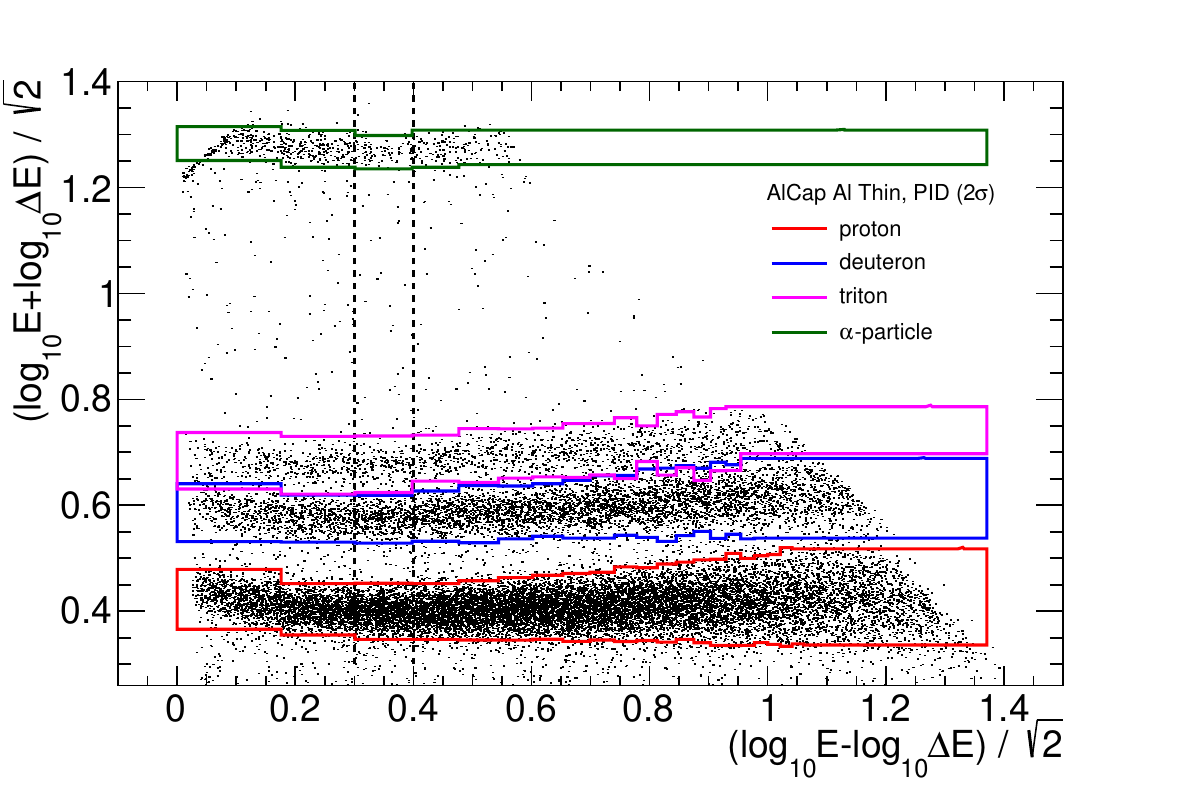}
\hspace*{-.39in}
\raisebox{0.06\height}{\includegraphics[angle=90,height=.637\columnwidth]{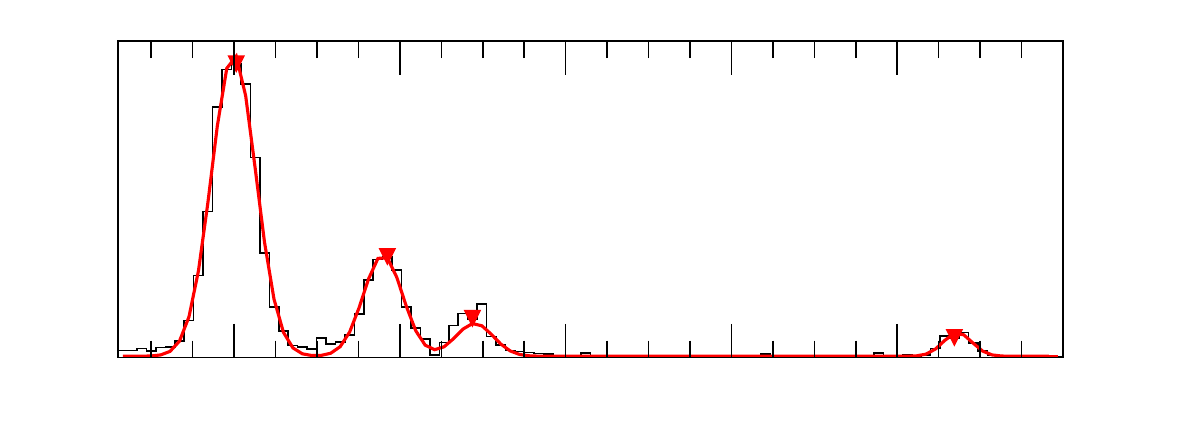} }
}\\
  \subfloat[][Geometric cuts obtained from the power-law function method. These bands are motivated by the Bethe formula.\label{fig:analysis:johns-pid-cut}]{\includegraphics[width=1.0\columnwidth]{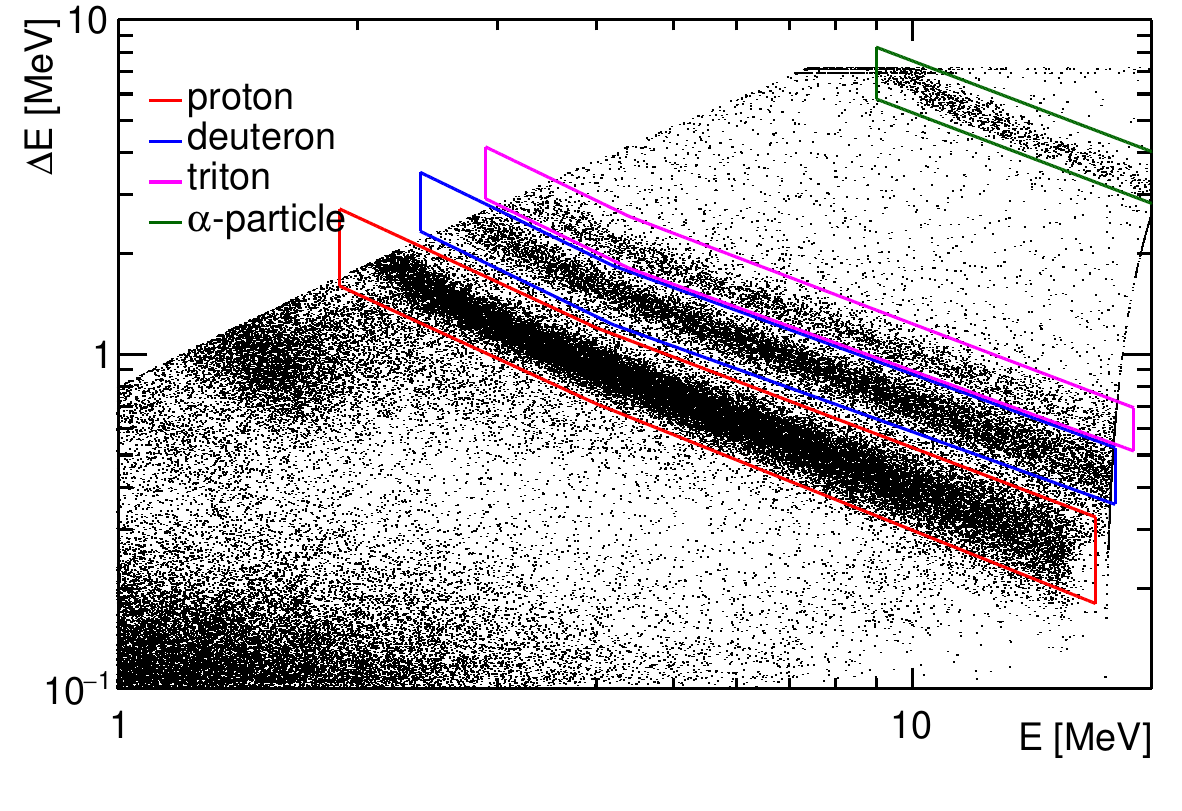}}
  \caption{Comparison between the two particle identification methods on the Al dataset: (top) data-driven method, and (bottom) power-law function method where $\Delta E = E_1$ and $E = E_1 + E_2$.}
  \label{fig:analysis:pid-cuts}
\end{figure}

In
the data-driven method (Fig.~\ref{fig:analysis:rotated-pid-cut}), the particle bands were linearized by taking the log of each axis, and then rotating the axes \SI{45}{\degree} counterclockwise. The linearized bands were then sliced along the x-axis and Gaussian functions were fitted in each slice (Fig.~\ref{fig:analysis:rotated-pid-cut}, right inset). In energy regions that had low statistics ($E>\SI{12}{MeV}$), the last stable fit parameters were used for subsequent slices. The fits define geometric cuts corresponding to $\pm3\sigma$ width, $\pm2\sigma$ width, etc. With this method, the efficiency of our band extraction is defined by the width. In this analysis, a width of $2\sigma$ was chosen as a compromise between efficiency and purity.

\newcommand{\y}{0.8}
\begin{figure*}[!htbp]
  \centering
  \includegraphics[width=\y\textwidth]{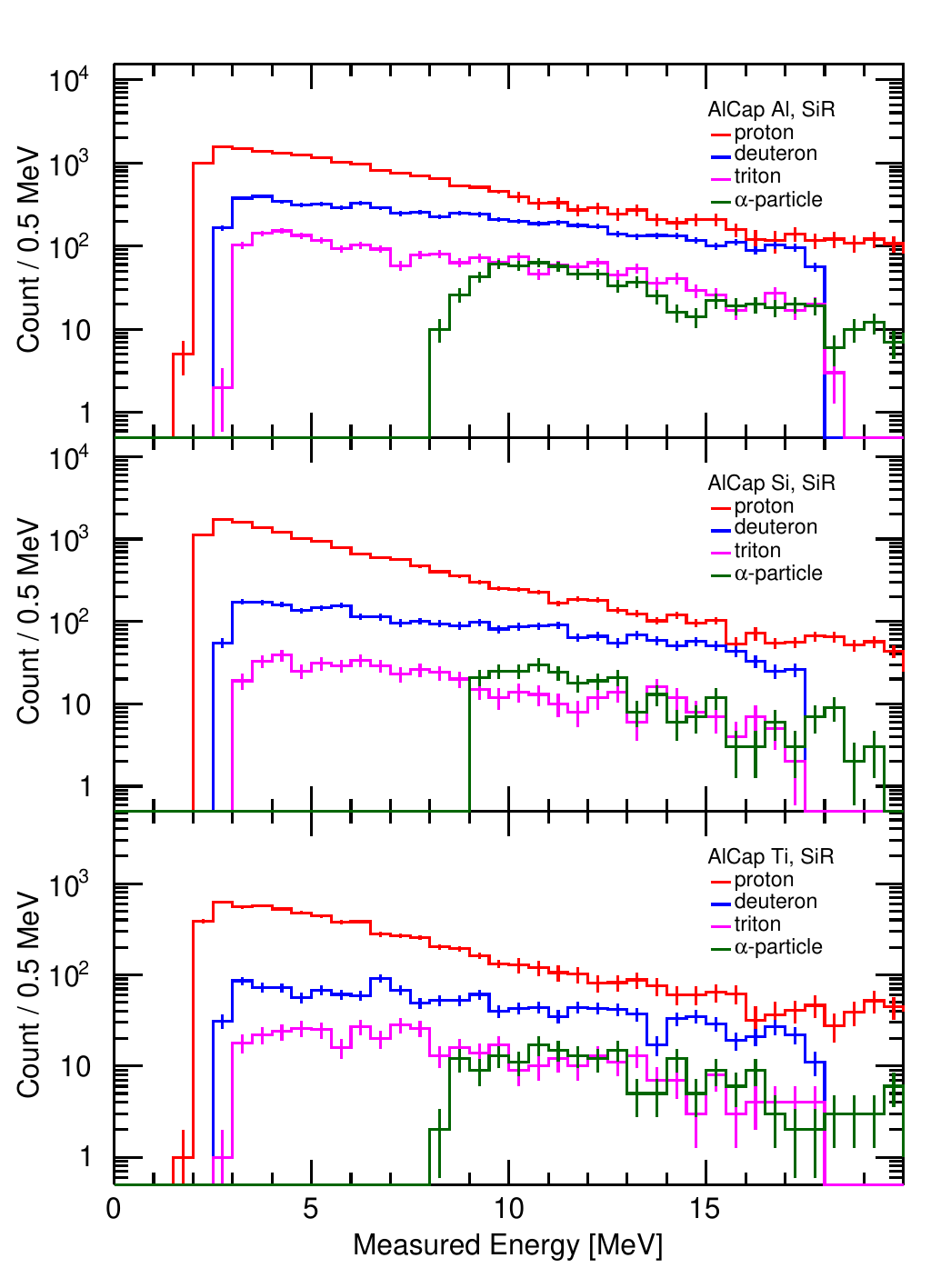}
  \caption{Measured charged particle energy spectra for aluminum (top), silicon (middle) and titanium (bottom) from the right detector package.} 
  \label{fig:analysis:charged-particles-folded}
\end{figure*}

In the power-law function method (Fig.~\ref{fig:analysis:johns-pid-cut}), a functional form, motivated by the Bethe-Bloch formula, parameterized each particle band. Two functions of the form $AE^{\alpha}$ were defined for each band: one corresponded to the upper edge, and the other corresponded to the lower edge of the band. $A$ and $\alpha$ are adjustable parameters and their values were determined by eye. The efficiencies and purities of this band extraction method were obtained from the Monte Carlo simulation. 

Figure~\ref{fig:analysis:charged-particles-folded} shows the measured energy spectra for each particle type as extracted with the data-driven method. The power-law function method applied to the same data produced nearly identical results.

\begin{figure}[!htbp]
  \centering
  \includegraphics[width=1.0\columnwidth]{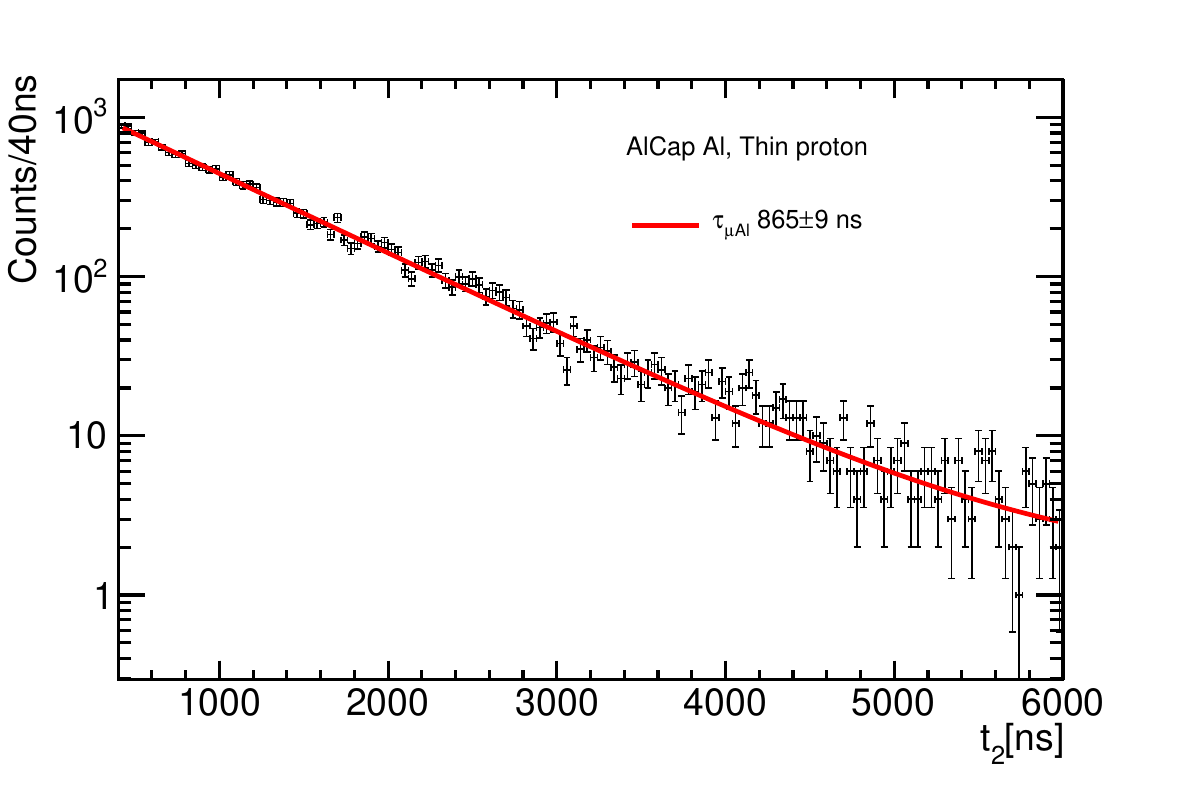}
  \caption{Arrival times of protons in the Al dataset. An exponential fit plus a constant (red line) obtains a muonic atom lifetime consistent with the literature.}
  \label{fig:analysis:time-fit}
\end{figure}

To ensure that the extracted particles are from muonic atom decays in the target, the arrival time distribution of the selected particles was fitted to an exponential function plus a constant. Figure~\ref{fig:analysis:time-fit} shows this fit for the aluminum dataset. There was good agreement between the fitted lifetimes ($\tau_{\text{Al}} = 865(9)$ ns, $\tau_{\text{Si}} = 757(6)$ ns, and $\tau_{\text{Ti}} = 326(4)$ ns) and those given in the literature (Table \ref{tab:muonic-atoms}).

Due to a gain non-linearity effect at high energies, the $\alpha$-particle band of the left-detector package was ignored.

\subsubsection{Unfolding}
\label{sec:analysis:unfolding}
\begin{figure*}[!htbp]
  \centering
  \captionsetup[subfigure]{justification=centering}
  \subfloat[][Simulated protons in the left detector\label{fig:analysis:unfold:tm_left_proton}]{\includegraphics[width=0.95\columnwidth]{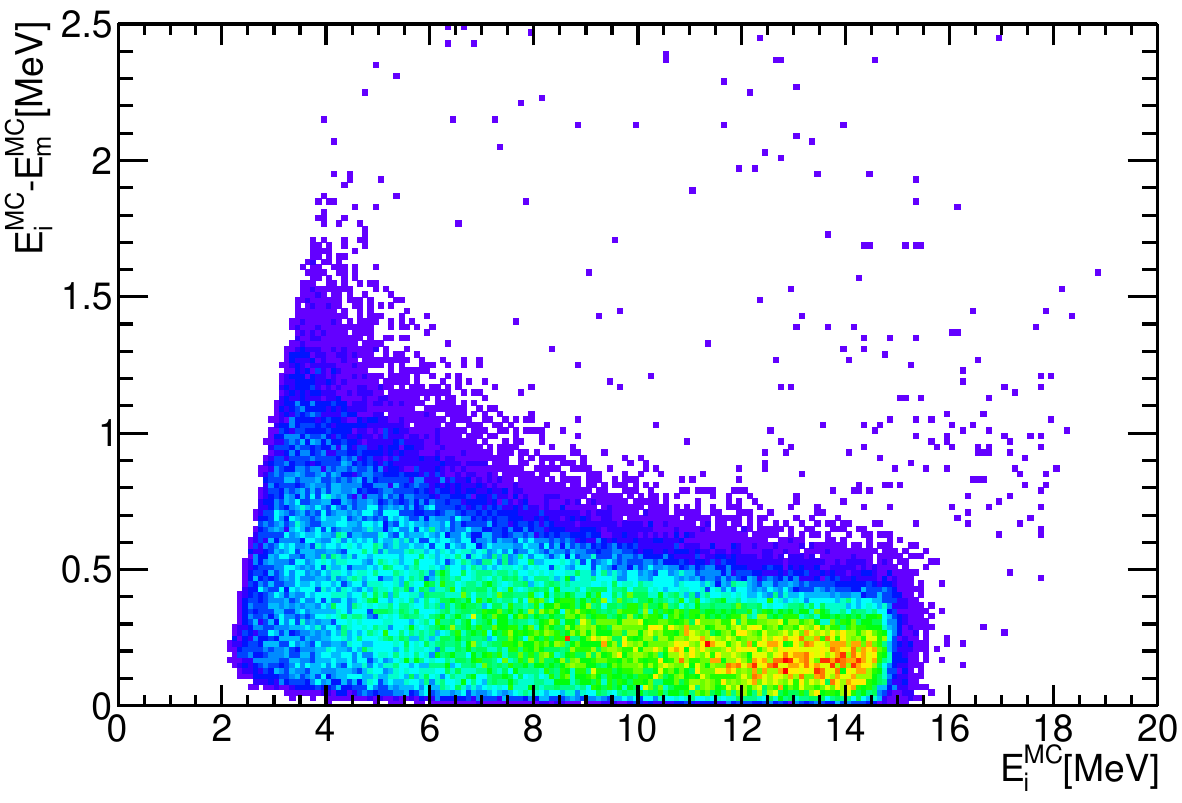}}
  ~\hfill
  \subfloat[][Simulated protons in the right detector\label{fig:analysis:unfold:tm_right_proton}]{\includegraphics[width=0.95\columnwidth]{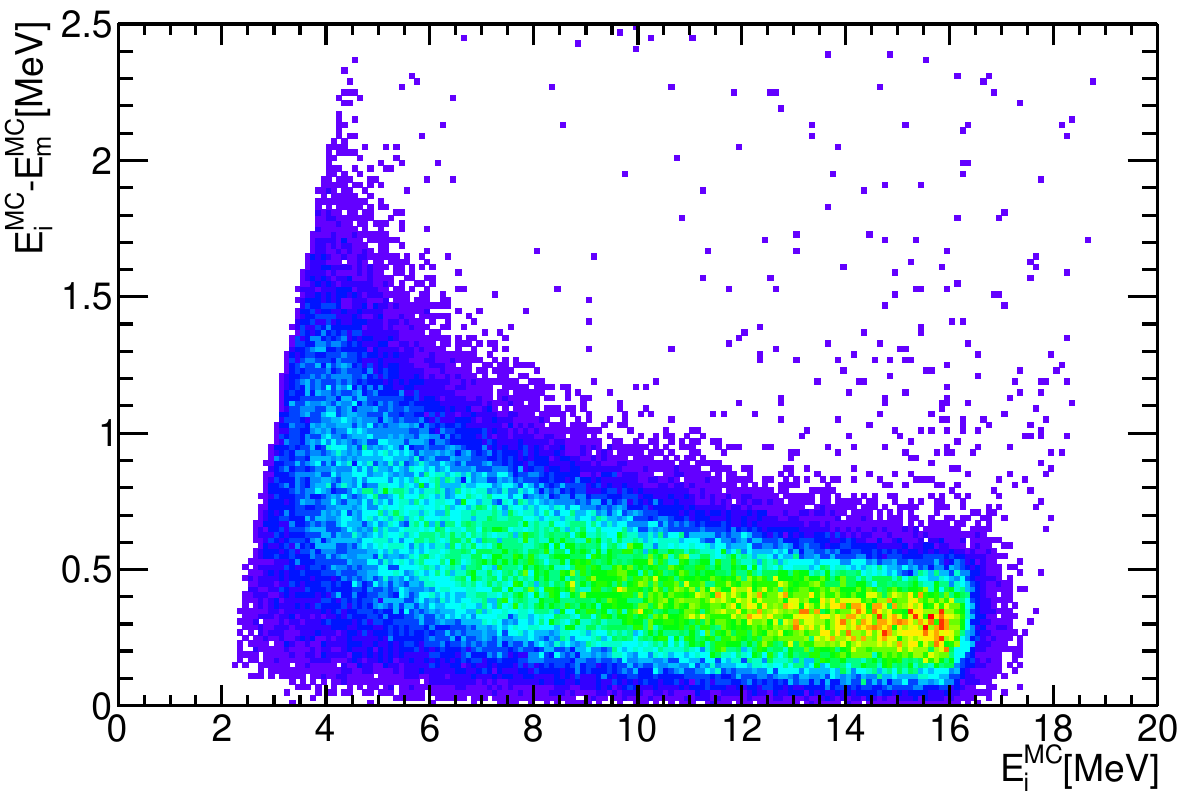}}
  \\
  \subfloat[][Simulated tritons in the left detector\label{fig:analysis:unfold:tm_left_triton}]{\includegraphics[width=0.95\columnwidth]{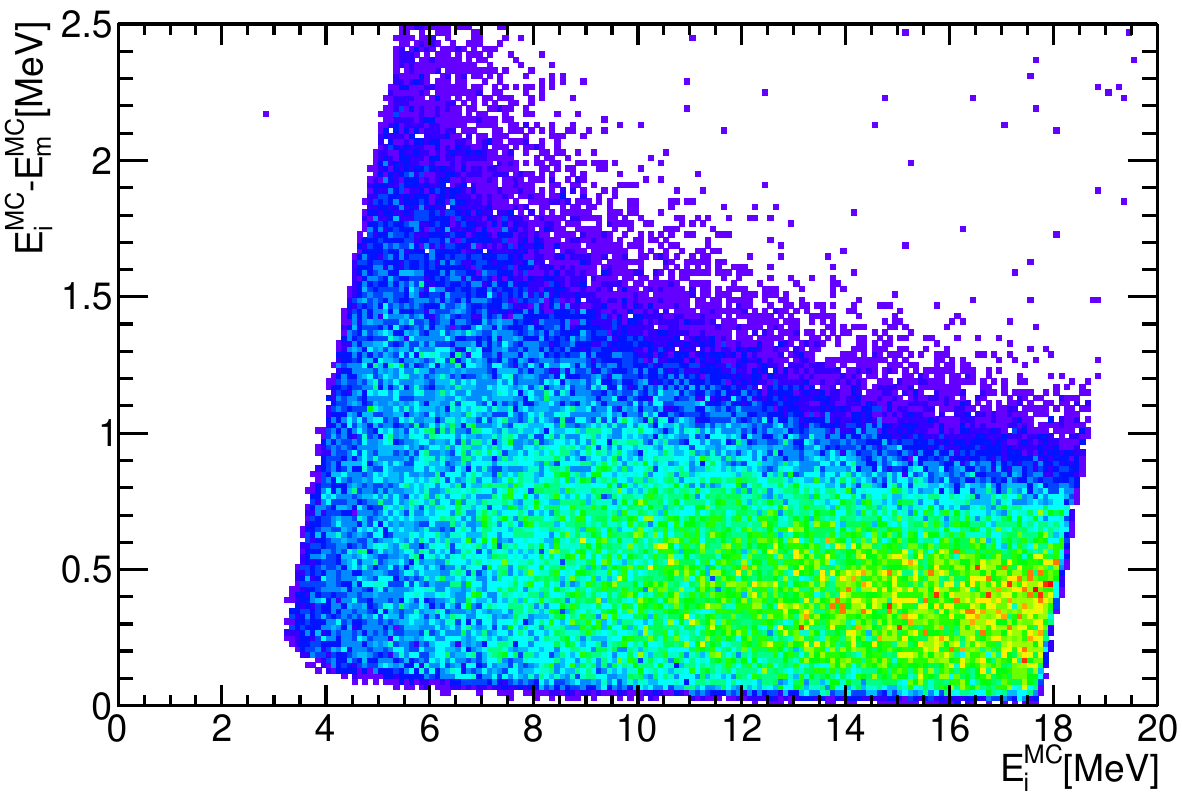}}
  ~\hfill
  \subfloat[][Simulated tritons in the right detector\label{fig:analysis:unfold:tm_right_triton}]{\includegraphics[width=0.95\columnwidth]{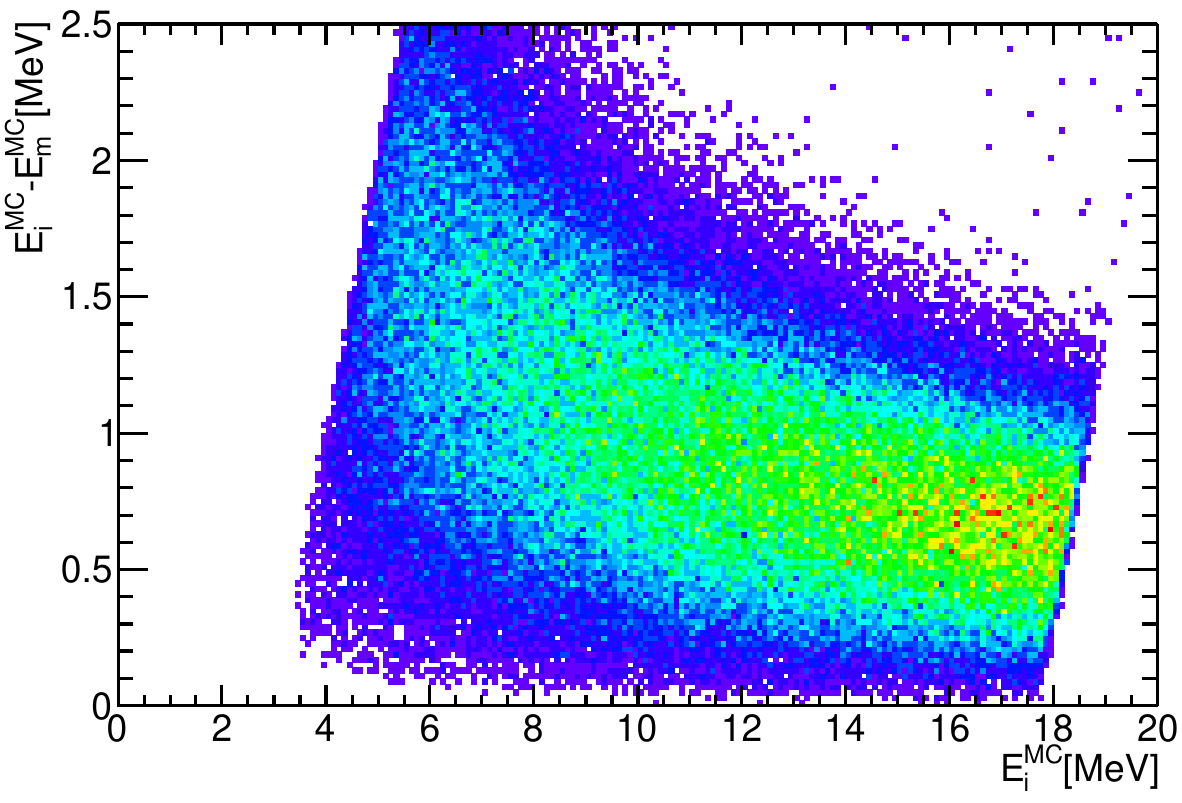}}
  \caption{Simulated energy loss, $E^{MC}_{i}-E^{MC}_{m}$, as a function of $E_{i}$, for protons (top) and tritons (bottom) emitted from the Al target that stop in the second layers of the left detector package (left) and right detector package (right). Note the difference in the upper $E^{MC}_{i}$ cutoff between protons and tritons is due to protons that punch-through the second layer, and the difference in the $E^{MC}_{i}-E^{MC}_{m}$ between the left and the right side is because the muon beam stops slightly closer to the back of the stopping target and so particles detected in the left detector package have passed through less of the target to reach the detector. This same simulation data is used to construct the response matrices for unfolding.}
\label{fig:analysis:unfold:tm}
\end{figure*}

An unfolding procedure corrects for the energy lost by charged particles as they leave the target. The amount of energy a particle loses for a given path length is strongly dependent on its energy and its species.  Figure~\ref{fig:analysis:unfold:tm} shows the simulated energy loss ($E^{MC}_{i} - E^{MC}_{m}$, where $E^{MC}_i$ is the simulated initial energy and $E^{MC}_{m}$ is the simulated measured energy) vs. simulated initial energy for particles simulated in the Monte Carlo simulation. The iterative unfolding method, as described by D'Agostini~\cite{DAgostini:1994fjx} and implemented in RooUnfold~\cite{Adye:2011gm} is used to unfold the measured spectra. 

Unfolding requires a response matrix that maps the measured energy to the initial energy of the particle. This was produced using the Monte Carlo simulation for heavy charged particles generated uniformly in the energy range \SIrange[]{0}{25}{MeV}. The response matrix also encoded the geometric acceptance of the detectors. 

Figure~\ref{fig:analysis:unfold:tm} shows that at higher energies the energy loss band is narrow and nearly energy independent; and at low energies the spread of energy losses
widens because the \dEdx\ increases. This induces 
larger uncertainties when unfolding the low-energy part of the spectrum.

In the iterative unfolding method, the regularization parameter is the number of iterations that are applied in the unfolding method. The default number of 4 iterations was used~\cite{Adye:2011gm}. Systematic uncertainties due to the unfolding are studied in Section~\ref{sec:systematics}.


\section{Systematic Uncertainties}
\label{sec:systematics}
The systematic uncertainties for the absolute yields of different particle species largely depend on common factors and are summarized
in Table~\ref{tab:systematics}. The bin-to-bin uncertainties affecting each energy spectrum have to be evaluated
separately and are included in the plots of the initial energy spectra.

\subsection{Hit Selection and Particle Identification}
To understand the systematic uncertainties associated with the hit selection, the cuts were changed and the full analysis chain was re-run. The arrival time cut ($t_{2}>\SI{400}{\nano\second}$) was changed to \SI{300}{\nano\second} and \SI{500}{\nano\second}; and the generous layer coincidence cut ($|t_{2}-t_{1}|<\SI{500}{\nano\second}$) was changed to \SI{200}{\nano\second}. In both cases, the differences in selected particles were within the allowed statistical differences due to one set of cuts being the subset of another and so is neglected as a systematic uncertainty.

Because of the dead layers between the second and third layers, a proton that punches through the second layer may stop in the dead layer before hitting the third layer and so would not trigger the veto. From simulations of the detectors with larger dead layers 10\% was added to the uncertainty in bins 15.5 -- 16.5 MeV.

For data where the simulation-derived punch-through correction was used (i.e. for the left-detector package, and protons that punch-through the third-layer), a systematic uncertainty of 10\% was added. This is derived from the statistics of the simulation-derived punch-through correction factors and by comparing the simulation-derived correction against the veto correction in the right detector package.

The systematic uncertainty obtained from the choice of particle identification method (data-driven vs. power-law function) is negligible because both methods produced almost identical raw spectra after efficiency and purity corrections. 

For the data-driven method, the spectra constructed from the $\pm2\sigma$ width cuts and the $\pm1\sigma$ width cuts agreed within the allowed difference for this set -- subset statistics. A tighter selection cut would result in negligible contamination from a neighboring particle band with a lower selection efficiency. 

For the power-law function method, the width of the cuts were increased by 20\%, a change in acceptance similar to the $1\sigma$--$2\sigma$ comparison of the data-driven method. Both wider and narrower selections were statistically consistent with the chosen cut values.

\subsection{Unfolding Procedure}
The number of iterations in the iterative unfolding procedure affects the result of the unfolding.
With more iterations, the unfolding errors as well as instabilities at the boundaries of the measured energy range increase. To quantify this, the number of iterations was increased and the resulting differences assigned as a systematic uncertainty. The unfolded spectrum was assumed to converge with 4 iterations and the difference in the integral between 4 and 20 iterations was taken as the uncertainty. The difference in the integral depended on the particle type. For protons, deuterons and tritons the differences were $<$\SI{3}{\percent}, while for $\alpha$-particles the uncertainty was 15\%. For the bin-by-bin uncertainties a conservative error was assigned as the quadratic sum of the error calculated by the unfolding plus the difference between the unfolded spectra with the two iteration parameters.

The bin width of the response matrix also affects the result. Unfolding with response matrices of bin widths in \SI{100}{keV} increments from \SI{100}{keV} to \SI{500}{keV} produced difference with the final result of, at most, 2\%. 

\subsection{Energy Miscalibration}
From comparison between data and simulation, a 2.5\% uncertainty on the energy calibration was assigned.
Response matrices were constructed with energy deposits 2.5\% higher and 2.5\% lower than the simulated truth. The difference between the bins in each unfolded spectrum is included as the systematic. The systematic uncertainty on the integrals depends on the particle. It ranges from 1\% for protons to 7.5\% for $\alpha$-particles due to the effect on the low-energy bins.

\subsection{Muon Beam and Detector Placement}
Small changes in the stopping position of the muon beam in the target changes the amount of material through which an emitted charged particle must transit. In order to study this effect, the muon beam was simulated with higher and lower energies than determined from the muon beam validation. The energy change was \SI{\pm100}{keV}, which was the uncertainty on the muon beam validation and corresponds to a stopping depth uncertainty of  \SI{\pm10}{\micro\meter}. New stopping positions and re-simulations of heavy charged particle emissions were used to obtain new response matrices for unfolding.

\begin{figure}[!htbp]
  \centering
  \includegraphics[width=1.0\columnwidth]{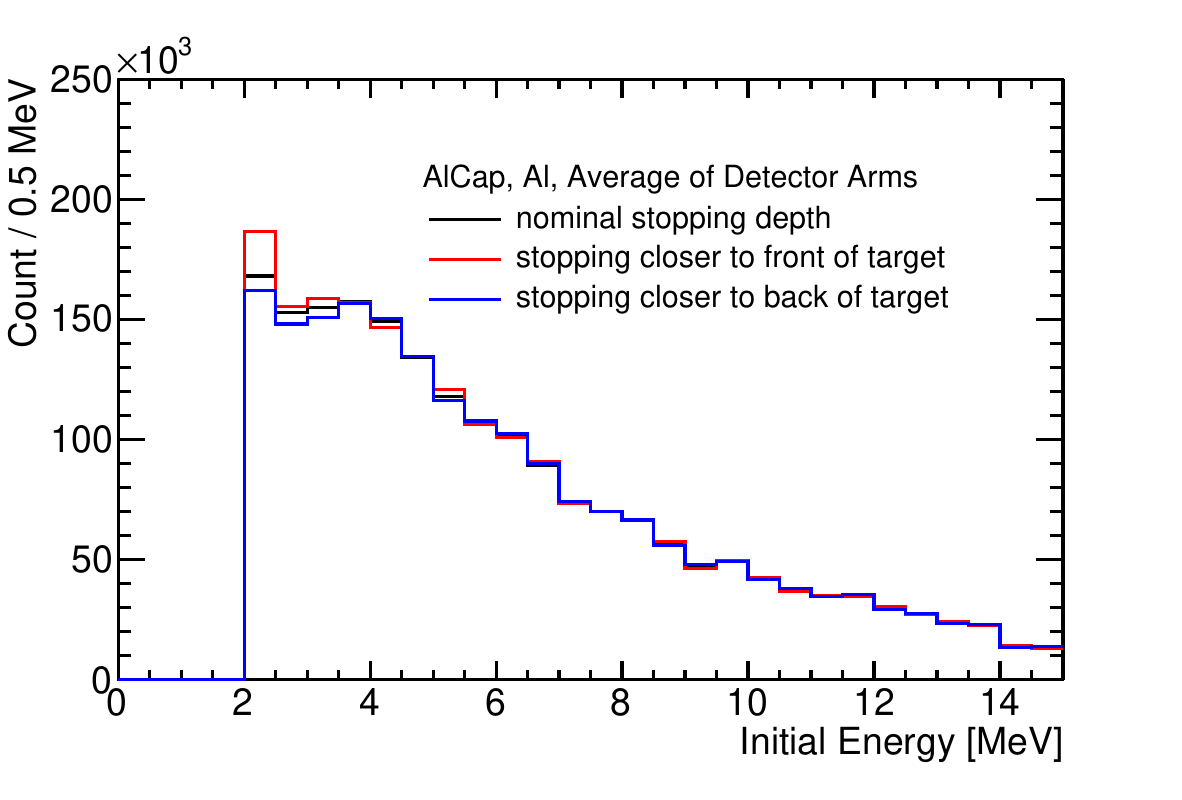}
  \caption{The average initial energy spectra combining left and right detector packages for different simulated muon stopping positions. As can be seen, the average is resistant to the assumed stopping depth.}
  \label{fig:avg-resistance}
\end{figure}

Figure~\ref{fig:avg-resistance} shows that the average of the left and right detector packages is insensitive to the muon stopping depth. Therefore, for the aluminum and titanium datasets, where detectors were placed symmetrically around the target with the results averaged, this systematic uncertainty is negligible. However, for the silicon dataset, where only one detector arm exists, the systematic uncertainty is 1\% on the integrated yield. All datasets used only one arm for the $\alpha$-particle spectra and the systematic uncertainty due to the muon stopping depth is significantly higher at $\approx$10\%

There is a \SI{\pm1.5}{mm} detector placement uncertainty. This changes the geometric acceptance as well as the effective material thickness for charged particle penetration. 
For the silicon dataset, where there was a single arm at a nominal \SI{122}{\milli\metre} away from the target, a solid-angle calculation showed that the uncertainty was \SI{2.2}{\%}. As a cross-check, a simulation of heavy charged particles in the Monte Carlo simulation with the detector arm placed \SI{1.5}{mm} closer and further from the stopping target produced results consistent with the solid-angle calculation. For datasets with both arms, the uncertainty is reduced by a factor of $\sqrt{2}$ because each arm was placed independently.

The layers could also be misaligned by $\pm$1 mm. This affects all datasets and would change the geometry acceptance of the detectors. From simulations, this effect adds a 2\% systematic uncertainty on the integral.

\begin{table}[!htbp]
  \caption{Contributions to the total yield systematic uncertainties. Uncertainties not given were determined to contribute $<$20\% to the total systematic uncertainty.}
  \label{tab:systematics}
  \begin{tabular}{p{5.0cm}>{\centering\arraybackslash}m{0.4cm}>{\centering\arraybackslash}m{0.4cm}>{\centering\arraybackslash}m{0.4cm}>{\centering\arraybackslash}p{0.4cm}}
  \hline
  \hline
  \multirow{3}{*}{Source} & \multicolumn{4}{c}{Systematic}\\
  & \multicolumn{4}{c}{Uncertainty [\%]}\\
    & $p$ & $d$ & $t$ & $\alpha$ \\
  \hline
  \multicolumn{5}{l}{\textbf{For Al dataset:}} \\
  energy miscalibration, $\pm$2.5\% & 1.1 & 1.0 & 1.5 & 7.5 \\
  unfolding iterations & 0.6 & 0.8 & 1.7 & 12.1\\ 
  response matrix bin width & 0.4 & 0.4 & 0.6 & 0.3  \\
  muon stopping depth, $\pm$10 $\mu$m & -- & -- & -- & 14.7\\
  detector misalignment, $\pm$1 mm & 2.0 & 2.0 & 2.0 & 2.0 \\
   detector placement, $\pm$1.5 mm & 1.6 & 1.6 & 1.6 & 2.2 \\
   \multicolumn{1}{l}{\textit{Quadrature sum:}} & 2.9 & 2.9 & 3.5 & 20.7 \\
  \hline
  \multicolumn{5}{l}{\textbf{For Si dataset:}} \\
  energy miscalibration, $\pm$2.5\% & 1.0 & 1.5 & 3.5 & 4.4 \\
  unfolding iterations & 1.0 & 2.2 & 3.5 & 15.0\\ 
  response matrix bin width & 0.7 & 0.6 & 2.1  & 0.8 \\
  muon stopping depth, $\pm$10 $\mu$m & 1.0 & 1.1 & 1.4 & 7.8\\
  detector misalignment, $\pm$1 mm & 2.0 & 2.0 & 2.0 & 2.0 \\
   detector placement, $\pm$1.5 mm & 2.2 & 2.2 & 2.2 & 2.2 \\
   \multicolumn{1}{l}{\textit{Quadrature sum:}} & 3.5 & 4.2 & 6.3 & 17.7 \\
  \hline
  \multicolumn{5}{l}{\textbf{For Ti dataset:}} \\
  energy miscalibration, $\pm$2.5\% & 1.3 & 2.3 & 2.7 & 3.3 \\
  unfolding iterations & 0.8 & 1.9 & 2.6 & 4.5\\ 
  response matrix bin width & 0.4 & 0.7 & 0.9 & 0.6  \\
  muon stopping depth, $\pm$10 $\mu$m & -- & -- & -- & 9.1\\
  detector misalignment, $\pm$1 mm & 2.0 & 2.0 & 2.0 & 2.0 \\
  detector placement, $\pm$1.5 mm & 1.6 & 1.6 & 1.6 & 2.2 \\
   \multicolumn{1}{l}{\textit{Quadrature sum:}} & 3.0 & 4.0 & 4.6 & 11.1 \\
  \hline
  \hline
  \end{tabular}
\end{table}


\section{Results}
\label{sec:results}

\newcommand{\x}{0.9}
\begin{figure*}[!htbp]
  \centering
  \includegraphics[width=\x\textwidth]{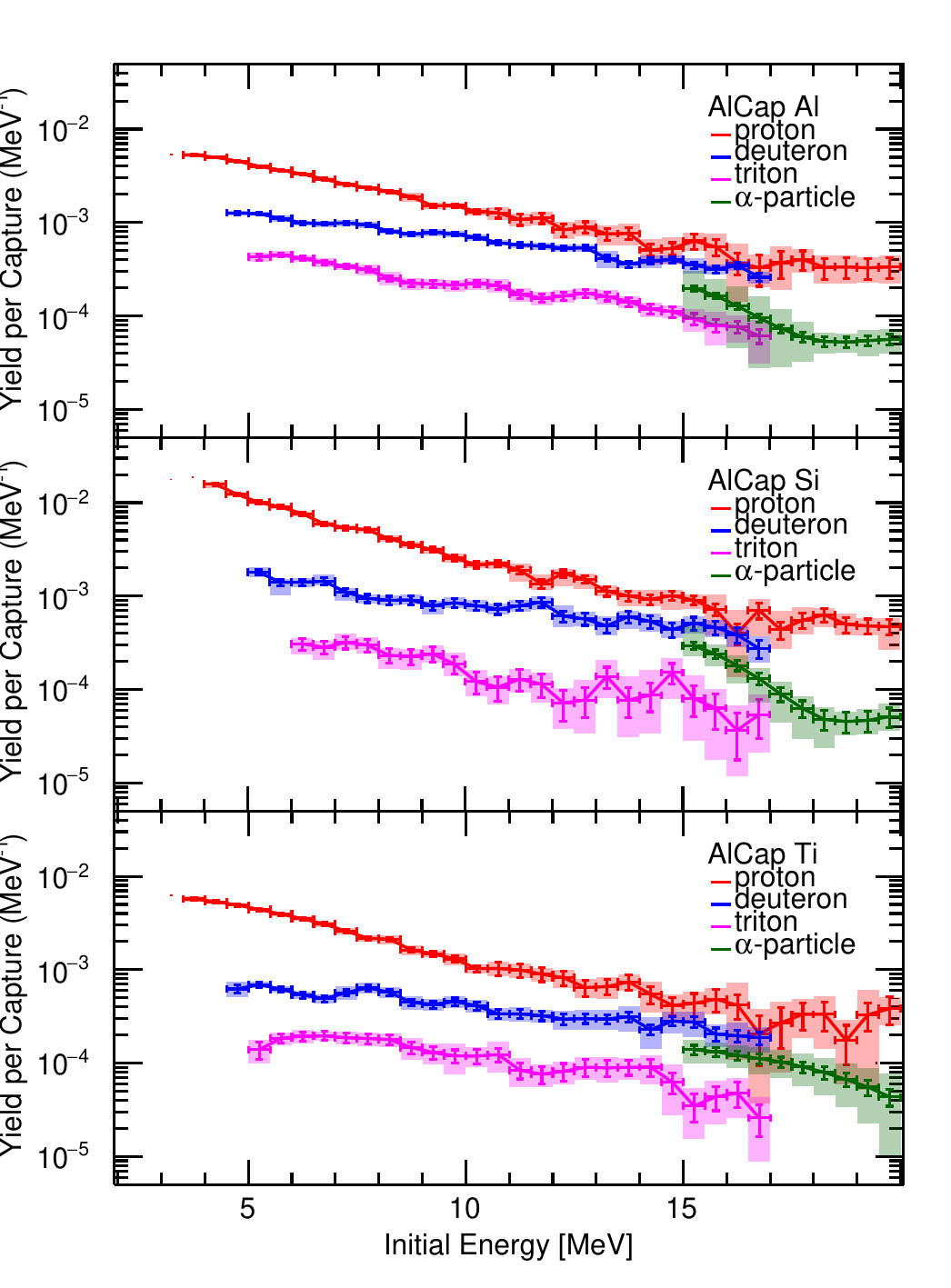}
   \caption{Initial energy spectra of charged particles emitted after nuclear muon capture on aluminum (top), silicon (middle) and titanium (bottom). The statistical uncertainties are represented by the error bars whereas the systematic uncertainties are the color-shaded areas for each energy bin. Overall systematic uncertainties due to detector misalignment and detector placement (see Table~\ref{tab:systematics}) are not included here.} 
   \label{fig:results}
\end{figure*}

\begin{table*}[!htbp]
  \begin{center}
  \captionsetup{justification=centering}
    \caption{Table of charged particle yields and spectral decay rates for the main AlCap datasets.}
    \label{tab:results}
    \begin{tabular}{ccccc}
      \hline
      \hline
      \multirow{2}{*}{Material} &  \multirow{2}{*}{Particle} & $E_{0}$ -- $E_{1}$ & $\int_{E_0}^{E_1} Y^{i}_{\eta}(E) dE$ & Decay Parameter, $T$ \\
                                &                            & [MeV]        & [per thousand muon captures] & [ MeV ]   \\
      \hline
      \multirow{4}{*}{aluminum} & protons   & \AlProtonBestRange   & \AlProtonBestRateFull   & \AlProtonExpo \\
                                & deuterons & \AlDeuteronBestRange & \AlDeuteronBestRateFull & \AlDeuteronExpo\\
                                & tritons   & \AlTritonBestRange   & \AlTritonBestRateFull   & \AlTritonExpo\\
                                & alphas    & \AlAlphaBestRange    & \AlAlphaBestRateFull    & \AlAlphaExpo\\
      \hline
      \multirow{4}{*}{silicon} & protons & \SiProtonBestRange & \SiProtonBestRateFull & \SiProtonExpo \\
      & deuterons & \SiDeuteronBestRange & \SiDeuteronBestRateFull & \SiDeuteronExpo  \\
      & tritons & \SiTritonBestRange & \SiTritonBestRateFull & \SiTritonExpo  \\
      & alphas & \SiAlphaBestRange & \SiAlphaBestRateFull & \SiAlphaExpo  \\
      \hline
      \multirow{4}{*}{titanium} & protons   & \TiProtonBestRange   & \TiProtonBestRateFull   & \TiProtonExpo \\
                                & deuterons & \TiDeuteronBestRange & \TiDeuteronBestRateFull & \TiDeuteronExpo\\
                                & tritons   & \TiTritonBestRange   & \TiTritonBestRateFull   & \TiTritonExpo\\
                                & alphas    & \TiAlphaBestRange    & \TiAlphaBestRateFull    & \TiAlphaExpo\\
      \hline
      \hline
    \end{tabular}
  \end{center}
\end{table*}

Figure~\ref{fig:results} presents the high-statistics, initial energy spectra of charged particles produced from the capture process in the AlCap experiment. This analysis did not use any theoretical assumption about their spectral shape. Initial energy spectra are only reported when at least 90\% of the energy response can be detected. 
For example, the response of a proton
emitted with $E_i=3$ MeV is mostly seen by the SiL detector (see Fig.~\ref{fig:analysis:unfold:tm}), while particles of smaller energy fall below the detector threshold and, thus, are not included in the figures. Accordingly, the valid range for unfolding depends on the particle type. The range for heavier particle spectra starts at higher energies but also extends to higher energies without punch through. However, due to detector saturation the deuteron and triton spectra are only valid up to 17 MeV.

For the datasets where both detector packages were operated (for aluminum and titanium targets), the data were combined by adding the initial energy spectra as an inverse variance weighted average accounting for their different statistical and systematic errors. The bin-by-bin variation of the initial energy spectra largely follow the measured data (Fig.~\ref{fig:analysis:charged-particles-folded}), with additional scatter induced by the unfolding process.

The main feature of all spectra is an exponential fall-off, which can be characterized by an evaporation spectrum $e^{-E/T}$ and
is consistent with emission predominantly from the compound-nucleus as predicted by Lifshitz and Singer~\cite{Lifshitz:1980gi}.

Table~\ref{tab:results} gives numerical results extracted from the initial energy spectra. The integral of the spectra over an energy region gives the total number of emitted particles per muon capture, and an exponential fit to the spectra gives the decay parameter, $T$. These fits can aid the extrapolation to energy ranges beyond the AlCap measurement. However, care must be taken in extending to lower energies since this data does not cover the expected low-energy cut-off due to the Coulomb barrier.

These results indicate significant nuclear structure dependence of charged particle emission after
muon capture. Both yields and spectral shapes depend on the target nucleus, with large differences observed even between the neighboring nuclei Al and Si. Further theoretical work could improve the understanding of these effects.

\section{Cross-Checks}
\label{sec:cross-checks}
\subsection{Aluminum:  Thick Target Analysis}
As a cross-check, data were collected with a thicker aluminum target (\SI{100}{\micro\meter} vs. \SI{50}{\micro\meter}). This
measurement required a higher beam energy to stop the muons in the target center, and the emitted particles lost more energy as they left. 
The increased energy-loss corrections provide a valuable cross-check of our simulation and unfolding technique.

\begin{figure}[!htbp]
  \centering
  \includegraphics[width=1.0\columnwidth]{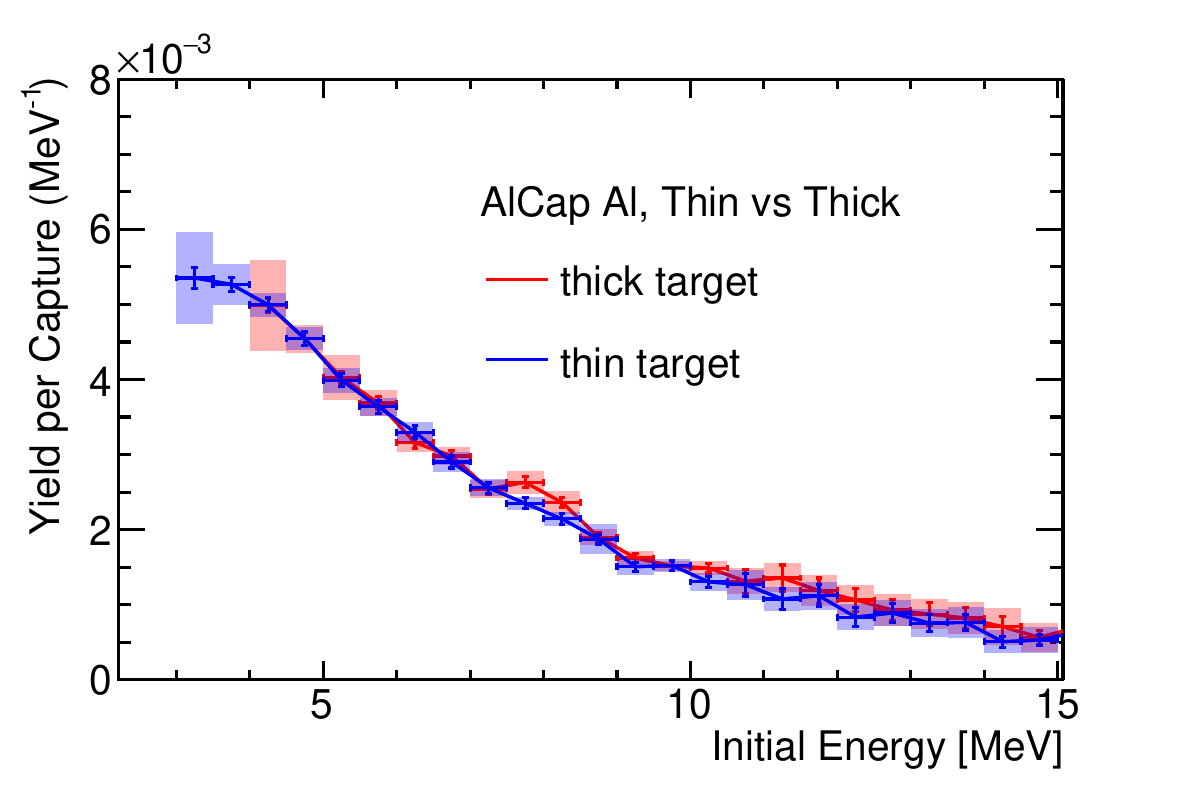}
  \caption{Initial energy spectra for protons measured with data from a thin (blue) and thick (red) aluminum stopping target.}
  \label{fig:cross-checks:thick-target-al}
\end{figure}

There are larger uncertainties in the analysis of the thicker target than in the thinner target. The lowest energies measured with the thin target cannot be measured with the thick target. Also there was no third layer veto available in either arm for the thick target and so the simulation-derived punch-through correction was applied. Despite this, Fig.~\ref{fig:cross-checks:thick-target-al} shows that the agreement between the thin and thick datasets is very good.

\subsection{Silicon: Active-Target Analysis}
\label{sec:cross-checks:silicon}

To validate the analysis technique, data were collected on a \SI{1.5}{mm}-thick active silicon target so that an analysis similar to that of Sobottka and Wills~\cite{Sobottka1968} could be performed. The energy spectrum of particles observed between \SI{3}{\micro\second} and \SI{4}{\micro\second} after the muon were analyzed, with an increased pile-up protection window of \SI{20}{\micro\second}. The total charged particle emission spectrum was compared to the sum of the Si spectra in Fig.~\ref{fig:results}. The number of muons stopped was \SiLGeHiGainNStoppedMuonsTab{}$\times10^{6}$ muons, corresponding to \SiLGeHiGainNCapturedMuonsTab{}$\times10^{6}$ captured muons. Between \SI{3}{\micro\second} and \SI{4}{\micro\second}, 1.4\% of muons in the silicon atoms either decay or are captured.

The standard pulse analysis is sufficient for the main analysis because the detectors have low hit rates (\SI{<100}{\Hz}). However, in the active-target analysis, the hits from ejected particles that occur after a muon stop was observed in the same detector were analyzed. Because the triggered and recorded waveforms were \SI{5}{\micro\second} in length, the muon stop hit and the ejected particle hit occurred in the same waveform. To resolve these hits, 
average pulse shape templates were constructed from well-separated pulses, and those templates were fitted to the waveform to find sub-pulses. The fitter was able to adequately resolve sub-pulses that were more than \SI{3}{\micro\second} apart, which was sufficient given the high statistics of this dataset.
Figure~\ref{fig:pulse-analysis:template} shows the template fitter resolving two hits in a single waveform. The energy and time of each sub-pulse was extracted with the standard pulse definitions.

Two backgrounds were subtracted from the measured spectrum. The low-energy accidental background was removed by subtracting the energy spectrum in the negative time region (i.e. between \SI{-4}{\micro\second} and \SI{-3}{\micro\second}). The decay electron background was removed by using the results of a simulation. The energy deposit of decay electrons in the simulation was validated by comparing to minimally-ionizing particle data from the main analysis.

\begin{figure}[!htbp]
  \centering
  \includegraphics[width=1.0\columnwidth]{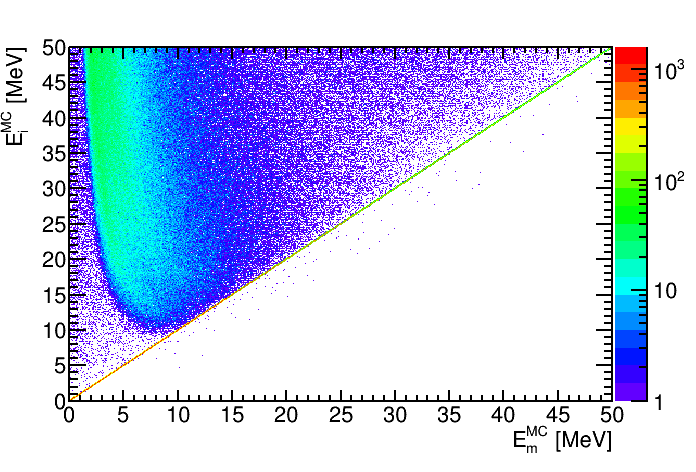}
  \caption{Proton response matrix for escape correction in active target analysis. For fully-contained protons, the initial and measured energies are equal. Above \SI{9}{MeV} protons can escape and the measured energies become smaller than the initial energies.}
  \label{fig:active-response-matrix}
\end{figure}
Two additional corrections accounted for the energy transferred to the recoiling nucleus and the effect of high-energy charged particles that escaped the stopping target. The recoil correction has not been discussed previously in the literature and accounted for a \SI{\approx20}{\percent} difference between the two silicon analyses at \SI{15}{MeV}.

The energy measured in the active target, $E_\text{meas}$, is the sum of the kinetic energy of the emitted charged particle, $E_\text{X}$ and the recoil energy of the nucleus in muon capture. According to theory~\cite{Lifshitz:1980gi, Lifshitz:1978qj} the majority of muon capture reactions can be factorized into the initial capture, where most of the energy is taken by the neutrino, and the slower equilibration of the formed compound nucleus. Thus, the total kinetic energy of the final nucleus consists of contributions from its recoil against the neutrino and the emitted charged particle, as well as any emitted neutrons. Without a detailed theoretical model, we limit our simplistic estimate to the effect of the kinetic energy of a nucleus with mass $m_{\text{N}}$ recoiling against the observed charged particle with mass $m_{\text{X}}$ from the decay of the intermediate compound nucleus. In that case $E_{\text{X}}  = E_{\text{meas}} / (1 + m_{\text{X}}/m_{\text{N}})$  in the center-of-mass frame. 
For protons, this is ~4\% smaller than the total observed energy $E_{\text{meas}}$. Since different particle types cannot be separated in this analysis, an average correction based on the particle ratios measured in the main silicon analysis is used. Between 15 MeV and 17 MeV, where all particles can be observed, the p:d:t:alpha fractions (with statistical and systematic uncertainties) are are \protonRatioCondensed, \deuteronRatioCondensed, \tritonRatioCondensed\ and \alphaRatioCondensed, respectively. This results in an average recoil correction of ~7\% that is applied to all hits in the active target. Additional simulations which include transformation into the laboratory frame and neutron emission suggest that this is a lower limit for the recoil energy imparted on the nucleus.

High-energy charged particles escape from the target and mimic low-energy, fully-contained events because they deposit a small amount of energy as shown by a simulated response matrix in Fig.~\ref{fig:active-response-matrix}. Thus the unobserved high energy part of the
spectrum can contribute to the low energy response, and unfolding becomes ill-defined. Therefore,  a shape of the form:
\begin{equation}
    N(E) = N_0 \left( 1 - \dfrac{T_{th}}{E}\right)^\alpha \left( e^{-E/T_0} + r e^{-E/T_1} \right)
\label{eq:shape}
\end{equation}
is assumed, where $N_{0}$ is a normalization constant. The first term, characterized by a threshold energy  $T_{th}$ and a shape parameter $\alpha$, 
provides a phenomenological description of the onset for charged particle emission above the Coulomb barrier. The second term allows for 
two exponential decay constants $T_{0}$ and $T_{1}$ suggested by the data, with $r$ parameterizing the relative fraction of the second 
exponential. In contrast to the main model-independent analysis, the chosen functional form imposes smoothness, eliminates fluctuations and enforces an exponential behavior. The analysis was performed by folding this functional form with the respective response matrix and fitting the result to the measured data.

\begin{figure*}[!htbp]
  \centering
   \subfloat[][The data (red) is reproduced by the fitted spectrum (green), which corresponds to the initial energy distribution (blue) folded with the detector response. \label{fig:active-results-selfcomp:explanation}]{\includegraphics[width=0.5\textwidth]{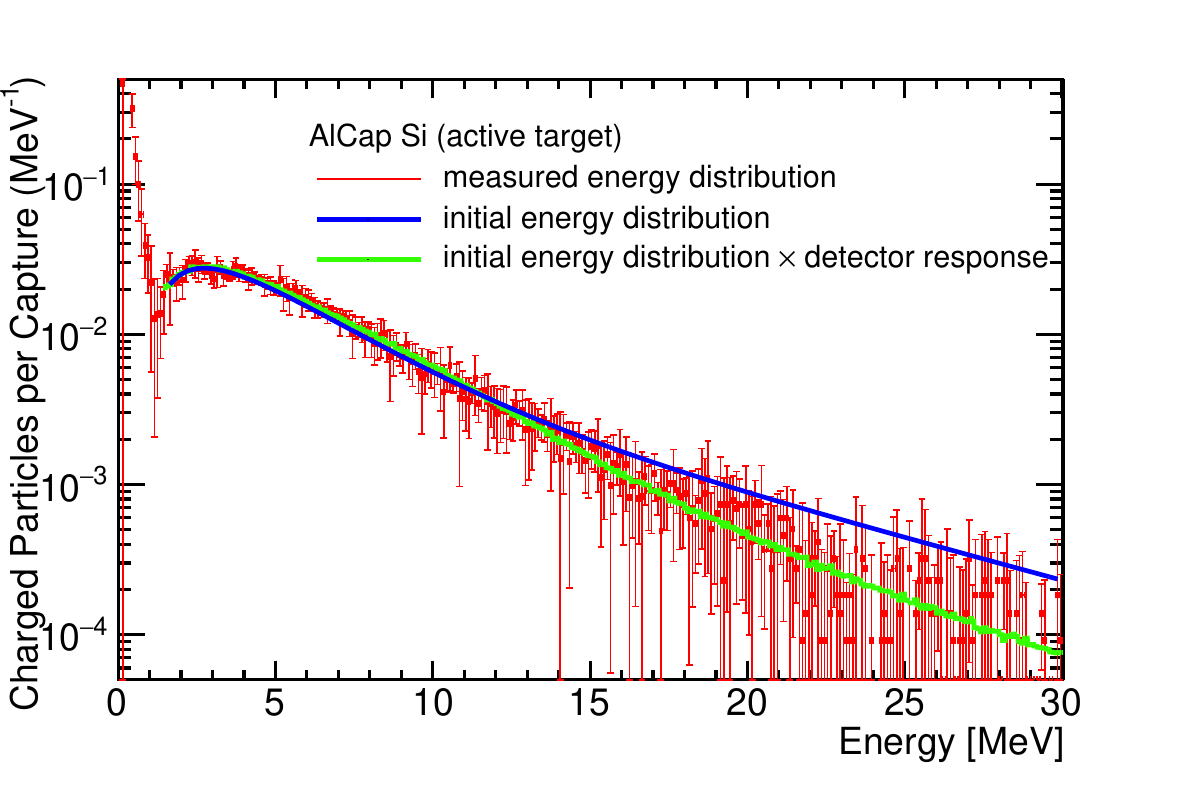}}
  \subfloat[][Comparison between the result of the active-target analysis (blue), and the sum of protons, deuterons and tritons (gray) and the sum of protons, deuterons, tritons and $\alpha$-particles (black) from the main silicon analysis.\label{fig:active-results-selfcomp:comp}]{\includegraphics[width=0.5\textwidth]{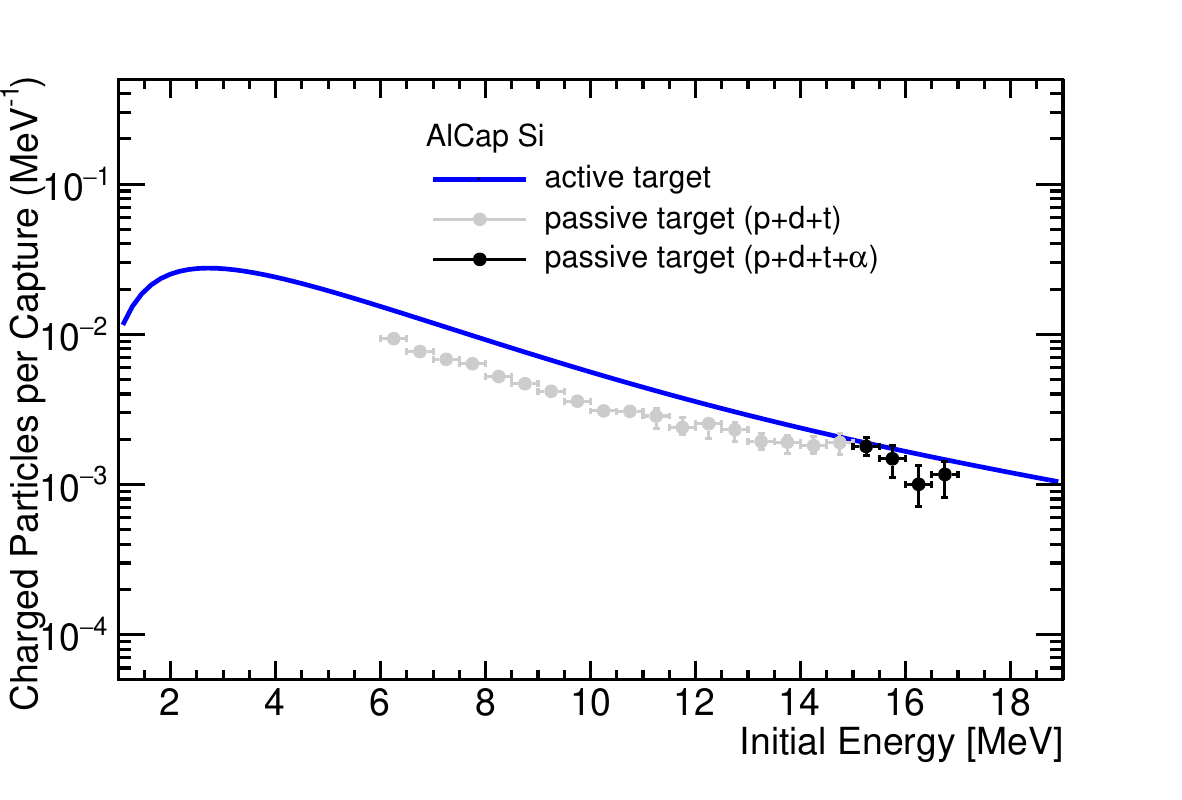}}
 \caption{Left: Figure showing fitting procedure used in the active-target analysis. Right: Comparison between the active-target and main silicon analyses.}
 \label{fig:active-results-selfcomp}
\end{figure*}

The steps of this procedure are shown in Fig.~\ref{fig:active-results-selfcomp:explanation}, where the red histogram is the raw data, the green line represents the result of the fit, which is obtained by folding the initial energy distribution (blue line) with the detector response.

Many of the systematic errors of this analysis are the same as obtained in the full analysis (e.g. simulation uncertainties). In addition, the particle composition uncertainty was negligible, and the dependence to the time slice used (subdivision analysis in \SI{500}{ns} time slices) added a 1\% error.  The energy miscalibraton systematic greatly affects the decay electron correction and adds significant uncertainty in the dip region.

Figure~\ref{fig:active-results-selfcomp:comp} shows the comparison between the passive and active silicon targets. There is good agreement between the two analyses in the energy region where all particle types are observed in the passive target. At lower energies, there is significant discrepancy, which is likely due to the fact that the passive analysis did not measure $\alpha$-particles at those energies.

\section{Discussion}
\label{sec:discussion}
\begin{table*}[!htbp]
\caption{Comparison of yields (in 10$^{-3}$ per capture) between AlCap and other work. The numbers in brackets denote the energy range in MeV used in the data selection. The inclusive yields are denoted by a tilde, and specific reactions are explicitly defined by their initial and final states.}
\label{tab:results_compare}
    \begin{center}
    \resizebox{1.0\textwidth}{!}{
      \begin{tabular}{c|ccc|cc|c}
        \hline
        \hline
          & \multicolumn{3}{c|}{aluminum} & \multicolumn{2}{c|}{silicon} & titanium \\
        \hline
         \multicolumn{1}{l|}{\textit{Direct measurements:}} & AlCap & TWIST~\cite{Gaponenko:2019efc} & Krane et.al.~\cite{Krane1979}  & AlCap & Sobottka-Wills~\cite{Sobottka1968} & AlCap \\
        \multirow{2}{*}{($\mu^{-}$, $\tilde{p}$)}& [\AlProtonMinE$\le$E$\le$\AlProtonMaxE] & [3.4$<$E] &  \multirow{2}{*}{--}  & [\SiProtonMinE$\le$E$\le$\SiProtonMaxE] & \multirow{2}{*}{--} &  [\TiProtonMinE$\le$E$\le$\TiProtonMaxE] \\
         &  \AlProtonTable & 32.2(2.3) &  &  \SiProtonTable &  & \TiProtonTable \\
         \multirow{2}{*}{($\mu^{-}$, $\tilde{d}$)}& [\AlDeuteronMinE$\le$E$\le$\AlDeuteronMaxE] & [4.5$<$E] & \multirow{2}{*}{--}  &  [\SiDeuteronMinE$\le$E$\le$\SiDeuteronMaxE] & \multirow{2}{*}{--} &  [\TiDeuteronMinE$\le$E$\le$\TiDeuteronMaxE] \\
         & \AlDeuteronTable & 12.2(1.1) & &  \SiDeuteronTable &  & \TiDeuteronTable\\
         \multirow{2}{*}{($\mu^{-}$, $\tilde{t}$)}& [\AlTritonMinE$\le$E$\le$\AlTritonMaxE] & \multirow{2}{*}{--} & \multirow{2}{*}{--} & [\SiTritonMinE$\le$E$\le$\SiTritonMaxE] & \multirow{2}{*}{--} &  [\TiTritonMinE$\le$E$\le$\TiTritonMaxE] \\
         & \AlTritonTable &  &  & \SiTritonTable &  & \TiTritonTable\\
         \multirow{2}{*}{($\mu^{-}$, $\tilde{\alpha}$)}& [\AlAlphaMinE$\le$E$\le$\AlAlphaMaxE] & \multirow{2}{*}{--} & \multirow{2}{*}{--} & [\SiAlphaMinE$\le$E$\le$\SiAlphaMaxE] & \multirow{2}{*}{--} &  [\TiAlphaMinE$\le$E$\le$\TiAlphaMaxE] \\
         & \AlAlphaTable &  &  &  \SiAlphaTable &  & \TiAlphaTable\\
        \multirow{2}{*}{($\mu^{-}$, all charged)} &\multirow{2}{*}{--} & \multirow{2}{*}{--} & [40$<$E] & [1.4$\le$E$\le$26] & [1.4$\le$E$\le$26] & \multirow{2}{*}{--}  \\
         &  &  & 1.38(9) & \SiLActiveRate & 150(20) &   \\
        \hline
         \multicolumn{1}{l|}{\textit{Activation measurements\footnote{These measurements did not directly measure the given reactions, but identified final nuclear states with $\gamma$-ray spectroscopy. Therefore, ($\mu^{-}$, $pn$) cannot be distinguished from ($\mu^{-}$, $d$).}:}} & Wyttenbach et. al.~\cite{wyttenbach1978probabilities} & Heusser-Kirsten~\cite{heusser1972radioisotope} & &  \multicolumn{2}{c|}{} \\
         $_{Z}^{A}X$($\mu^{-}$, $pn$)$_{Z-2}^{A-2}X$ & 28(4) & -- & &  \multicolumn{2}{c|}{} \\
         $_{Z}^{A}X$($\mu^{-}$, $p2n$)$_{Z-2}^{A-3}X$ & -- & 35(8) & &  \multicolumn{2}{c|}{} \\
         $_{Z}^{A}X$($\mu^{-}$, $\alpha$)$_{Z-3}^{A-4}X$ & 7.6(1.1) & -- & &  \multicolumn{2}{c|}{} \\
        \hline
         \multicolumn{1}{l|}{\textit{Theory:}} & \multicolumn{3}{c|}{Lifshitz-Singer~\cite{Lifshitz:1980gi}} & \multicolumn{2}{c|}{Lifshitz-Singer~\cite{Lifshitz:1980gi}} \\
         ($\mu^{-}$, $\tilde{p}$) & \multicolumn{3}{c|}{40} & \multicolumn{2}{c|}{--} \\
          ($\mu^{-}$, $\tilde{d}$) & \multicolumn{3}{c|}{12}& \multicolumn{2}{c|}{21} & \\
           ($\mu^{-}$, $\tilde{\alpha}$) & \multicolumn{3}{c|}{20}& \multicolumn{2}{c|}{34} \\
           ($\mu^{-}$, all charged) & \multicolumn{3}{c|}{--}& \multicolumn{2}{c|}{144} \\
           $_{Z}^{A}X$($\mu^{-}$, $p$)$_{Z-2}^{A-1}X$ & \multicolumn{3}{c|}{9.7} & \multicolumn{2}{c|}{32} \\
           $_{Z}^{A}X$($\mu^{-}$, $d$)$_{Z-2}^{A-2}X$ & \multicolumn{3}{c|}{6.0} & \multicolumn{2}{c|}{8.2} \\
           $_{Z}^{A}X$($\mu^{-}$, $\alpha$)$_{Z-3}^{A-4}X$ & \multicolumn{3}{c|}{7.3} & \multicolumn{2}{c|}{17} \\
        \hline
        \hline
      \end{tabular}
      }
    \end{center}
  \end{table*}

AlCap has significantly expanded the information on charged particle emission in muon capture. There is only data for Al from one other contemporary experiment, TWIST, which can be partially compared~\cite{Gaponenko:2019efc}. The total charged particle emission in Si can be compared to the pioneering Sobottka-Wills measurement ~\cite{Sobottka1968} which used a silicon detector as an active target.

TWIST was designed to measure the Michel decay parameters by stopping $\mu^{+}$ in a thin aluminum target. It also collected data with a $\mu^{-}$ beam. With this data TWIST measured proton and deuteron spectra after muon capture in this target.
AlCap and TWIST are complementary both in technique and analysis. AlCap measured the total energy of emitted particles and provided
direct particle identification, while TWIST measured their momenta in a precision magnet spectrometer. 
AlCap used a model-independent, stable unfolding procedure and cut off regions that do not pass the 90\% detected energy response requirement. 
The TWIST analysis relied on a more sophisticated unfolding method which augmented the data with physically-motivated exponential models above their detector's cutoff energy. TWIST also modeled the triton and $\alpha$-particle spectra with a theoretical model, as those were not directly observable.

\subsection{Charged particle yields}
Table~\ref{tab:results_compare} compares the charged particle yield per capture reaction with the existing literature. The experimental results were obtained
with different techniques, namely detection with silicon detectors, the TWIST spectrometer or the activation method. The first two methods cover an
energy range specific for each experiment, as indicated in the table.  

The activation method reconstructs the emission rate from the number of radioactive daughter nuclei produced in muon capture. As not all potential daughters can be detected, it only establishes lower limits on the inclusive rate. This method also lacks the PID resolving capabilities of AlCap and TWIST which creates ambiguities for example between proton-after-neutron and deuteron emission. Theoretical calculations of inclusive proton and deuteron yields~\cite{Lifshitz:1980gi} can be compared with experiment. 

The TWIST and AlCap results agree and so confirm that the 
charged particle yield in Al is more than a factor 2 smaller than in Si. This is important good news for the muon-to-electron conversion experiments, which have previously
based their background models on Si.
The AlCap yield in the active-target Si measurement agrees within uncertainties with Ref.\cite{Sobottka1968}. 
The lower limits from the activation experiments are respected, but they are rather weak. Theory does remarkably well. However, the calculations were based on partially phenomenological models of the 1970s, and so it would be interesting to explore whether this agreement survives scrutiny with modern many-body methods,
like the shell model and random phase approximation. These theoretical frameworks have been employed for muon capture~\cite{Simkovic2020}  and neutrino reactions~\cite{Suzuki2018}, the latter including charged particle
emission, to study the $g_A$ quenching problem, double beta decay matrix elements and neutrino detection.

\subsection{Energy spectra}

\begin{figure}[!htbp]
  \centering
 \includegraphics[width=1.0\columnwidth]{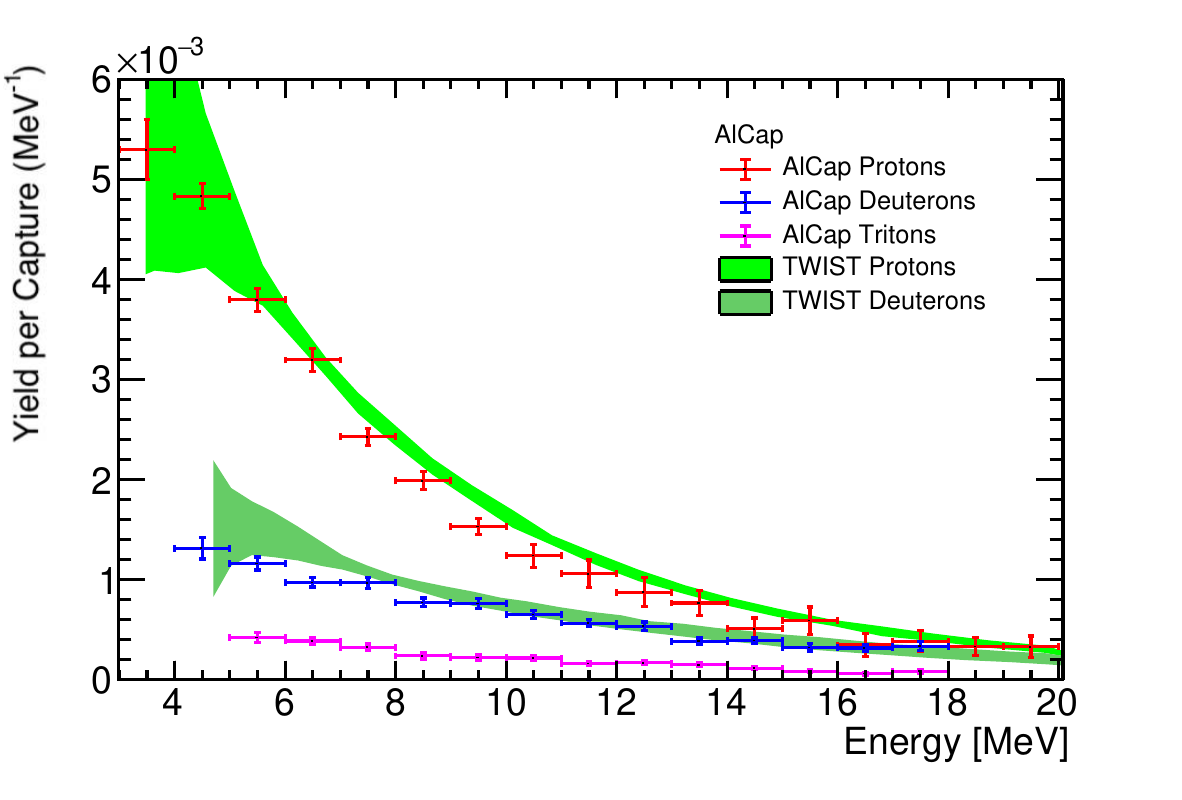}
  \caption{Energy spectra of protons, deuterons and tritons from the AlCap experiment (points) and TWIST experiment (shaded regions). AlCap overall systematic uncertainties due to detector misalignment and detector placement, typically 2.6\% for protons and deuterons (see Table~\ref{tab:systematics}) are not included.}
  \label{fig:discussion:twist-comparison}
\end{figure}

\begin{figure}[!htbp]
  \centering
\includegraphics[width=0.5\textwidth]{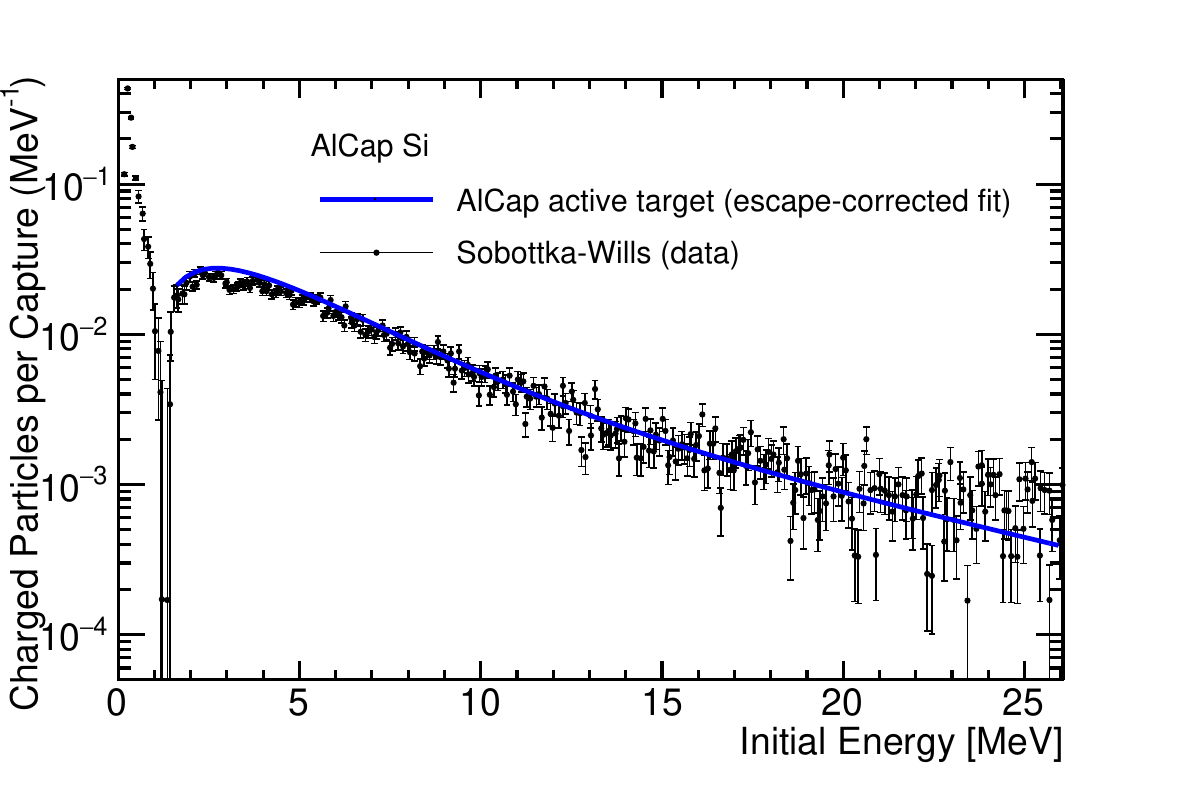}
  \caption{Comparison between AlCap's active-target analysis and the result of Sobottka and Wills~\cite{Sobottka1968} with statistical error bars added as inferred from their paper.}
 \label{fig:active-results-litcomp}
\end{figure}

A few AlCap energy spectra can be compared to other measurements. Figure~\ref{fig:discussion:twist-comparison}  demonstrates that the proton and deuteron spectra agree well between TWIST and AlCap. The AlCap measurements are more precise in the low energy
region relevant for muon-to-electron conversion experiments, while the TWIST measurement extends to higher energies than shown here.

The high-energy ($E>40$~MeV) charged particle spectrum measured by Krane et. al.~\cite{Krane1979} showed an exponential decay constant of 7.5(4)~MeV, which they describe as "mainly protons". 

In Fig.~\ref{fig:active-results-litcomp}, AlCap active-target spectrum is compared to the Sobottka-Wills measurement~\cite{Sobottka1968}. In this figure, statistical error bars have been added to the previous results as inferred from their paper. 
The previous experiment used a 3-mm thick active Si target which reduces the escape corrections compared to AlCap's  1.5-mm thick target. Note that a recoil correction is not mentioned; presumably this correction was not made. Regardless, the AlCap active-target result is consistent with the majority of the spectrum.
As regards the low energy region, both experiments indicate a drop-off towards the Coulomb barrier energy. However, the earlier paper mentions that the statistical uncertainty at the threshold energy is large, and AlCap cannot determine the threshold parameters precisely because of systematic uncertainties incurred from subtracting decay-electron background.


\section{Summary}
\label{sec:summary}
AlCap provides the most comprehensive set of measurements on yields and spectra of low-energy protons, deuterons, tritons and $\alpha$-particles from muon capture on aluminum, titanium, and silicon. 

The experimental program was driven by the need to quantify standard model background interfering in searches for ultra-rare muon-to-electron conversion events
predicted by extension of the standard model.
The new generation of such experiments, COMET and Mu2e,
have already used preliminary results from AlCap in their simulations. As these results were lower than had previously been assumed, COMET decided to forgo a proton absorber for their Phase-I experiment, and the proton absorber for Mu2e will be thinner and shorter than originally planned. Both measures will improve the energy resolution of their detectors.

We have made the first measurement of the heavy charged particle spectra from titanium. This is a promising stopping target material that could be used in those experiments
or in future generations of charged lepton flavor violation searches. 
We have also made the first measurements of triton emission after nuclear muon capture. This measurement for silicon is useful for direct dark matter experiments which use silicon detectors to search for rare dark matter signals~\cite{Saldanha:2020ubf}. Cosmic ray muons captured in the detectors can produce tritons, which then beta-decay and are a source of background.

We hope that the new data of muon capture with charged particle emission will stimulate renewed theoretical interest as a test of modern theoretical
frameworks and as consistency checks for related neutrino-nucleus reactions.


\section{Acknowledgments}
\label{sec:ack}
The collaboration would like to thank the PSI scientific and support staff for their help during data collection, as well as Frederik Wauters and Ran Hong.

The work was supported in part by the Japan Society for the Promotion of Science (JSPS) KAKENHI Grant Nos. 25000004 and 18H0523; %
the  US  Department  of  Energy  Office  of  Science,  Office  of  Nuclear  Physics  under  Award  No.  DE-FG02-97ER41020;
 the U.S. Department of Energy, Office of Science, Office of High Energy Physics  under Contract No. DE-AC02-05CH11231; 
  and, the Science and Technology Facilities Council, UK.

\appendix
\section{Data Tables}
In this appendix we provide the data included in Fig.~\ref{fig:results} with the statistical and systematic bin uncertainties separated out.

\begin{table*}
\caption{Particle yields, $Y^{i}_{\eta}$, for particles emitted after nuclear muon capture on Al. The energies quoted are the bin centers. The first error is the statistical uncertainty and the second error is the systematic uncertainty. Overall systematic uncertainties due to detector placement and misalignment (see Table~\ref{tab:systematics}) are not included.}
\label{tab:data-Al}
\begin{center}
\begin{tabular}{S[table-format=3.2]S[table-format=3.2]S[table-format=3.2]S[table-format=3.2]S[table-format=3.2]}
\hline
\hline
\multicolumn{1}{c}{\multirow{2}{*}{$E_{i}$[MeV]}} & \multicolumn{1}{c}{$Y^{i}_{p}$}& \multicolumn{1}{c}{$Y^{i}_{d}$}& \multicolumn{1}{c}{$Y^{i}_{t}$}& \multicolumn{1}{c}{$Y^{i}_{\alpha}$}\\
 & \multicolumn{4}{c}{[per thousand muon captures per MeV]} \\
\hline
3.75  &~~$5.27 \pm 0.09 \pm 0.25$~~& \multicolumn{1}{c}{no result} & \multicolumn{1}{c}{no result} & \multicolumn{1}{c}{no result} \\
4.25  &~~$4.99 \pm 0.09 \pm 0.13$~~& \multicolumn{1}{c}{no result} & \multicolumn{1}{c}{no result} & \multicolumn{1}{c}{no result} \\
4.75  &~~$4.55 \pm 0.09 \pm 0.13$~~ &~~$1.27 \pm 0.06 \pm 0.03$~~& \multicolumn{1}{c}{no result} & \multicolumn{1}{c}{no result} \\
5.25  &~~$3.99 \pm 0.09 \pm 0.14$~~ &~~$1.24 \pm 0.04 \pm 0.05$~~ &~~$0.43 \pm 0.03 \pm 0.04$~~& \multicolumn{1}{c}{no result} \\
5.75  &~~$3.64 \pm 0.09 \pm 0.08$~~ &~~$1.10 \pm 0.04 \pm 0.07$~~ &~~$0.45 \pm 0.02 \pm 0.03$~~& \multicolumn{1}{c}{no result} \\
6.25  &~~$3.30 \pm 0.08 \pm 0.11$~~ &~~$0.99 \pm 0.04 \pm 0.03$~~ &~~$0.41 \pm 0.02 \pm 0.02$~~& \multicolumn{1}{c}{no result} \\
6.75  &~~$2.90 \pm 0.08 \pm 0.11$~~ &~~$0.97 \pm 0.04 \pm 0.07$~~ &~~$0.37 \pm 0.02 \pm 0.04$~~& \multicolumn{1}{c}{no result} \\
7.25  &~~$2.56 \pm 0.07 \pm 0.09$~~ &~~$0.99 \pm 0.04 \pm 0.06$~~ &~~$0.34 \pm 0.02 \pm 0.02$~~& \multicolumn{1}{c}{no result} \\
7.75  &~~$2.35 \pm 0.07 \pm 0.05$~~ &~~$0.94 \pm 0.04 \pm 0.06$~~ &~~$0.31 \pm 0.02 \pm 0.03$~~& \multicolumn{1}{c}{no result} \\
8.25  &~~$2.14 \pm 0.07 \pm 0.07$~~ &~~$0.80 \pm 0.04 \pm 0.02$~~ &~~$0.25 \pm 0.02 \pm 0.04$~~& \multicolumn{1}{c}{no result} \\
8.75  &~~$1.87 \pm 0.07 \pm 0.19$~~ &~~$0.75 \pm 0.04 \pm 0.03$~~ &~~$0.22 \pm 0.02 \pm 0.03$~~& \multicolumn{1}{c}{no result} \\
9.25  &~~$1.51 \pm 0.06 \pm 0.09$~~ &~~$0.78 \pm 0.04 \pm 0.05$~~ &~~$0.22 \pm 0.02 \pm 0.02$~~& \multicolumn{1}{c}{no result} \\
9.75  &~~$1.52 \pm 0.07 \pm 0.05$~~ &~~$0.75 \pm 0.04 \pm 0.04$~~ &~~$0.21 \pm 0.02 \pm 0.02$~~& \multicolumn{1}{c}{no result} \\
10.25  &~~$1.30 \pm 0.07 \pm 0.09$~~ &~~$0.69 \pm 0.04 \pm 0.04$~~ &~~$0.22 \pm 0.02 \pm 0.03$~~& \multicolumn{1}{c}{no result} \\
10.75  &~~$1.27 \pm 0.15 \pm 0.12$~~ &~~$0.61 \pm 0.03 \pm 0.02$~~ &~~$0.21 \pm 0.02 \pm 0.03$~~& \multicolumn{1}{c}{no result} \\
11.25  &~~$1.08 \pm 0.15 \pm 0.08$~~ &~~$0.58 \pm 0.03 \pm 0.02$~~ &~~$0.17 \pm 0.02 \pm 0.02$~~& \multicolumn{1}{c}{no result} \\
11.75  &~~$1.12 \pm 0.15 \pm 0.12$~~ &~~$0.56 \pm 0.03 \pm 0.04$~~ &~~$0.15 \pm 0.02 \pm 0.02$~~& \multicolumn{1}{c}{no result} \\
12.25  &~~$0.84 \pm 0.13 \pm 0.12$~~ &~~$0.53 \pm 0.03 \pm 0.04$~~ &~~$0.16 \pm 0.02 \pm 0.02$~~& \multicolumn{1}{c}{no result} \\
12.75  &~~$0.89 \pm 0.13 \pm 0.12$~~ &~~$0.54 \pm 0.03 \pm 0.04$~~ &~~$0.18 \pm 0.02 \pm 0.02$~~& \multicolumn{1}{c}{no result} \\
13.25  &~~$0.76 \pm 0.11 \pm 0.14$~~ &~~$0.41 \pm 0.03 \pm 0.08$~~ &~~$0.16 \pm 0.02 \pm 0.02$~~& \multicolumn{1}{c}{no result} \\
13.75  &~~$0.76 \pm 0.11 \pm 0.17$~~ &~~$0.36 \pm 0.03 \pm 0.02$~~ &~~$0.14 \pm 0.02 \pm 0.01$~~& \multicolumn{1}{c}{no result} \\
14.25  &~~$0.51 \pm 0.07 \pm 0.13$~~ &~~$0.39 \pm 0.03 \pm 0.05$~~ &~~$0.12 \pm 0.01 \pm 0.01$~~& \multicolumn{1}{c}{no result} \\
14.75  &~~$0.53 \pm 0.08 \pm 0.15$~~ &~~$0.40 \pm 0.03 \pm 0.02$~~ &~~$0.11 \pm 0.01 \pm 0.01$~~& \multicolumn{1}{c}{no result} \\
15.25  &~~$0.64 \pm 0.12 \pm 0.07$~~ &~~$0.35 \pm 0.03 \pm 0.06$~~ &~~$0.09 \pm 0.01 \pm 0.02$~~ &~~$0.20 \pm 0.01 \pm 0.10$~~\\
15.75  &~~$0.55 \pm 0.11 \pm 0.19$~~ &~~$0.31 \pm 0.03 \pm 0.03$~~ &~~$0.08 \pm 0.01 \pm 0.03$~~ &~~$0.16 \pm 0.01 \pm 0.08$~~\\
16.25  &~~$0.37 \pm 0.10 \pm 0.21$~~ &~~$0.35 \pm 0.03 \pm 0.03$~~ &~~$0.08 \pm 0.01 \pm 0.02$~~ &~~$0.13 \pm 0.01 \pm 0.08$~~\\
16.75  &~~$0.33 \pm 0.12 \pm 0.04$~~ &~~$0.26 \pm 0.03 \pm 0.03$~~ &~~$0.06 \pm 0.01 \pm 0.03$~~ &~~$0.10 \pm 0.01 \pm 0.07$~~\\
17.25  &~~$0.37 \pm 0.12 \pm 0.13$~~& \multicolumn{1}{c}{no result} & \multicolumn{1}{c}{no result}  &~~$0.07 \pm 0.01 \pm 0.04$~~\\
17.75  &~~$0.40 \pm 0.10 \pm 0.06$~~& \multicolumn{1}{c}{no result} & \multicolumn{1}{c}{no result}  &~~$0.06 \pm 0.01 \pm 0.03$~~\\
18.25  &~~$0.33 \pm 0.08 \pm 0.07$~~& \multicolumn{1}{c}{no result} & \multicolumn{1}{c}{no result}  &~~$0.05 \pm 0.01 \pm 0.01$~~\\
18.75  &~~$0.33 \pm 0.09 \pm 0.08$~~& \multicolumn{1}{c}{no result} & \multicolumn{1}{c}{no result}  &~~$0.05 \pm 0.01 \pm 0.01$~~\\
19.25  &~~$0.33 \pm 0.08 \pm 0.06$~~& \multicolumn{1}{c}{no result} & \multicolumn{1}{c}{no result}  &~~$0.05 \pm 0.01 \pm 0.02$~~\\
19.75  &~~$0.33 \pm 0.08 \pm 0.07$~~& \multicolumn{1}{c}{no result} & \multicolumn{1}{c}{no result}  &~~$0.06 \pm 0.01 \pm 0.01$~~\\
\end{tabular}
\end{center}
\end{table*}

\begin{table*}
\caption{Particle yields, $Y^{i}_{\eta}$, for particles emitted after nuclear muon capture on Si. The energies quoted are the bin centers. The first error is the statistical uncertainty and the second error is the systematic uncertainty. Overall systematic uncertainties due to detector placement and misalignment (see Table~\ref{tab:systematics}) are not included.}
\label{tab:data-Si}
\begin{center}
\begin{tabular}{S[table-format=3.2]S[table-format=3.2]S[table-format=3.2]S[table-format=3.2]S[table-format=3.2]}
\hline
\hline
\multicolumn{1}{c}{\multirow{2}{*}{$E_{i}$[MeV]}} & \multicolumn{1}{c}{$Y^{i}_{p}$}& \multicolumn{1}{c}{$Y^{i}_{d}$}& \multicolumn{1}{c}{$Y^{i}_{t}$}& \multicolumn{1}{c}{$Y^{i}_{\alpha}$}\\
 & \multicolumn{4}{c}{[per thousand muon captures per MeV]} \\
\hline
3.75 & \multicolumn{1}{c}{no result} & \multicolumn{1}{c}{no result} & \multicolumn{1}{c}{no result} & \multicolumn{1}{c}{no result} \\
4.25  &~~$15.79 \pm 0.50 \pm 0.97$~~& \multicolumn{1}{c}{no result} & \multicolumn{1}{c}{no result} & \multicolumn{1}{c}{no result} \\
4.75  &~~$12.31 \pm 0.43 \pm 0.42$~~& \multicolumn{1}{c}{no result} & \multicolumn{1}{c}{no result} & \multicolumn{1}{c}{no result} \\
5.25  &~~$10.08 \pm 0.38 \pm 0.27$~~ &~~$1.81 \pm 0.12 \pm 0.21$~~& \multicolumn{1}{c}{no result} & \multicolumn{1}{c}{no result} \\
5.75  &~~$9.01 \pm 0.36 \pm 0.08$~~ &~~$1.41 \pm 0.11 \pm 0.06$~~& \multicolumn{1}{c}{no result} & \multicolumn{1}{c}{no result} \\
6.25  &~~$7.63 \pm 0.33 \pm 0.32$~~ &~~$1.41 \pm 0.11 \pm 0.09$~~ &~~$0.31 \pm 0.04 \pm 0.02$~~& \multicolumn{1}{c}{no result} \\
6.75  &~~$5.95 \pm 0.29 \pm 0.17$~~ &~~$1.45 \pm 0.12 \pm 0.24$~~ &~~$0.28 \pm 0.04 \pm 0.02$~~& \multicolumn{1}{c}{no result} \\
7.25  &~~$5.39 \pm 0.28 \pm 0.25$~~ &~~$1.10 \pm 0.10 \pm 0.07$~~ &~~$0.32 \pm 0.05 \pm 0.06$~~& \multicolumn{1}{c}{no result} \\
7.75  &~~$5.12 \pm 0.27 \pm 0.11$~~ &~~$0.95 \pm 0.10 \pm 0.05$~~ &~~$0.30 \pm 0.05 \pm 0.04$~~& \multicolumn{1}{c}{no result} \\
8.25  &~~$4.10 \pm 0.24 \pm 0.32$~~ &~~$0.90 \pm 0.09 \pm 0.07$~~ &~~$0.23 \pm 0.04 \pm 0.01$~~& \multicolumn{1}{c}{no result} \\
8.75  &~~$3.55 \pm 0.23 \pm 0.07$~~ &~~$0.91 \pm 0.10 \pm 0.09$~~ &~~$0.23 \pm 0.04 \pm 0.02$~~& \multicolumn{1}{c}{no result} \\
9.25  &~~$3.14 \pm 0.21 \pm 0.08$~~ &~~$0.79 \pm 0.09 \pm 0.03$~~ &~~$0.24 \pm 0.04 \pm 0.03$~~& \multicolumn{1}{c}{no result} \\
9.75  &~~$2.55 \pm 0.19 \pm 0.22$~~ &~~$0.85 \pm 0.09 \pm 0.08$~~ &~~$0.19 \pm 0.04 \pm 0.04$~~& \multicolumn{1}{c}{no result} \\
10.25  &~~$2.17 \pm 0.18 \pm 0.11$~~ &~~$0.79 \pm 0.09 \pm 0.06$~~ &~~$0.12 \pm 0.03 \pm 0.02$~~& \multicolumn{1}{c}{no result} \\
10.75  &~~$2.23 \pm 0.18 \pm 0.05$~~ &~~$0.72 \pm 0.09 \pm 0.06$~~ &~~$0.11 \pm 0.03 \pm 0.01$~~& \multicolumn{1}{c}{no result} \\
11.25  &~~$1.90 \pm 0.16 \pm 0.27$~~ &~~$0.79 \pm 0.09 \pm 0.07$~~ &~~$0.13 \pm 0.03 \pm 0.01$~~& \multicolumn{1}{c}{no result} \\
11.75  &~~$1.35 \pm 0.14 \pm 0.34$~~ &~~$0.86 \pm 0.10 \pm 0.10$~~ &~~$0.11 \pm 0.03 \pm 0.02$~~& \multicolumn{1}{c}{no result} \\
12.25  &~~$1.75 \pm 0.16 \pm 0.09$~~ &~~$0.61 \pm 0.08 \pm 0.16$~~ &~~$0.07 \pm 0.03 \pm 0.02$~~& \multicolumn{1}{c}{no result} \\
12.75  &~~$1.52 \pm 0.15 \pm 0.18$~~ &~~$0.57 \pm 0.08 \pm 0.08$~~ &~~$0.08 \pm 0.03 \pm 0.04$~~& \multicolumn{1}{c}{no result} \\
13.25  &~~$1.12 \pm 0.13 \pm 0.14$~~ &~~$0.48 \pm 0.08 \pm 0.09$~~ &~~$0.14 \pm 0.04 \pm 0.06$~~& \multicolumn{1}{c}{no result} \\
13.75  &~~$1.00 \pm 0.12 \pm 0.12$~~ &~~$0.60 \pm 0.09 \pm 0.05$~~ &~~$0.08 \pm 0.03 \pm 0.06$~~& \multicolumn{1}{c}{no result} \\
14.25  &~~$0.92 \pm 0.11 \pm 0.22$~~ &~~$0.54 \pm 0.08 \pm 0.08$~~ &~~$0.09 \pm 0.03 \pm 0.07$~~& \multicolumn{1}{c}{no result} \\
14.75  &~~$1.01 \pm 0.12 \pm 0.09$~~ &~~$0.44 \pm 0.08 \pm 0.08$~~ &~~$0.15 \pm 0.04 \pm 0.05$~~& \multicolumn{1}{c}{no result} \\
15.25  &~~$0.91 \pm 0.11 \pm 0.09$~~ &~~$0.51 \pm 0.08 \pm 0.06$~~ &~~$0.08 \pm 0.03 \pm 0.05$~~ &~~$0.29 \pm 0.03 \pm 0.18$~~\\
15.75  &~~$0.72 \pm 0.10 \pm 0.30$~~ &~~$0.46 \pm 0.08 \pm 0.05$~~ &~~$0.06 \pm 0.03 \pm 0.03$~~ &~~$0.24 \pm 0.03 \pm 0.06$~~\\
16.25  &~~$0.40 \pm 0.10 \pm 0.29$~~ &~~$0.39 \pm 0.07 \pm 0.05$~~ &~~$0.04 \pm 0.02 \pm 0.03$~~ &~~$0.18 \pm 0.02 \pm 0.05$~~\\
16.75  &~~$0.71 \pm 0.14 \pm 0.18$~~ &~~$0.28 \pm 0.06 \pm 0.08$~~ &~~$0.05 \pm 0.02 \pm 0.02$~~ &~~$0.13 \pm 0.02 \pm 0.03$~~\\
17.25  &~~$0.44 \pm 0.09 \pm 0.24$~~& \multicolumn{1}{c}{no result} & \multicolumn{1}{c}{no result}  &~~$0.09 \pm 0.02 \pm 0.03$~~\\
17.75  &~~$0.55 \pm 0.10 \pm 0.19$~~& \multicolumn{1}{c}{no result} & \multicolumn{1}{c}{no result}  &~~$0.06 \pm 0.01 \pm 0.02$~~\\
18.25  &~~$0.63 \pm 0.11 \pm 0.04$~~& \multicolumn{1}{c}{no result} & \multicolumn{1}{c}{no result}  &~~$0.05 \pm 0.01 \pm 0.01$~~\\
18.75  &~~$0.50 \pm 0.09 \pm 0.12$~~& \multicolumn{1}{c}{no result} & \multicolumn{1}{c}{no result}  &~~$0.05 \pm 0.01 \pm 0.00$~~\\
19.25  &~~$0.48 \pm 0.09 \pm 0.05$~~& \multicolumn{1}{c}{no result} & \multicolumn{1}{c}{no result}  &~~$0.05 \pm 0.01 \pm 0.01$~~\\
19.75  &~~$0.47 \pm 0.09 \pm 0.06$~~& \multicolumn{1}{c}{no result} & \multicolumn{1}{c}{no result}  &~~$0.05 \pm 0.01 \pm 0.01$~~\\
\end{tabular}
\end{center}
\end{table*}

\begin{table*}
\caption{Particle yields, $Y^{i}_{\eta}$, for particles emitted after nuclear muon capture on Ti. The energies quoted are the bin centers. The first error is the statistical uncertainty and the second error is the systematic uncertainty. Overall systematic uncertainties due to detector placement and misalignment (see Table~\ref{tab:systematics}) are not included.}
\label{tab:data-Ti}
\begin{center}
\begin{tabular}{S[table-format=3.2]S[table-format=3.2]S[table-format=3.2]S[table-format=3.2]S[table-format=3.2]}
\hline
\hline
\multicolumn{1}{c}{\multirow{2}{*}{$E_{i}$[MeV]}} & \multicolumn{1}{c}{$Y^{i}_{p}$}& \multicolumn{1}{c}{$Y^{i}_{d}$}& \multicolumn{1}{c}{$Y^{i}_{t}$}& \multicolumn{1}{c}{$Y^{i}_{\alpha}$}\\
 & \multicolumn{4}{c}{[per thousand muon captures per MeV]} \\
\hline
3.75  &~~$5.72 \pm 0.15 \pm 0.35$~~& \multicolumn{1}{c}{no result} & \multicolumn{1}{c}{no result} & \multicolumn{1}{c}{no result} \\
4.25  &~~$5.34 \pm 0.14 \pm 0.13$~~& \multicolumn{1}{c}{no result} & \multicolumn{1}{c}{no result} & \multicolumn{1}{c}{no result} \\
4.75  &~~$4.92 \pm 0.14 \pm 0.10$~~ &~~$0.62 \pm 0.06 \pm 0.10$~~& \multicolumn{1}{c}{no result} & \multicolumn{1}{c}{no result} \\
5.25  &~~$4.41 \pm 0.13 \pm 0.06$~~ &~~$0.69 \pm 0.04 \pm 0.03$~~ &~~$0.14 \pm 0.03 \pm 0.03$~~& \multicolumn{1}{c}{no result} \\
5.75  &~~$3.94 \pm 0.13 \pm 0.11$~~ &~~$0.62 \pm 0.04 \pm 0.03$~~ &~~$0.18 \pm 0.02 \pm 0.02$~~& \multicolumn{1}{c}{no result} \\
6.25  &~~$3.51 \pm 0.12 \pm 0.15$~~ &~~$0.54 \pm 0.04 \pm 0.03$~~ &~~$0.19 \pm 0.02 \pm 0.02$~~& \multicolumn{1}{c}{no result} \\
6.75  &~~$3.08 \pm 0.12 \pm 0.15$~~ &~~$0.49 \pm 0.04 \pm 0.04$~~ &~~$0.20 \pm 0.02 \pm 0.01$~~& \multicolumn{1}{c}{no result} \\
7.25  &~~$2.59 \pm 0.11 \pm 0.18$~~ &~~$0.57 \pm 0.04 \pm 0.08$~~ &~~$0.19 \pm 0.02 \pm 0.02$~~& \multicolumn{1}{c}{no result} \\
7.75  &~~$2.18 \pm 0.10 \pm 0.15$~~ &~~$0.64 \pm 0.05 \pm 0.08$~~ &~~$0.18 \pm 0.02 \pm 0.02$~~& \multicolumn{1}{c}{no result} \\
8.25  &~~$2.12 \pm 0.10 \pm 0.21$~~ &~~$0.57 \pm 0.05 \pm 0.06$~~ &~~$0.18 \pm 0.02 \pm 0.02$~~& \multicolumn{1}{c}{no result} \\
8.75  &~~$1.63 \pm 0.09 \pm 0.18$~~ &~~$0.45 \pm 0.04 \pm 0.06$~~ &~~$0.15 \pm 0.02 \pm 0.03$~~& \multicolumn{1}{c}{no result} \\
9.25  &~~$1.48 \pm 0.09 \pm 0.06$~~ &~~$0.43 \pm 0.04 \pm 0.05$~~ &~~$0.13 \pm 0.02 \pm 0.03$~~& \multicolumn{1}{c}{no result} \\
9.75  &~~$1.30 \pm 0.09 \pm 0.18$~~ &~~$0.46 \pm 0.04 \pm 0.05$~~ &~~$0.12 \pm 0.02 \pm 0.04$~~& \multicolumn{1}{c}{no result} \\
10.25  &~~$1.03 \pm 0.08 \pm 0.07$~~ &~~$0.41 \pm 0.04 \pm 0.05$~~ &~~$0.12 \pm 0.02 \pm 0.02$~~& \multicolumn{1}{c}{no result} \\
10.75  &~~$1.03 \pm 0.17 \pm 0.07$~~ &~~$0.34 \pm 0.04 \pm 0.04$~~ &~~$0.12 \pm 0.02 \pm 0.04$~~& \multicolumn{1}{c}{no result} \\
11.25  &~~$0.99 \pm 0.16 \pm 0.07$~~ &~~$0.34 \pm 0.04 \pm 0.05$~~ &~~$0.08 \pm 0.02 \pm 0.02$~~& \multicolumn{1}{c}{no result} \\
11.75  &~~$0.89 \pm 0.15 \pm 0.05$~~ &~~$0.32 \pm 0.04 \pm 0.03$~~ &~~$0.08 \pm 0.02 \pm 0.01$~~& \multicolumn{1}{c}{no result} \\
12.25  &~~$0.82 \pm 0.15 \pm 0.11$~~ &~~$0.30 \pm 0.04 \pm 0.06$~~ &~~$0.08 \pm 0.02 \pm 0.01$~~& \multicolumn{1}{c}{no result} \\
12.75  &~~$0.64 \pm 0.13 \pm 0.08$~~ &~~$0.30 \pm 0.03 \pm 0.06$~~ &~~$0.09 \pm 0.02 \pm 0.02$~~& \multicolumn{1}{c}{no result} \\
13.25  &~~$0.66 \pm 0.13 \pm 0.12$~~ &~~$0.30 \pm 0.03 \pm 0.06$~~ &~~$0.09 \pm 0.02 \pm 0.02$~~& \multicolumn{1}{c}{no result} \\
13.75  &~~$0.74 \pm 0.14 \pm 0.04$~~ &~~$0.31 \pm 0.04 \pm 0.09$~~ &~~$0.09 \pm 0.02 \pm 0.02$~~& \multicolumn{1}{c}{no result} \\
14.25  &~~$0.55 \pm 0.11 \pm 0.17$~~ &~~$0.23 \pm 0.03 \pm 0.08$~~ &~~$0.09 \pm 0.02 \pm 0.03$~~& \multicolumn{1}{c}{no result} \\
14.75  &~~$0.41 \pm 0.09 \pm 0.08$~~ &~~$0.28 \pm 0.03 \pm 0.07$~~ &~~$0.06 \pm 0.02 \pm 0.03$~~& \multicolumn{1}{c}{no result} \\
15.25  &~~$0.44 \pm 0.12 \pm 0.09$~~ &~~$0.27 \pm 0.04 \pm 0.08$~~ &~~$0.03 \pm 0.01 \pm 0.02$~~ &~~$0.14 \pm 0.02 \pm 0.04$~~\\
15.75  &~~$0.48 \pm 0.13 \pm 0.08$~~ &~~$0.20 \pm 0.03 \pm 0.04$~~ &~~$0.04 \pm 0.01 \pm 0.02$~~ &~~$0.13 \pm 0.02 \pm 0.03$~~\\
16.25  &~~$0.42 \pm 0.12 \pm 0.28$~~ &~~$0.20 \pm 0.03 \pm 0.07$~~ &~~$0.05 \pm 0.01 \pm 0.02$~~ &~~$0.12 \pm 0.01 \pm 0.03$~~\\
16.75  &~~$0.21 \pm 0.11 \pm 0.13$~~ &~~$0.19 \pm 0.03 \pm 0.05$~~ &~~$0.03 \pm 0.01 \pm 0.01$~~ &~~$0.11 \pm 0.01 \pm 0.03$~~\\
17.25  &~~$0.27 \pm 0.13 \pm 0.14$~~& \multicolumn{1}{c}{no result} & \multicolumn{1}{c}{no result}  &~~$0.10 \pm 0.01 \pm 0.03$~~\\
17.75  &~~$0.33 \pm 0.11 \pm 0.10$~~& \multicolumn{1}{c}{no result} & \multicolumn{1}{c}{no result}  &~~$0.09 \pm 0.01 \pm 0.03$~~\\
18.25  &~~$0.33 \pm 0.11 \pm 0.15$~~& \multicolumn{1}{c}{no result} & \multicolumn{1}{c}{no result}  &~~$0.08 \pm 0.01 \pm 0.03$~~\\
18.75  &~~$0.17 \pm 0.08 \pm 0.08$~~& \multicolumn{1}{c}{no result} & \multicolumn{1}{c}{no result}  &~~$0.07 \pm 0.01 \pm 0.03$~~\\
19.25  &~~$0.33 \pm 0.11 \pm 0.25$~~& \multicolumn{1}{c}{no result} & \multicolumn{1}{c}{no result}  &~~$0.06 \pm 0.01 \pm 0.03$~~\\
19.75  &~~$0.38 \pm 0.13 \pm 0.03$~~& \multicolumn{1}{c}{no result} & \multicolumn{1}{c}{no result}  &~~$0.04 \pm 0.01 \pm 0.03$~~\\
\end{tabular}
\end{center}
\end{table*}


\end{document}